\documentstyle[12pt,aps,psfig,graphics]{revtex}
\tighten

\begin{document}

\title{Kinetic Theory Estimates for the Kolmogorov-Sinai Entropy, and the
Largest Lyapunov Exponents for Dilute, Hard-Ball Gases and for
Dilute, Random Lorentz Gases}

\author{H.\ van Beijeren, and R.\ van Zon}
\address{Institute for Theoretical Physics, University of Utrecht, \\
Princetonplein 5, 3584 CC Utrecht,
The Netherlands}
\author{J.\ R.\ Dorfman}
\address{Institute for Physical
Science and Technology, and Department of Physics, \\ University of Maryland,
College Park, MD, 20742, USA.}

\date{\today}

\maketitle

\newcommand{\percent}{\%}
\newcommand{\T}{\tiny T}
\newcommand{\average}[1]{\left\langle #1 \right\rangle}
\newcommand{\vect}[1]{\vec{#1}}
\newcommand{\vecd}[1]{\delta\vec{#1}}
\newcommand{\matr}[1]{{\bf{#1}}}
\newcommand{\eql}[1]{\label{#1}}
\newcommand{\figl}[1]{\label{fig:#1}}
\newcommand{\dd}[2]{\frac{\partial #1}{\partial #2}}
\newcommand{\ddh}[2]{\frac{d #1}{d #2}}
\newcommand{\ddt}[2]{\frac{\partial_{#1}#2}{\partial t}}
\newcommand{\fake}[2]{\left#2\begin{picture}(0,#1)\end{picture}\right.}
\newcommand{\dif}{\delta}
\newcommand{\myexp}[1]{\exp\left[#1\right]}
\newcommand{\mylog}[1]{\log\left[#1\right]}
\newcommand{\abslog}[1]{\log\left\|#1\right\|}
\newcommand{\vvrp}{\vec{r}^{\,\prime}}
\newcommand{\vvp}{\vec{v}^{\,\prime}}
\newcommand{\vvr}{\vec{r}}
\newcommand{\vv}{\vec{v}}
\newcommand{\vm}{\vec{m}}
\newcommand{\vp}{\vec{p}}
\newcommand{\vdrp}{\dif\vec{r}^{\,\prime}}
\newcommand{\vdvp}{\dif\vec{v}^{\,\prime}}
\newcommand{\vdr}{\dif\vec{r}}
\newcommand{\vdv}{\dif\vec{v}}
\newcommand{\vdV}{\dif\vec{V}}
\newcommand{\vdR}{\dif\vec{R}}
\newcommand{\vhat}[1]{\hat{#1}}
\newcommand{\blinp}[2]{#1\cdot#2}
\newcommand{\inp}[2]{(#1\cdot#2)}
\newcommand{\hv}{\vhat{v}}
\newcommand{\hw}{\vhat{w}}
\newcommand{\norm}[1]{\left\| #1\right\|}
\newcommand{\eq}[1]{Eq.~(\ref{#1})}
\newcommand{\fig}[1]{Fig.~(\ref{fig:#1})}
\newcommand{\identity}{\matr{1}}
\newcommand{\ho}{\mbox{h.o.}}
\newcommand{\Reg}{\gamma_R}
\newcommand{\Img}{\gamma_I}
\newcommand{\dg}{\delta\gamma}
\newcommand{\Redg}{\dg_R}
\newcommand{\Imdg}{\dg_I}
\newcommand{\dw}{\delta w}
\newcommand{\ro}{\mbox{\boldmath{$\rho$}}}
\newcommand{\F}{{\cal{F}}}
\newcommand{\ma}{{\bf{a}}}
\newcommand{\Eq}{Eq.}
\newcommand{\Eqs}{Eqs.}
\newcommand{\Fig}{Fig.}
\newcommand{\Figs}{Figs.}
\newcommand{\tilden}{\tilde n}
\newcommand{\hn}{\hat\sigma}
\newcommand{\he}{\hat{e}}
\newcommand{\realnumbers}{{{\mathrm I}\!{\mathrm R}}}
\newlength{\myha}
\newlength{\myhb}
\newlength{\myhc}
\renewcommand{\inp}[2]{   \settoheight{\myha}{$\displaystyle #1\big.$}
\settoheight{\myhb}{$\displaystyle #2\big.$}   \ifdim\myha>\myhb{
\left.\left\langle#1\big.\right|#2\big.\right\rangle    }\else{
\left\langle#1\big.\left|#2\big.\right\rangle\right.   }\fi}
\newcommand{\mel}[3]{   \settoheight{\myha}{$\displaystyle #1\big.$}
\settoheight{\myhb}{$\displaystyle #2\big.$}
\settoheight{\myhc}{$\displaystyle #3\big.$}   \ifdim\myha>\myhb{
\ifdim\myhc>\myha{         \left\langle#1\big.\left|#2\big.\left|
#3\big.\right\rangle\right.\right.      }\else{
\left.\left.\left\langle#1\big.
\right|#2\big.\right|#3\big.\right\rangle      }\fi   }\else{
\ifdim\myhc>\myhb{         \left\langle#1\big.\left|#2\big.\left|
#3\big.\right\rangle\right.\right.      }\else{
\left\langle#1\big.\left|#2\big.\right|         #3\big.\right\rangle      }\fi
 }\fi} \newcommand{\sfrac}[2]{\mbox{$\frac{#1}{#2}$}}
\newcommand{\ds}{\displaystyle}

\begin{abstract} The kinetic theory of gases provides methods for  calculating
Lyapunov exponents and other quantities, such as Kolmogorov-Sinai entropies,
that characterize the chaotic behavior of hard-ball gases. Here we illustrate
the use of these methods for calculating the Kolmogorov-Sinai entropy, and the
largest positive Lyapunov exponent, for dilute hard-ball gases in equilibrium.
The calculation of the largest Lyapunov exponent makes interesting connections
with the theory of propagation of hydrodynamic fronts. Calculations are also
presented for the Lyapunov spectrum of dilute, random  Lorentz gases in two and
three dimensions, which are considerably simpler  than the corresponding
calculations for hard-ball gases. The article concludes with a brief discussion
of some interesting open problems.

\end{abstract}

\section{Introduction}

The purpose of this article is to demonstrate that familiar methods from the
kinetic theory of gases can be extended in order to provide good estimates for
the chaotic properties of dilute, hard-ball gases \footnote{We will use the
term hard-ball to denote hard core systems in any number of dimensions, rather
than using the term hard-disk for two dimensional, hard-core systems, etc.}.
The kinetic theory of gases has a long history, extending over a period of a
century and a half, and is responsible for many central insights into, and
results for the properties of gases, both in and out of thermodynamic
equilibrium~\cite{brush}. Strictly speaking, there are two  familiar versions
of kinetic theory, an informal version and a formal version. The informal
version is based upon very elementary considerations of the collisions suffered
by molecules in a gas, and upon elementary probabilistic notions regarding the
velocity and free path distributions of the molecules. In the hands of Maxwell,
Boltzmann, and others, the informal version of kinetic theory led to such
important predictions as the independence of the viscosity of a gas on its
density at low densities, and to qualitative results for the equilibrium
thermodynamic  properties, the transport coefficients, and the structure of
microscopic boundary layers in a dilute gas. The more formal theory is also due
to Maxwell and Boltzmann, and may be said to have had its beginning with the
development of the Boltzmann transport equation in 1872~\cite{boltz}.  At that
time Boltzmann obtained, by heuristic arguments, an equation for the time
dependence of the  spatial and velocity distribution function for particles in
the gas. This equation provided a formal foundation for the informal methods of
kinetic theory. It leads directly to the Maxwell-Boltzmann velocity
distribution for the gas in equilibrium. For non-equilibrium systems, the
Boltzmann equation leads to a version of the Second Law of Thermodynamics (the
Boltzmann H-theorem), as well as to the Navier-Stokes equations of fluid
dynamics, with explicit expressions for the transport coefficients in terms of
the intermolecular potentials governing the interactions between the particles
in the gas~\cite{chapcow}.  It is not an exaggeration to state that the kinetic
theory of gases was one of the  great successes of nineteenth century physics.
Even now, the Boltzmann equation  remains one of the main cornerstones of our
understanding of nonequilibrium processes in fluid as well as solid systems,
both classical and quantum mechanical. It continues to be a subject of
investigation in both the mathematical and physical literature, and its
predictions often serve as a way of distinguishing  different molecular models
employed to calculate gas properties. However, there is still not a rigorous
derivation of the Boltzmann equation that demonstrates its validity over the
long  time scales typically used in applications. Nevertheless, the Boltzmann
equation has been generalized to higher density, and in so far as they are
available, the predictions of the generalized Boltzmann equation are in accord
with experiments and with numerical simulations of the properties of moderately
dense gases~\cite{jrdhvb}.

In spite of the many successes of the Boltzmann equation it cannot be a strict
consequence of mechanics, since it is not time reversal invariant, as are the
equations of mechanics, and it does not exhibit other mechanical phenomena,
such as Poincar\'e recurrences~\cite{eheren}. Boltzmann  realized that the
equation has to be understood as representing the typical behavior of a gas, as
sampled from an ensemble of similarly prepared gases, rather than the exact
behavior of a particular laboratory gas. He also understood that the
fluctuations about the typical behavior should be very small, and not important
for laboratory experiments. To support his arguments, he developed the
foundations of statistical mechanics, introducing what we now call the
micro-canonical ensemble. This ensemble is described by giving all systems in
it a fixed total energy, and then assuming that the probability of finding a
system in a certain small region on the constant energy surface is proportional
to the dynamically invariant measure of the small region, given by the Lebesgue
measure of the region divided by the magnitude of the gradient of the
Hamiltonian function at the point of interest on the surface. As we know, this
micro-canonical ensemble forms the starting point for all statistical
calculations of the thermodynamic properties of fluid and many other systems.

In his attempt to provide a mechanical argument for the effectiveness of the
micro-canonical ensemble for calculating the thermodynamic properties of
fluids, Boltzmann formulated the ergodic hypothesis, which, in its modern form,
states that the time average of any Lebesgue-integrable dynamic quantity of an
isolated, many particle system approaches the ensemble average of  the
quantity, taken with respect to the micro-canonical ensemble~\cite{uhlfo}. This
hypothesis is the subject of several articles in this Volume, but its value for
the foundations of statistical thermodynamics is often questioned~\cite{lebo }.
The questionable points can be summarized in a few items: (1) The ergodic
hypothesis applies to classical systems while nature is fundamentally quantum
mechanical. (2) Even granting the approximate validity of classical mechanics
for many purposes, no laboratory system is truly isolated from the rest of the
universe. Instead, laboratory systems are constantly perturbed by outside
influences, and these sources of randomness, together with the simple  laws of
large numbers for systems with many degrees of freedom, may be responsible for
the  utility of the micro-canonical ensemble for the calculation of equilibrium
properties. (3) Even granting the ability to isolate a laboratory system from
the rest of the universe, the time it would take for a system's phase space
trajectory to sample all of the available phase space on the constant energy
surface is just too long for ergodic behavior to be a physically reasonable
explanation for the  effectiveness of the micro-canonical ensemble. It seems
more likely that the equilibrium behavior of a system of many particles depends
on a number of these factors, including perturbations from the environment, the
fact that thermodynamic systems have a large number of degrees of freedom, and
the fact that the physically relevant quantities are projections of phase space
quantities onto a subspace of a few dimensions. Thus even though the time scale
for ergodic behavior on the full phase space may  be unreasonably long, the
projected behavior on relevant subspaces may not take very much time for the
establishment of equilibrium values of the measurable quantities, such as
pressure, temperature, etc. Pending the further clarification of these and
similar issues, it seems fair to say that the complete understanding of the
reasons for the validity of the micro-canonical ensemble as the basic
understructure for statistical thermodynamics has yet to be achieved.

In addition to resolving the various issues  needed to complete our
understanding of the equilibrium behavior of fluids, we would also like to
understand the dynamics of the {\it approach} to thermodynamic equilibrium on
as deep a level as possible. Such an understanding would enable us to provide a
justification of the Boltzmann equation and its many generalizations, as well
as of the successful use of hydrodynamic or stochastic equations in
nonequilibrium situations. The counterpart of Boltzmann's ergodic hypothesis
for nonequilibrium phenomena is the assumption that a dynamical system should
be mixing, in the sense of Gibbs. That is, given some initial distribution of
points on the constant energy surface in phase space,  in a region of positive
measure, a system is mixing if the distribution of the points eventually
becomes uniform over the energy surface, with respect to the invariant
measure~\cite{arnav}. If one can prove that an isolated dynamical system with a
large number of degrees of freedom is mixing, then one can show that the phase
space distribution function for the system will approach an equilibrium
distribution at long times, and consequently, quantities averaged with respect
to this distribution function will approach their equilibrium values. Needless
to say, the same concerns listed above suggest that the approach to equilibrium
may be a very complicated affair with a number of possible factors acting in
concert or individually as circumstances require. One might also ask why a
phase space distribution function is needed at all, since a laboratory system
corresponds to a point in phase space at any given time, and not to a
distribution of points in phase space. The usual argument given in statistical
mechanics texts is that it is easier to describe the average behavior of an
ensemble of points than to solve the complete set of the equations of motion of
a single system, and to draw conclusions from such a solution. Of course,
computer simulations of fluid systems are often attempts to solve the equations
of motion for an individual system, but they too are influenced by noise in the
form of round-off errors, and so do not really describe an isolated dynamical
system, unless one resorts to certain lattice-type models that can be treated
by integer arithmetic on the computer. Our hope, often unstated, is that the
properties we explore using the methods of statistical mechanics are somehow
typical of the behavior of an individual system in the laboratory, even if we
know that this cannot be strictly true. We hope, and occasionally can prove,
that the deviations from typical behavior are small.

In any case, it is clearly important to know what role the dynamics of the
fluid system might play in the approach to equilibrium. Certainly the
equilibrium and nonequilibrium properties of the system are sensitive to the
underlying molecular structure of the fluid and  to the interactions between
the particles of which it is  composed\cite{gaspbook,jrdbook}. Our experience
with the Boltzmann equation also assures us that the role of molecular
collisions cannot be underestimated, even if we are not entirely certain why
the Boltzmann equation works for an individual laboratory system.  Therefore,
when faced with a plausible model of a fluid, one would like to know if the
dynamics of the model is ergodic, mixing, K, or Bernoulli. It is of
considerable interest to establish the dynamical properties of a large isolated
system of particles as the starting point for our investigation of the
foundations of nonequilibrium statistical mechanics, and then to look at the
consequences of: (a) external noise on the system, and (b) the restriction of
our interest to only a small class of functions of the dynamical variables
needed for physical applications.

As a step in this direction we show here that kinetic theory can be used to
demonstrate (not prove) that isolated hard-ball systems are chaotic dynamical
systems~\cite{hvbddp}. We will show, in fact, that, at low  densities at least,
hard-ball systems have a positive Kolmogorov-Sinai entropy, which we can
estimate, and that the largest positive Lyapunov exponent can also be
estimated. These estimates, in fact, are in good agreement with the results of
numerical simulations. What these estimates are unable to tell us is whether or
not the systems are indeed ergodic, mixing, K, or Bernoulli, since we cannot
use these kinetic theory techniques to show that the phase space consists of
only one invariant region and not a countable number of them. In fact there is
some evidence that if the intermolecular potential is not discontinuous but
smoother than a hard sphere potential, then there may be some elliptic islands
of positive measure  in the phase space\cite{RomKedar}. Strictly speaking, the
ergodic hypothesis is not valid for such potentials. It remains to be seen
whether or not this phenomenon is ultimately of some importance for statistical
mechanics, especially for systems with large numbers of particles, and at
temperatures where quantum effects may be neglected.

As a by-product of the analysis given here we will also be able to calculate
the largest Lyapunov exponents and Kolmogorov-Sinai Entropy for a dilute
Lorentz gas with one moving moving particle in an array of randomly placed (but
non-overlapping) fixed hard ball scatterers~\cite{vbd,vbld}. This system is
much easier to analyze than a  gas where all the particles are in motion, and
was the first type of  hard-ball system whose chaotic dynamics were studied in
detail, either by  rigorous methods or by kinetic theory.

The plan of this article is as follows:  In Section II we will set up the
equations of motion for hard-ball systems that will enable us to analyze the
separation of initially close (infinitesimally close, actually) trajectories in
phase space. The dynamics of the separation of trajectories in phase space is,
of course, an essential ingredient in analyzing Lyapunov exponents and
Kolmogorov-Sinai (KS) entropies. In Section III we will apply these results to
a calculation of the KS entropy of a hard-ball gas using informal kinetic
theory rather than a formal Boltzmann equation approach. This informal method
gives the leading density behavior of the KS entropy, but more formal methods
are needed to go further. We will outline the more formal method based upon an
extension of the Boltzmann equation, but we will not go too deeply into its
solution, since it rapidly becomes very technical. In Section IV we outline a
method for estimating the largest Lyapunov exponent for a dilute hard-ball gas
using a mean field theory based upon the Boltzmann equation\cite{myself,leid}.
In Section V we apply these methods, both informal and formal, to the
calculation of the KS entropy and largest Lyapunov exponents for a dilute
Lorentz gas with fixed hard-ball scatterers.  We conclude in Section VI with
remarks and a discussion of outstanding open problems.

\section{The dynamics of hard-ball systems}

In this section we will present a method due to Dellago, Posch and Hoover
\cite{Dellago} for describing the dynamical behavior of infinitesimally   close
trajectories in phase space for hard-ball systems. We begin with  a
consideration of a system of $N$ identical  hard balls in $d$  dimensions, each
of mass $m$, and diameter $a$. Their positions and  velocities are denoted by
$\vec{r}_i$ and $\vec{v}_i$, respectively,  where $i=1,2,\ldots N$ labels the
particles. For simplicity we can  imagine that the particles are all placed in
some cubical volume  $V=L^d$ and that periodic boundary conditions are applied
at the  faces of the cube. The dynamics consists of periods of free motion of
the particles  separated by instantaneous binary collisions between some pair
of  particles. During free motion, the equations of the system are
\begin{equation}
	\dot{\vec{r}}_i = \vec{v}_i ,\quad
	\dot{\vec{v}}_i = 0.
\end{equation}
At the instant of collision between particles $i$ and $j$, say, there is  an
instantaneous change in the velocities. It is convenient to write the  dynamics
in terms of the center of mass motion ($\vec R_{ij}$,$\vec  V_{ij}$) and
relative motion ($\vec r_{ij}$, $\vec v_{ij}$),
\begin{eqnarray*}
	\vec R_{ij} = (\vec r_i+\vec r_j)/2 &	,
	\quad&
	\vec V_{ij} = (\vec v_i+\vec v_j)/2
,
\\
	\vec r_{ij} = \vec r_i-\vec r_j &	,
	\quad&
	\vec v_{ij} = \vec v_i-\vec v_j .
\end{eqnarray*}
The change can be described by the equations
\begin{eqnarray*}
	\vec R_{ij}' = \vec R_{ij} &	,
	\quad&
	\vec V_{ij}' = \vec V_{ij}
,
\\
	\vec r_{ij}' = \vec r_{ij} \equiv a \hat\sigma &	,
	\quad&
	\vec v_{ij}' = \matr{M}_{\hat \sigma}\cdot\vec v_{ij},
\end{eqnarray*}
where the matrix $\matr{M}_{\hat\sigma}$ describes a specular reflection on a
plane with normal $\hat\sigma$, i.e., the unit vector in the direction from
particle $j$ to $i$ at the instant of  the $(i,j)$ collision,
\begin{equation}
	\matr{M}_{\hat\sigma} \equiv \identity - 2\hat\sigma\hat\sigma .
\label{eq:M}
\end{equation}
Nondotted products of vectors are dyadic products and $\identity$ is the
identity matrix.  In terms of the individual particle velocities, the  dynamics
is described by
\begin{eqnarray}
	\vec v_i' &=& \vec v_i - (\vec v_{ij}\cdot\hat\sigma)\hat\sigma
,
\nonumber\\
	\vec v_j' &=& \vec v_j + (\vec v_{ij}\cdot\hat\sigma)\hat\sigma .
\eql{vdynamics}
\end{eqnarray}
These equations (plus the  dynamics at the boundaries) are all that one needs
in order to determine the trajectory of the system in phase space. However, in
order to determine Lyapunov exponents and other chaotic properties of  the
system, we need to examine two infinitesimally close trajectories,  and obtain
the equations that govern their rate of separation.

Equations for the rate of separation of two phase space trajectories of
hard-ball systems have been developed by Sinai using differential
geometry\cite{gaspbook,sinai}. This leads to an expression for the rate of
separation in terms of an operator that is expressed as a continuous  fraction.
Here we adopt a somewhat different but equivalent approach that  is nicely
suited to kinetic theory calculations. To obtain the equations  we need, we
consider two infinitesimally close phase space points,  $\Gamma$ (the reference
point) and $\Gamma+\delta\Gamma$ (the adjacent point),  given by  $(\vec
r_1,\vec v_1,\vec r_2,\vec v_2,\ldots,\vec r_N,\vec v_N)$ and by $(\vec
r_1+\delta\vec r_1,\vec v_1+\delta\vec v_1,\vec r_2+\delta\vec  r_2,\vec
v_2+\delta\vec v_2,\ldots, \vec r_N+\delta\vec r_N,\vec  v_N+\delta\vec v_N)$,
respectively. The $2N$ infinitesimal deviation  vectors $\delta \vec r_i$,
$\delta \vec v_i$  describe the displacement in  phase space  between the two
trajectories. The velocity deviation vectors are not all independent since we
will restrict their values by  requiring that the total momentum and total
kinetic  energy of the two trajectories be the same. That is,
\begin{equation}
	\sum_{i=1}^{N} \delta \vec v_i=0 ,\quad
	\sum_{i=1}^{N} \vec v_i \cdot \delta \vec v_i = 0 .
\label{cons}
\end{equation}
where, in the energy equation, we have neglected second order terms in the
velocity deviations. We will not use these equations in any serious way since
we will be interested in the limit of large $N$, in which case they are not
very  important. In the case of the Lorentz gas, to be discussed in a later
section, the conservation of momentum equation is not relevant, but we will
require that both of the two trajectories have the same energy, so that the
velocity deviation vector is orthogonal to the velocity itself.

Between collisions on the displaced trajectory, the deviations satisfy
\begin{equation}
	\delta\dot{\vec r}_i
	= \delta\vec v_i 	,
	\quad
	\delta\dot{\vec v}_i = 0 .
\label{eq:ff}
\end{equation}
The treatment of the effect of the collisions on the deviation vectors is more
complicated. One assumes that the two trajectories  are so close together that
the same sequence of binary collisions takes  place on each trajectory over
arbitrarily long times. Let us suppose that  we consider a collision between
particle $i$ and $j$ on each trajectory.  Now, since the trajectories are
slightly displaced, the $(i,j)$ collision  will take place at slightly
different times on each trajectory. This  slight (in fact infinitesimal) time
displacement must be included in the  analysis of  the dynamical behavior of
the deviation vectors at a collision.

The dynamics of the deviations can be considered separately for the
center-of-mass coordinates and the relative coordinates. The center-of   mass
coordinates behave as in a free flight, so that
\[
	\delta \vec R_{ij}' = \delta \vec R_{ij} 	,
	\quad
	\delta \vec V_{ij}' = \delta \vec V_{ij} .
\]
For the relative coordinates, we consider the relative coordinates of two
infinitesimally close trajectories, $\vec r_{ij},\vec v_{ij}$ and $\vec
r^*_{ij},\vec v^*_{ij}$:  \begin{center}
\begin{tabular}{@{\ }l@{\ }||@{\ }l@{\ }}
	\hfil reference trajectory \hfil &
	\hfil adjacent trajectory  \hfil \\
	\hfil (collision at $t=0$) \hfil &
	\hfil (collision at $t=\delta t$) \hfil \\\hline
	\hspace{2in} & 	\hspace{2in} \vspace{-1em}\\
	$t<0$:		&
	$t<\delta t$:	\\
	$\quad\vec v_{ij}(t)$ constant &
	$\quad\vec v^*_{ij}(t)$ constant\\
	$\quad\vec r_{ij}(t) = \vec v_{ij} t + a \hat\sigma$ &
	$\quad\vec r^*_{ij}(t) = \vec v^*_{ij}(t-\delta t)
		+ a \hat\sigma^*$ \\
	\hspace{2in} & \hspace{2in} \vspace{-1em}\\
	$t>0$:	&
	$t>\delta t$: \\
	$\quad\vec v_{ij}(t) = \matr{M}_{\hat\sigma}\cdot\vec v_{ij}$ &
	$\quad\vec v^*_{ij}(t) = \matr{M}_{\hat\sigma^*}\cdot\vec v^*_{ij}$\\
	$\quad\vec r_{ij}(t) = \matr{M}_{\hat\sigma}\cdot\vec v_{ij}t
		+a \hat\sigma $ &
	$\quad\vec r^*_{ij}(t) = \matr{M}_{\hat\sigma^*}\cdot\vec v^*_{ij}
		(t-\delta t) + a \hat\sigma^*$
\end{tabular}
\end{center}  The transformations at a collision in terms of the deviation
vectors  are  found with the aid of
\begin{eqnarray*}
	\delta\vec r_{ij} = \vec r^*_{ij}(0^-)-\vec r_{ij}(0^-)
&,
\quad&
	\delta\vec v_{ij} = \vec v^*_{ij}(0^-)-\vec v_{ij}(0^-)
,
\\
	\delta\vec r_{ij}' = \vec r^*_{ij}(\delta t^+)
				-\vec r_{ij}(\delta t^+)
&,
\quad&
	\delta\vec v_{ij}' = \vec v^*_{ij}(\delta t^+)
				-\vec v_{ij}(\delta t^+) ,
\end{eqnarray*}
where the superscripts $^+$ and $^-$ indicate immediately after and immediately
 before the collision, respectively.  We should use Eq.~(\ref{eq:ff}) in
between collisions, i.e., from the last collision up to $t=0^-$. Then we use
the collision rule, to be derived in this section, that links $\delta\vec r'_i$
and $\delta\vec v'_i$ to $\delta\vec r_i$ and $\delta\vec v_i$.
Eq.~(\ref{eq:ff}) is used again from $t=\delta t^+$, starting with $\delta\vec
r'_i$ and $\delta\vec v_i'$. It is also possible to use Eq.~(\ref{eq:ff}) from
$t=0^+$, as if the values of $\delta\vec r_{ij}'$ and $\delta\vec v_{ij}'$ in
the above equation were valid at $t=0^+$. The difference is an erroneous
additional $\delta t\delta\vec v'_{ij}$ for $\delta\vec r_{ij}$, when we apply
Eq.~(\ref{eq:ff})  from $t=0^+$, but this additional term is quadratic in the
deviations and therefore negligible.  In this way, the collision  may also be
viewed as instantaneous for the deviation vectors.

We can now write
\[
	\delta \vec r_{ij} = \hat\sigma^* a
				- \vec v^*_{ij}\delta t
				- \hat\sigma a
			   = \delta\hat\sigma a - \vec v_{ij}\delta t,
\]
where $\delta\hat\sigma=\hat\sigma^*-\hat\sigma$, and in the last equality we
neglect  terms quadratic in the deviations.  Because
$|\hat\sigma|=|\hat\sigma^*|=1$, we have $\delta\hat\sigma\cdot\hat\sigma=0$.
Taking the  inner product with  $\hat\sigma$ of the above formula gives
\[
  \delta t = -\frac{\hat\sigma\cdot\delta r_{ij}}
	{\hat\sigma\cdot\vec v_{ij}} 	,
	\quad
  \delta\hat\sigma = \left[\frac{(\hat\sigma\cdot\vec v_{ij})\identity-
	\vec v_{ij}\hat\sigma}{a(\hat\sigma\cdot\vec v_{ij})}\right]\cdot
	\delta\vec r_{ij}.
\]
Substitution of these  results into $\delta\vec r_{ij}' =
\hat\sigma^*a-\hat\sigma a- \matr{M}_{\hat\sigma}\cdot\vec{v}_{ij}\,\delta t$
and $ \delta\vec v_{ij}'= \matr{M}_{\hat\sigma^*}\cdot \vec
v^*_{ij}-\matr{M}_{\hat\sigma}\cdot \vec  v_{ij}$, yields
\begin{eqnarray*}
	\delta\vec r_{ij}' &=& \matr{M}_{\hat\sigma}\cdot\delta \vec r_{ij},
\\
	\delta\vec v_{ij}' &=& -2
\matr{Q}_{\hat\sigma}(i,j)
		\cdot\delta
		\vec r_{ij}
		+ \matr{M}_{\hat\sigma}\cdot\delta \vec v_{ij},
\end{eqnarray*}
where  $\matr{Q}_{\hat\sigma}(i,j)$ is the matrix
\begin{equation}
	\matr{Q}_{\hat\sigma}(i,j)
	= \frac{[(\hat\sigma\cdot\vec v_{ij})\identity
			+\hat\sigma\vec v_{ij}]\cdot
			 [(\hat\sigma\cdot\vec v_{ij})\identity
			-\vec v_{ij}\hat\sigma]
	}{a(\hat\sigma\cdot\vec v_{ij})}
{}.
\label{eq:Q}
\end{equation}
In terms of the individual particle deviations, the collision dynamics  reads
\begin{eqnarray}
	\delta\vec r_i'&=&\delta\vec r_i
			-(\delta
			\vec r_{ij}\cdot\hat\sigma)\hat\sigma
,
\nonumber\\
	\delta\vec r_	j'&=&\delta\vec r_j
			+(\delta\vec r_{ij}\cdot\hat\sigma)\hat\sigma
,
\nonumber
\\
	\delta\vec v_i'&=&\delta\vec v_i
			-(\delta\vec v_{ij}\cdot\hat\sigma)\hat\sigma
			-
			\matr{Q}_{\hat\sigma}(i,j)
			\cdot\delta\vec r_{ij}
,
\nonumber\\
	\delta\vec v_			j'&=&\delta\vec v_j
			+(\delta\vec v_{ij}\cdot\hat\sigma)\hat\sigma
			+
			\matr{Q}_{\hat\sigma}(i,j)
			\cdot\delta\vec r_{ij}
{}.
\label{eq:dvd}
\end{eqnarray}

\Eqs~(\ref{eq:ff}), (\ref{eq:Q}) and (\ref{eq:dvd}) are the dynamical equations
that govern the time dependence of the deviation vectors, $\{\delta\vec
r_i,\delta\vec v_i \}$.  They have to be solved together with the equations for
the $\{\vec r_i,\vec v_i\}$ in order to have a completely determined system.
That is, in order to follow the deviation vectors in time, one needs to know
when, where, and with what velocities the various collisions take place in the
gas.

In the next section, we will use these equations to provide a first  estimate
of the Kolmogorov-Sinai entropy for a dilute gas of  hard balls in $d$
dimensions.

\section{Estimates of the Kolmogorov-Sinai Entropy for a Dilute Gas}

We consider  hard-ball system from the previous section when the  the gas is
dilute, {\it i.e.}  $na^{d}\ll 1$,  with $n$ the density $N/V$, and it is in
equilibrium with no external forces acting on it. Since the hard-ball system is
a conservative, Hamiltonian system, one can easily show that all nonzero
Lyapunov exponents are paired with a corresponding exponent of identical
magnitude but of opposite sign\cite{gaspbook,jrdbook,Ott}. Obviously there are
an equal number of positive and negative Lyapunov exponents, and the sum of all
the  Lyapunov exponents must be equal to zero. For the system we consider here,
the Kolmogorov-Sinai (KS) entropy is equal to the sum of the positive Lyapunov
exponents, by Pesin's theorem\cite{Pesin}. We will compute the KS entropy per
particle in the  thermodynamic limit, for a dilute hard-ball gas, using methods
of kinetic theory~\cite{hvbddp,dlvb}.

To carry out this calculation we will use the fact that  when an
infinitesimally small $2Nd$-dimensional volume in phase phase is projected onto
the $Nd$-dimensional subspace corresponding to  the velocity directions, the
volume of this projection must grow  exponentially with time $t$, in the long
time asymptotic limit, with an  exponent that is the sum of all of the positive
Lyapunov exponents.  That is,  when we denote this projected volume element by
$\delta{\cal{V}}_{v}(t)$, as $t\rightarrow \infty$,
\begin{equation}
\frac{\delta{\cal{V}}_{v}(t)}{\delta{\cal{V}}_{v}(0)}=\exp
\left\{t\,\sum_{\lambda_{i}>0}\lambda_{i}\right\}.
\label{ks1}
\end{equation}
The same result also holds for another small volume element, denoted by
${\delta\cal{V}}_{r}(t)$, that is the projection onto the  $Nd$-dimensional
subspace corresponding to the position directions\footnote{It is worth pointing
out that while both $\delta{\cal{V}}_v$ and $\delta{\cal{V}}_{r}$ grow
exponentially in time, their combined volume, i.e. the original
$2Nd$-dimensional volume,  stays constant. This seemingly paradoxical statement
can be understood by realizing that almost all projections of the
$2Nd$-dimensional volume onto $Nd$-dimensional subpaces will grow exponentially
in time with an exponent given by the sum of the largest $Nd$  Lyapunov
exponents.}. The advantage of using an element in velocity space resides in the
fact that the velocity deviation vectors do not change during the time
intervals between  collisions in the gas, but only at the instants of
collisions. We will make use of this fact shortly.

Our first step will be to obtain a general formula for the KS entropy of a
hard-ball system, which, in principle, should describe the complete density
dependence of this quantity. Then we will apply this result to the low density
case, using both informal and formal kinetic theory methods.

\subsection{The KS Entropy as an Ensemble Average}

Our object here is to express the KS entropy for a hard-ball system as an
equilibrium ensemble average of an appropriate microscopic quantity, which in
turn can be evaluated by standard methods of statistical mechanics. To do this
we rewrite Eq. (\ref{ks1}) as a time average of a dynamical quantity as
\begin{eqnarray}
\sum_{\lambda_{i}>0}\lambda_{i} =
  \lim_{t\rightarrow\infty}\frac{1}{t} \ln \frac{\delta
{\cal{V}}_{v}(t)}{\delta {\cal{V}}_{v}(0)},
\nonumber \\
= \lim_{t\rightarrow\infty}\frac{1}{t}\int_{0}^{t}d\tau
\frac{d}{d\tau} \ln\frac{\delta {\cal{V}}_{v}(\tau)}{\delta {\cal{V}}_{v}(0)},
\nonumber \\
=
\left<\frac{d}{d\tau}\ln\frac{\delta{\cal{V}}_{v}(\tau)}{\delta{\cal{V}}_{v}(0)}
\right>,
\label{ks2}
\end{eqnarray}
where the angular brackets denote an average over an appropriate equilibrium
ensemble to be specified further on. Here we have {\it assumed} that the
hard-ball system under consideration is ergodic so that long time averages may
be replaced by ensemble averages. Now we can use elementary kinetic theory
arguments to give a somewhat more explicit form to the ensemble average
appearing in Eq. (\ref{ks2}). Since the volume element in velocity space does
not change during the time between any two binary collisions, and since the
binary collisions in a hard-ball gas are instantaneous, the ensemble average of
the time derivative may be written as
\begin{equation}
\sum_{\lambda_{i}>0}\lambda_{i} =\sum_{i<j} \left< a^{d-1}\int
d\hat{\sigma}
\,
\Theta(-\vv_{ij}\cdot\hat{\sigma})|\vv_{ij}\cdot\hat{\sigma}|
\delta(\vvr_{ij}-a\hat{\sigma})\ln\left[\frac{\delta
{\cal{V}}_{v}
^{\prime}
}{\delta{\cal{V}}_{v}}\right]\right>.
\label{ks3}
\end{equation}
Here the step function, $\Theta(x)$, is equal to unity for $x>0$, and zero
otherwise. The prime on the velocity space volume denotes its value immediately
after the collision between particles $i$ and $j$, while the unprimed quantity
is its value immediately before collision. In the derivation of Eq. (\ref{ks3})
we consider the rate at which binary collisions take place in the gas, and then
calculate the change  in velocity space volumes at each collision.  Thus, we
have integrated over all allowed values of the collision vector $\hat{\sigma}$
for a collision between particles $i$ and $j$ with relative velocity
$\vv_{ij}$, and included, by means of a delta function, the condition that the
two particles must be separated by a distance $a$ at collision. The other
factors in Eq. (\ref{ks3}) take into account the rate at which collisions
between two particles take place in the gas. A more formal derivation in terms
of binary collision operators is easy to construct, but not necessary for our
purpose here\cite{dlvb}.

Suppose, for the moment, that the $Nd$-dimensional velocity deviation vector
immediately after the  $(i,j)$ collision,
$(\delta\vv_1,\delta\vv_2,\ldots,\delta\vv_i',\ldots,\delta\vv_j',\ldots,
\delta\vv_n)$  is related to the velocity deviation vector immediately before
collision  through the matrix equation
\begin{equation}
\left( \begin{array}{c}
  \delta\vv_1 \\
  \delta\vv_2 \\
  \vdots \\
  \delta\vv_i'\\
  \vdots \\
  \delta\vv_j'\\
  \vdots \\
   \delta\vv_n \end{array}
  \right )
= {\bf{A}}_{ij}\cdot \left( \begin{array}{c}
  \delta\vv_1 \\
  \delta\vv_2 \\
  \vdots \\
  \delta\vv_i
\\
  \vdots \\
  \delta\vv_j
\\
  \vdots \\
   \delta\vv_n \end{array}
  \right ).
\label{ks4}
\end{equation}
It then would follow that due to the occurrence of the  $(i,j)$ collision
\begin{equation}
 \frac{\delta {\cal{V}}_{v}
'
}{\delta{\cal{V}}_{v}}=|\det{\bf{A}}_{ij}|,
\label{ks5}
\end{equation}
and the sum of the positive Lyapunov exponents would be given by
\begin{equation}
\sum_{\lambda_{i}>0}\lambda_{i} = \frac{N(N-1)}{2}\left< a^{d-1}\int
d\hat{\sigma}\Theta(-\vv_{12}\cdot\hat{\sigma})|\vv_{12}\cdot\hat{\sigma}|
\delta(\vvr_{12}-a\hat{\sigma})\ln|\det{\bf{A}}_{12}|\right>.
\label{ks6}
\end{equation}
Here we have assumed that the ensemble average is symmetric in the particle
indices, and chosen a particular pair of particles, denoted by particle indices
$1,2$. We now must argue for the validity of Eq. (\ref{ks4}) and then calculate
the determinant of the matrix  ${\bf{A}}_{12}$.

If we examine Eqs. (\ref{eq:dvd}), we see that we can obtain an equation of the
form of Eq. (\ref{ks4}) if we can relate the spatial deviation vectors
$\delta\vvr_i$ and $\delta\vvr_j$ immediately before an  $(i,j)$ collision to
the velocity deviation vectors immediately before collision. Such a relation
must be linear since we are keeping terms only to first order in the
deviations. We therefore make the {\it Ansatz}  that, in general, the spatial
deviation vectors are related to the velocity deviation vectors through a set
of $d\times d$ matrices, called radius of curvature (ROC) matrices (even though
they have the dimension of time), $\ro_{kl}$, via
\begin{equation}
\delta\vvr_{k}(t)=\sum_{l}\ro_{kl}\cdot\delta\vv_{l}.
\label{ks7}
\end{equation}
Then some simple algebra shows that
\begin{equation}
\det{\bf{A}}_{12} =
\det\left[{\bf{M}}_{\hat{\sigma}}(1,2)
-
{\bf{Q}}
_{\hat\sigma}
(1,2)\cdot(\ro_{11}+\ro_{22}
-\ro_{12} -\ro_{21})\right].
\label{ks8}
\end{equation}
We can carry the calculation of the KS entropy  one step further now by
inserting Eq. (\ref{ks8}) into the expression for the KS entropy, Eq.
(\ref{ks6}), to obtain
\begin{eqnarray}
\sum_{\lambda_{i}>0}\lambda_{i} =
\frac{1}{2}
a^{d-1}\int dx_1\,dx_2
\,d\ro_{11}d\ro_{12}d\ro_{21}d\ro_{22}
d\hat{\sigma}
\,
\Theta(-\vv_{12}\cdot\hat{\sigma})|\vv_{12}\cdot\hat{\sigma}|
\delta(\vvr_{12}-a\hat{\sigma}) \times \nonumber \\
\ln|\det\left[{\bf{M}}_{\hat{\sigma}}(1,2)
- {\bf{Q}}_{\hat\sigma}
(1,2)\cdot(\ro_{11}+\ro_{22}
-\ro_{12}
-\ro_{21})\right]|{\cal{F}}_{2}(x_1,x_2,\ro_{11},\ro_{12},\ro_{21},\ro_{22}),
\label{ks81}
\end{eqnarray}
where $x_{i}=(\vvr_i,\vv_i)$, $dx_i = d\vvr_i\,d\vv_i$, and we have defined a
new pair distribution function,
${\cal{F}}_{2}(x_1,x_2,\ro_{11},\ro_{12},\ro_{21},\ro_{22})$ by
\begin{equation}
{\cal{F}}_{2}(x_1,x_2,\ro_{11},\ro_{12},\ro_{21},\ro_{22})
=N(N-1)\int dx_3 ... dx_N \left.\prod\right.'
d{\ro_{ij}}{\cal{F}}_{N}(x_1,x_2,...,x_N,\ro_{11},...,\ro_{NN}).
\label{ks82}
\end{equation}
Here ${\cal{F}}_{N}$ is the normalized ensemble distribution function for the
positions and velocities of all of the particles and for all of the ROC
matrices. The prime on the product means that integrations are not carried out
over the four ROC matrices whose indices are $11,12,21$ and $22$.

To proceed further, we will need to determine the pair distribution function,
${\cal{F}}_2$, and then evaluate the averages needed for the KS entropy. We
will do this using both informal and formal kinetic theory arguments. We
mention here that so far we have made no low density approximations, so that
Eq. (\ref{ks81}) is a general expression for the KS entropy of a hard-ball
system, given the validity of the  {\it Ansatz},  Eq. (\ref{ks7}).

\subsection{Informal Evaluation of the KS Entropy at Low Densities}

A very simple and useful approximation to the KS entropy~\cite{hvbddp}  for a
dilute hard-ball gas can be obtained by noticing that the spatial deviation
vectors for particles  $1$ and $2$,  immediately before their mutual collision
can be written in the form
\begin{equation}
\delta\vvr_i =t_i\delta\vv_i +\delta\vvr_{i}(\tau_{i,0}) \,\,\,{\rm
for}\,\, i=1,2.
\label{ks9}
\end{equation}
Here $\tau_{i,0}$ is the time of the previous collision of particle $i$ with
some other particle in the gas, and $t_{i}$ is the time interval between that
previous collision and the  $(1,2)$ collision. The quantity,
$\delta\vvr_{i}(\tau_{i,0})$ is the spatial deviation vector for particle $i$
immediately after the previous collision.

The following argument is correct for dispersing billiard systems such as the
Lorentz gas, and gives the leading order in the density for the hard-ball gas,
system which is only semi-dispersing \footnote{A dispersing billiard takes any
infinitesimal, parallel pencil of trajectories in configuration space before a
collision into a defocusing pencil in all directions. A semi-dispersing
billiard leaves rays still parallel after collision in at least one plane
through the pencil. A simple dispersing billiard is the outer surface of a
sphere, and a simple semi-dispersing billiard is the outer surface of a
circular cylinder.}.  Comparing the two terms on the  right-hand side of Eq.
(\ref{ks9}),  one finds that their ratio scales, on the average, as the mean
free time which is proportional to the inverse first power of the gas density.
Therefore it seems reasonable that for low densities, the terms in Eq.
(\ref{ks9}) proportional to the $t_i$ should be dominant and that the terms
$\delta\vvr_i$ can be neglected.  If we do this, we can use the approximations
\begin{eqnarray}
\ro_{12} & = & \ro_{21} = 0 \nonumber \\
\ro_{ii} & = & t_{i}\identity \,\,\,\, {\rm for}\,\,i=1,2.
\label{ks10}
\end{eqnarray}
If we insert this approximation in Eq. (\ref{ks8}), we find that
\begin{equation}
\det{\bf{A}}_{12} \approx
\det\left[{\bf{M}}_{\hat{\sigma}}(1,2)
-{\bf{Q}}_{\hat\sigma}
(1,2)(t_1+t_2)\right].
\label{ks11}
\end{equation}
The determinant can be evaluated using the explicit expressions for ${\bf{M}}$
and ${\bf{Q}}$ in Eqs. (\ref{eq:M}) and  (\ref{eq:Q}). For two dimensional
systems we find
\begin{equation}
|\det\left[{\bf{M}}_{\hat{\sigma}}(1,2)
-{\bf{Q}}_{\hat\sigma}
(1,2)(t_1+t_2)\right]| =
1 +\frac{|\vv_{12}|(t_1 +t_2)}{a\cos\phi},
\label{ks12}
\end{equation}
where $\cos\phi = |\hat{\sigma}\cdot\vv_{12}|/|\vv_{12}|$, with $-\pi/2 \leq
\phi \leq \pi/2$. For the three dimensional case we find
\begin{equation}
|\det\left[{\bf{M}}_{\hat{\sigma}}(1,2)
-{\bf{Q}}_{\hat\sigma}
(1,2)(t_1+t_2)\right]| =
1+\frac{|\vv_{12}|(t_1 +t_2)}{a\cos\phi}(1+\cos^{2}\phi)
+\left(\frac{|\vv_{12}|(t_1 + t_2)}{a}\right)^{2}.
\label{ks13}
\end{equation}
We can now insert these expressions into the right-hand side of Eq.
(\ref{ks81}) to obtain explicit expressions for the KS entropy per particle of
dilute hard-ball gases, provided we can specify the pair distribution function,
${\cal{F}}_2$. With the approximations made above in Eq. (\ref{ks10}), a
consistent equilibrium approximation for the pair function at low densities is
provided by the form\footnote{We note that some delta functions are needed to
convert $\ro_{ii}$ from a $2\times 2$ matrix to a scalar quantity $t_i$, but
this is simple to fix, and we do not provide the details here.}
\begin{equation}
{\cal{F}}_{2}(x_1,x_2,\ro_{11},\ro_{12},\ro_{21},\ro_{22})=n^{2}\varphi_0(\vv_1)
\varphi_0(\vv_2)\nu(\vv_1)\nu(\vv_2)
e^{
-(\nu(\vv_1)t_1
+\nu(\vv_2)t_2)
}
\delta(\ro_{12})\delta(\ro_{21}).
\label{ks14}
\end{equation}
Here $\varphi_0(\vv_i)$ are the normalized  Maxwellian velocity distributions,
$\nu(\vv_i)$ is the collision frequency for a particle with velocity $\vv_{i}$,
\begin{equation}
\nu(\vv_i) = na^{d-1} \int d\vv_{3}\int
d\hat{\sigma}
|\vv_{i3}\cdot\hat{\sigma}|\Theta(-\vv_{i3}\cdot\hat{\sigma})
\varphi_0(\vv_3),
\label{141}
\end{equation}
and we have used the low density expression for the distribution of free flight
times, $t_i$ between collisions. \Eq~(\ref{ks14}) is equivalent to the
approximation
\begin{equation}
\F_2(1,2) \approx \F_1(1)\F_{1}(2)\delta(\ro_{12})\delta(\ro_{21}),
\label{ks142}
\end{equation}
where
\begin{equation}
\F_1(i)=n\varphi_0
(\vv_i)\nu(\vv_i)
\exp
[-t_i\nu(\vv_i)].
\label{ks143}
\end{equation}

\begin{figure}[htb]
\centerline{\psfig{file=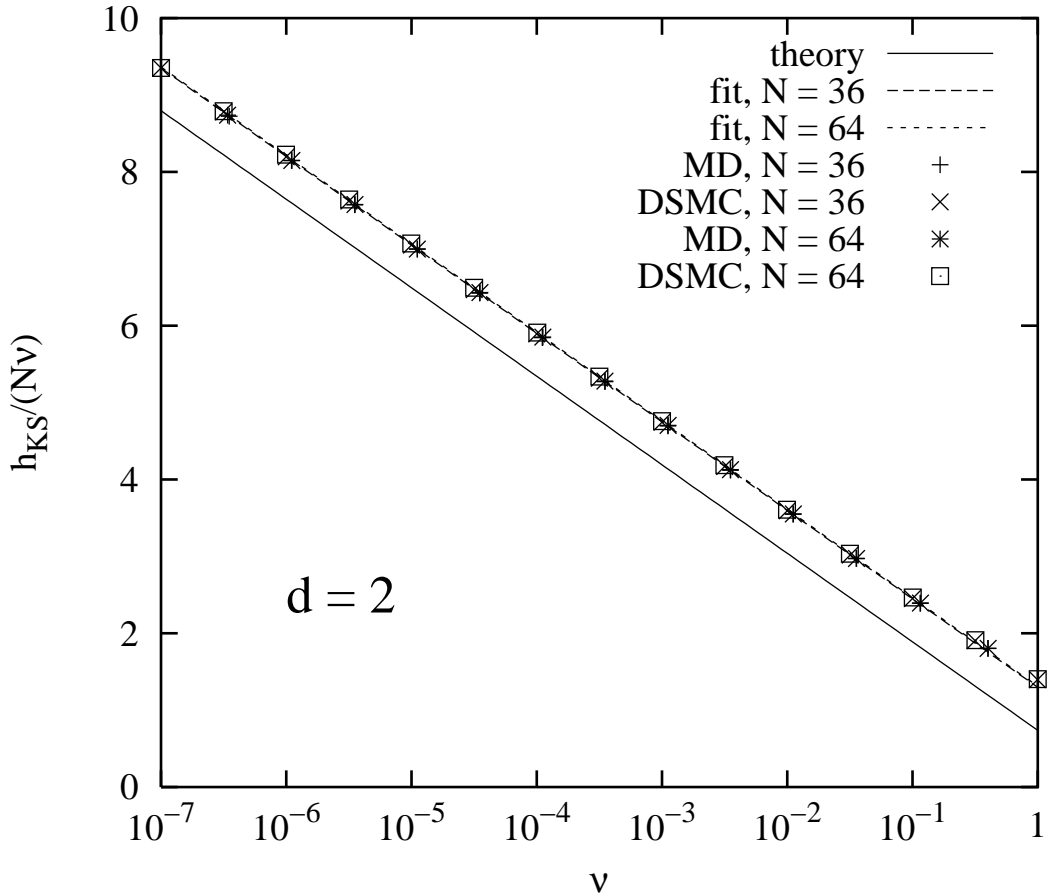}}
\caption{A plot of the results for the KS entropy per particle for a
dilute gas of two-dimensional hard balls, as a function of the collision
frequency $\nu$ in units of $\sqrt{k_BT/(ma)}$. The solid line is the result
given in Eq. (\ref{ks17a}), and the data points are the numerical results of
Dellago and Posch. Here $N$ is the number of particles used in the
simulations. The data points are labeled according to the computational
method, molecular dynamics (MD) or direct-simulation Monte-Carlo (DSMC).
Also plotted, as the dashed curves, are fits to the data points, to
functions of the form (\ref{ks16}), with $A_d$ and $B_d$ as fitting
parameters.}
\label{hks2}
\end{figure}

\begin{figure}[htb]
\centerline{\psfig{figure=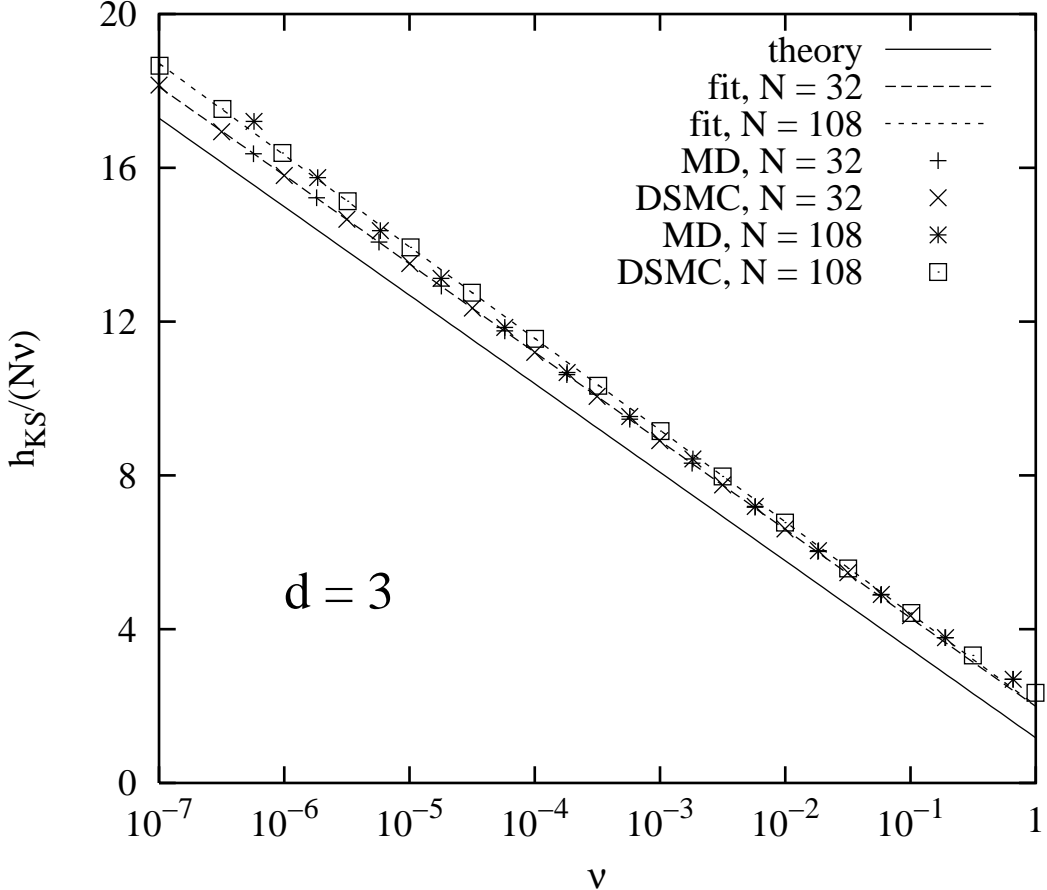}}
\caption{A plot of the results for the KS entropy per particle for a
dilute gas of three-dimensional hard balls, as a function of the
collision frequency $\nu$ in units of $\sqrt{\frac{3}{2}k_BT/(ma)}$. The
solid line is the result given in Eq. (\ref{ks17}), the data points are
results of Dellago and Posch and the dashed curves are fits. The notation
is the same as in the previous figure.}
\label{hks3}
\end{figure}

It is now a simple matter to calculate $h_{KS}/N$. To lowest order in density
we can expand the logarithm of the determinant about the term with the highest
power of the time, which is linear in time for $d=2$, and quadratic in time for
$d=3$. By the usual scaling arguments, the additional terms in the expansion of
the logarithm appear to be at least one order higher in the density than the
terms we keep. Thus, for two dimensional systems
\begin{eqnarray}
h_{KS}/N = \frac{\sum_{\lambda_i >0}\lambda_{i}}{N}
\approx \frac{na}{2}\int d\vv_1\,d\vv_2
\int_0^{\infty}dt_1\,\int_0^{\infty}dt_2
\,\varphi_0(\vv_1)\varphi_0(\vv_2)
\int_{-\pi/2}^{\pi/2}d\phi\,\cos\phi|\vv_{12}|
\times \nonumber \\
\nu(\vv_1)\nu(\vv_2)\exp\left[-(\nu(\vv_1)t_1
+\nu(\vv_2)t_2) \right]\ln\left[\frac{|\vv_{12}|(t_1
+t_2)}{a\cos\phi}\right].
\label{ks15}
\end{eqnarray}
The integral may now be performed, in part, numerically, by using the
equilibrium values for the velocity dependent  collision frequencies. A similar
calculation may be done for the three dimensional case as well. In each case we
find that~\cite{hvbddp}
\begin{equation}
h_{KS}/N = A_d\nu_d[-\ln \tilde{n}_d +B_d +
{\cal O}(\tilde n_d)
].
\label{ks16}
\end{equation}
Here  $\nu_d$ is the average collision frequency per particle at equilibrium,
\begin{equation}
\nu_2 =
(2\pi^{1/2}na)/(\beta m)^{1/2}
\mbox{ and }
\nu_3
=
(4\pi^{1/2}na^2)/(\beta m)^{1/2}
,\label{nud}
\end{equation}
where $\beta =(k_{B}T)^{-1}$, $T$ is the temperature of the gas, and $k_B$ is
Boltzmann's constant. In addition, $\tilde{n}_d$ is a reduced density given by
\begin{equation}
\tilde{n}_2
= na^2
\mbox{, and }
\tilde{n}_3 =na^{3}\pi
{}.
\end{equation}
The quantities $A_d,B_d$ are numerical factors given by
\begin{eqnarray}
\label{ks17a}
 A_2 =& 1/2,& \,\,\,{\rm and}\,\,B_2  =  0.209, \\
 A_3 =& 1,&\,\,\, {\rm and}\,\, B_3  =  0.562.
\label{ks17}
\end{eqnarray}
The values for $A_d$ are in excellent agreement with computer simulations by
Dellago and Posch, but the values for the $B_d$ are too small by factors of $3$
or so. These results are illustrated in Figs. (\ref{hks2}) and (\ref{hks3})
together with the results of numerical simulations by Dellago, and Posch.

The source of the  discrepancy  between the values of $B_d$ can be attributed
to our neglect of the spatial deviation vectors in Eq.  (\ref{ks9}). For
semi-dispersing billiards, the spatial  deviations in certain directions remain
comparable to the corresponding  component of $t_i\vdv_i$. Therefore, they may
contribute to the first correction to the density logarithm in Eq.
(\ref{ks16}). In the next subsection we will pursue this point a bit further.

\subsection{Toward the Formal Evaluation of the KS Entropy at Low Densities}

As we have seen above, the KS entropy of an equilibrium hard-ball system may be
calculated if one knows the pair distribution function  ${\cal{F}}_2$. The
simple approximation to this function, Eq. (\ref{ks14}), correctly gives the
leading order term at low density, but not the next term in a density
expansion. At low densities, one might expect that the two colliding particles,
$1,2$ in the above expressions are uncorrelated just before their collision, so
that a useful approximation to the pair function would still be of the form
\begin{equation}
{\cal{F}}_{2}(x_1,x_2,\ro_{11},\ro_{12},\ro_{21},\ro_{22})={\cal{F}}_1(x_1,\ro_{11}){\cal{F}}_1(x_2,\ro_{22})\delta(\ro_{12})\delta(\ro_{21}),
\label{ks18}
\end{equation}
but we should use a better approximation to  the one particle distribution
function, ${\cal{F}}_1$, than that used above. More specifically, we consider
the set of reduced distribution functions, ${\cal{F}}_1,{\cal{F}}_2,...$, which
satisfy a set of BBGKY hierarchy equations, and then use this set to find a
good, low density approximation to the equation for ${\cal{F}}_1$, in much the
same way as the  BBGKY hierarchy is used to derive the standard Boltzmann
transport equation and its higher density corrections~\cite{dlvb}. However as
we now consider a set of extended distribution functions which include the
$\ro$ matrices as variables in addition to the positions and velocities of the
particles, the hierarchy equations must be extended to include these new
variables as well. The full hierarchy technique requires the use of somewhat
cumbersome cluster expansion methods, and has been outlined
elsewhere\cite{dlvb}. Nevertheless, the low density approximation to the
equation for $\F_1$ is physically plausible since it is a direct extension of
the Boltzmann equation to include the additional ROC matrices as variables.
This extended Boltzmann equation is given by
\begin{eqnarray}
\left[\frac{\partial}{\partial t}
+{\cal{L}}_0(1)\right]\F_{1}(x_1,\ro_{11},t) =
\int dx_2
d\ro_{12}d\ro_{21}d\ro_{22}\bar{\cal{T}}_{-}(1,2)\F_1(x_1,\ro_{11},t)\F_1(x_2,
\ro_{22},t)\times
\nonumber \\
\delta(\ro_{12})\delta(\ro_{21}).
\label{ks19}
\end{eqnarray}
Here we have included a possible time dependence of the distribution functions,
the operator ${\cal{L}}_0(1)$ is the free streaming operator which accounts for
the change in $\F_1$ due to the free motion of the particle. It is given by
\begin{equation}
{\cal{L}}_0(1) =
\vv_1\cdot\frac{\partial}{\partial\vvr_1}+\sum_{\alpha=1}^{d}\frac{\partial}
{\partial\ro_{11,\alpha\alpha}}.
\label{ks20}
\end{equation}
Here we assume that there are no external forces acting on the particles, so
that in the course of free motion for a time interval, ($\tau,t+\tau $), the
position of the particle changes according to $\vvr_1(t+\tau )=\vvr_1(\tau
)+t\vv_1(\tau ) $, and the ROC matrix, according to $\ro_1(t+\tau )=\ro_1(\tau
)+t\identity $. The derivatives with respect to the position variable and with
respect to the diagonal  components of $\ro_{11}$ reflect this free flight
motion. The  right-hand side of the extended Boltzmann equation, Eq.
(\ref{ks19}), expresses the change in $\F_1$ due to binary collisions, and we
have supposed that the two colliding particles are uncorrelated before their
collision, as is normally done in the derivations of the Boltzmann equation. In
this term we have defined a binary collision operator by
\begin{eqnarray}
\bar{\cal{T}}_{-}(1,2) =a^{d-1}\int
d\hat{\sigma}|\vv_{12}\cdot\hat{\sigma}|\Theta(\vv_{12}\cdot\hat{\sigma})
\Big[
\delta(\vvr_{12}-a\hat{\sigma})\int
d\ro'_{11}d\ro'_{12}d\ro'_{21}d\ro'_{22} \nonumber \\
\times \prod_{i,j=1,2}
\delta(\ro_{ij}-\ro_{ij}(\ro'_{11},...,\ro'_{22}))
{\cal{P}}'_{\hat\sigma}(1,2)
 \nonumber  \\
-\delta(\vvr_{12}+a\hat{\sigma})\Big].
\label{ks21}
\end{eqnarray}
Here  ${\cal{P}}'_{\hat\sigma}$  is a substitution operator that replaces ROC
matrices to its right by the corresponding primed matrices, and velocities to
its right by restituting velocities, namely those which for a given
$\hat{\sigma}$ produce $\vv_{1},\vv_{2}$ after a binary collision. The delta
functions in the ROC matrices require that the primed ROC matrices be
restituting ones, namely, those which produce the unprimed matrices after the
$1,2$ binary collision.

The restituting values of the ROC matrices can be found by means of Eqs.
(\ref{eq:Q}) and (\ref{eq:dvd}), and the  {\em Ansatz}, Eq. (\ref{ks7}). To  do
this we consider only the ROC matrices involving the two colliding particles,
and ignore the other ROC matrices involving other particles, such as
$\ro_{13}$, etc. In the equations below we will denote with primes the values
of  restituting quantities, i.e., the quantities {\em before} collision will be
denoted with primes, and those {\em after}  collisions without primes. Then
before collision
\begin{eqnarray}
\vdrp_{1} &=&\ro'_{11}\cdot\vdvp_1 +\ro_{12}'\cdot\vdvp_2,\nonumber \\
\vdrp_{2} &=& \ro'_{22}\cdot\vdvp_2+\ro_{21}'\cdot\vdvp_1,
\label{ks22}
\end{eqnarray}
and it follows that
\begin{eqnarray}
\delta{\vec{R}}'_{12}
&=&
\ro'_a\cdot\delta{\vec{V}}'_{12}+\ro'_b\cdot\vdvp_{12}
,
\nonumber \\
\vdrp_{12}
&=&
\ro'_c\cdot\vdvp_{12} + \ro'_{d}\cdot\delta{\vec{V}}'_{12}
,
\label{ks23}
\end{eqnarray}
where\begin{eqnarray}
\ro_a & = & \frac{1}{2}\left(\ro_{11}+\ro_{12}+\ro_{21}+\ro_{22}\right)
,
\nonumber \\
\ro_b & = & \frac{1}{4}\left(\ro_{11}-\ro_{12}+\ro_{21}-\ro_{22}\right)
,
\nonumber \\
\ro_c & = &  \frac{1}{2}\left(\ro_{11}-\ro_{12}-\ro_{21}+\ro_{22}\right)
,
\nonumber \\
\ro_d & = &  \left(\ro_{11}+\ro_{12}-\ro_{21}-\ro_{22}\right).
\label{ks24}
\end{eqnarray}

Now one writes the collision equations for the spatial deviations of the center
of mass and relative coordinates in terms of the variables before (primed) and
after collision (unprimed),
\begin{eqnarray}
\ro_a\cdot\delta{\vec{V}}_{12}+\ro_b\cdot\vdv_{12} & = &
\ro'_a\cdot\delta{\vec{V}}'_{12}+\ro'_b\cdot\vdv'_{12}
,
\nonumber \\
\ro_c\cdot\vdv_{12}+\ro_d\cdot\delta{\vec{V}}_{12} & = &
{\bf{M}}_{\hat{\sigma}}\cdot[
\ro'_c\cdot\vdv'_{12}+\ro'_d\cdot\delta{\vec{V}}'_{12}],
\label{ks25}
\end{eqnarray}
as well as the collision equation for the deviation of the relative and center
of mass velocities,
\begin{eqnarray}
\vdv_{12} & = & {\bf{M}}_{\hat{\sigma}}\cdot\vdv'_{12}
-2{\bf{Q}}'_{\hat\sigma}
\cdot[
\ro'_c\cdot\vdv'_{12}+\ro'_d\cdot\delta{\vec{V}}'_{12}], \nonumber \\
\delta{\vec{V}}_{12} & = & \delta{\vec{V}}'_{12}.
\label{ks26}
\end{eqnarray}
Then we use the fact that the deviations of the relative and center of mass
velocities before collision are independent of each other, and that there are
no orthogonalization or normalization constraints on the components of these
vectors that would make some of them dependent on the others\footnote{The
constraints mentioned earlier, Eq. (\ref{cons}), on the  deviations of the
total momentum and energy of the $N$-body system do not have any significant
effect on the dynamics of two particles if $N \gg 2$. }. Therefore, upon
inserting Eq. (\ref{ks26}) into Eq. (\ref{ks25}), and comparing coefficients of
$\delta{\vec{V}}'_{12}$ and of $\vdv'_{12}$, we obtain
\begin{eqnarray}
\ro_a
-
2\ro_b\cdot{\bf{Q}}'_{\hat\sigma}
\cdot\ro'_d & = &  \ro'_a
,
\nonumber \\
\ro_b\cdot{\bf{M}}_{\hat{\sigma}}
-
2\ro_b\cdot{\bf{Q}}'_{\hat\sigma}
\cdot\ro'_c &
= &
\ro'_b
,
\nonumber \\
\ro_c\cdot{\bf{M}}_{\hat{\sigma}}
-
2\ro_c\cdot{\bf{Q}}'_{\hat\sigma}
\cdot\ro'_c &
= &
{\bf{M}}_{\hat{\sigma}}\cdot\ro'_c
,
\nonumber \\
\ro_d
-
2\ro_c\cdot{\bf{Q}}'_{\hat\sigma}
\cdot\ro'_d &  = &
{\bf{M}}_{\hat{\sigma}}\cdot\ro'_d.
\label{ks27}
\end{eqnarray}

In order to complete the specification of the restituting ROC matrices, we use
the fact that the particles are taken to be dynamically uncorrelated before
collision so that  $\ro'_{12} =\ro'_{21}=0$,  and we can also express the
matrix  ${\bf{Q}}'_{\hat\sigma}$ in terms of the unprimed relative velocity
after collision: ${\bf{Q}}'_{\hat\sigma}={\bf{Q}}_{\hat\sigma}(\vv_{12}')  = -
{\bf{Q}}_{\hat\sigma}^T(\vv_{12})$. Thus, the primed variables before collision
can be expressed in terms of the unprimed variables after collision, provided
the Eqs. (\ref{ks27}) for the  $\ro'$ matrices can be solved. It is important
to point out that, as we will see in Section V, the solution of these equations
is very different for dispersive billiards such as encountered in the Lorentz
gas, and semi-dispersive billiards as occur in the hard-ball gas where all of
the particles move. The difference resides in the fact that for dispersive
billiards the elements of ROC matrices after collision are of the order of
$a/v$ and therefore much smaller than the typical matrix element before
collision, which is on the order of  the mean free time. For semi-dispersing
billiards this is no longer true, and some of the matrix elements remain large
after collision as well. This is equivalent to the statement that the spatial
deviation vectors immediately after collision are not negligible for
semi-dispersing billiard systems.

It is an elementary matter now to write expressions for the low density KS
entropy in terms of the distribution functions and the elements of the ROC
matrices. One need only replace $\F_{2}$ in Eq. (\ref{ks81}) by
$\F_{1}(x_1,\ro_{11})\F_{1}(x_2,\ro_{22})\delta(\ro_{12})\delta(\ro_{21})$, and
 use the leading order term in the expression for the determinants, namely
\begin{equation}
|\det\left[{\bf{M}}_{\hat{\sigma}}(1,2)
-{\bf{Q}}_{\hat\sigma}(1,2)(\ro_{11}+\ro_{22})
\right]|
\approx \frac{2|\vv_{12}|\rho_{\perp\perp}}{a\cos\phi},
\label{ks29}
\end{equation}
for two dimensional systems. Here the matrix element $\rho_{\perp\perp}$ is
defined by
\begin{equation}
\rho_{\perp\perp}
=\frac{1}{2}(\hat{v}_{12\perp}\cdot(\ro_{11}+\ro_{22})\cdot\hat{v}_{12\perp}).
\label{ks30}
\end{equation}
Here the subscript $\perp$ on the two dimensional unit vector
$\hat{v}_{12\perp}$  denotes a unit vector in a direction perpendicular to
$\vv_{12}$.  For the three dimensional system, we obtain
\begin{equation}
|\det\left[{\bf{M}}_{\hat{\sigma}}(1,2)
-{\bf{Q}}_{\hat\sigma}(1,2)(\ro_{11}+\ro_{22})
\right]|
\approx
\left(\frac{2|\vv_{12}|}{a}\right)^{2}[\rho_{\perp\perp}^{11}\rho_{\perp\perp}^{22}-\rho_{\perp\perp}^{12}\rho_{\perp\perp}^{21}].
\label{ks31}
\end{equation}
Here
\begin{equation}
\rho_{\perp\perp}^{ij}
=\frac{1}{2}(\hat{v}_{12\perp}^{i}\cdot(\ro_{11}+\ro_{22})\cdot\hat{v}_{12\perp}
^{j}),
\label{ks32}
\end{equation}
where the unit vectors
$\hat{v}_{12},\hat{v}_{12\perp}^{1},\hat{v}_{12\perp}^{2}$ form an orthonormal
coordinate basis in three dimensions.

It now remains to solve the extended Boltzmann equation for $\F_{1}$. This
appears to be somewhat difficult due to the complications inherent in the
restituting, or gain term. If one ignores this term entirely, for whatever
reason, the solution becomes simple. In fact one recovers the approximate form
for $\F_1$ given by Eq. (\ref{ks143}), and  therefore one recovers the same
expressions for the KS entropy as given by Eqs.
(\ref{ks16}--\ref{ks17})\cite{dlvb}. To do better, one has to include the
restituting term in the equation for $\F_1$. How to (approximately) solve the
extended Boltzmann equation then, is still under investigation.

\section{The Largest Lyapunov Exponent for the Hard-Ball System at Low
Density}

In this section, we will obtain an estimate for the largest Lyapunov  exponent
$\lambda_{max}$. This exponent determines the asymptotic  growth of all the
deviation vectors: $\delta \vec v_i(t),\delta\vec  r_i(t)\sim
e^{\lambda_{max}t}$. We will calculate it in the low density  regime,
$\tilden_d\ll 1$, when a Boltzmann equation gives an appropriate description of
the behavior of the one-particle distribution function. The results will be
checked against simulations.

\subsection{Kinetic Theory Approach}

At low density, two colliding particles, $i$ and $j$, have spent a long  time
in free flight before the collision. Therefore, in  \Eq~(\ref{ks9}), just
before collision we keep only the dominant  term involving the free flight time
$t_i$. We saw in the previous section  that for the KS entropy this
approximation turned out to be  unsatisfactory  for describing the behavior of
the  ROC matrices. For the  leading behavior of the velocity deviations,
however, it will be a valid  approximation.

We insert this in \Eq~(\ref{eq:dvd}) and keep only the leading terms  in the
free flight times,
\begin{eqnarray}
        \vdr_i' \approx -\vdr_j' & \approx &
	-\sfrac{1}{2}\matr{M}_{\!\hn}(t_i\vdv_i-t_j\vdv_j) ,
\nonumber\\
        \vdv_i' \approx -\vdv_j' & \approx &
	\matr{Q}_{\hat\sigma}(t_i\vdv_i-t_j\vdv_j).
\eql{dvdynamics}
\end{eqnarray}
Let us denote the typical velocity of a particle in the gas by
$v_0=\sqrt{k_BT/m}$. One sees that $\vdv'_i/v_0$ and $\vdr'_i/\sigma$ are of
the same order order of magnitude, which justifies the approximation that
$\vdr_i\ll t_i\vdv_i$. Notice we have eliminated the position deviation
vectors, but now we need to know the free flight times between collisions. To
compare different contributions in \Eq~(\ref{dvdynamics}), we want to know the
order  of magnitude of the velocity deviation vectors. We define clock values
$k_i$ to represent their order of magnitude measured in inverse powers of
$\tilde n_d$:,
\begin{equation}
        \vdv_i \equiv \delta v_0 \left(\frac{1}{\tilde n_d}\right)^{k_i}
\he_i ,
\eql{defk}
\end{equation}
where $\he_i$ is a unit vector and $\delta v_0$ is a fixed infinitesimal
velocity. To obtain the largest Lyapunov exponent, it is enough to know the
time evolution of these clock values. They will grow linearly with time,
\[
        k_i \sim w \nu_d t .
\]
The average collision frequency $\nu_d$ sets a time scale proportional to the
density. It is extracted so that the clock speed $w$ can be interpreted as the
average increase of a clock value in a collision. We see from \Eq~(\ref{defk})
that the clock speed is related to the largest Lyapunov exponent via
$\lambda_{max} = - w \nu_d \ln \tilde n_d$.

After the collision, the two particles have equal clock value, given by
\begin{eqnarray}
        k'_i = k'_j &=&
\frac{\ln|\matr{Q}_{\hat\sigma}(i,j)(t_i\vdv_i-t_j\vdv_j)/\delta v_0|
}{-\ln\tilden_d}
\eql{8ca}
\\
        &=&\max(k_i,k_j)+1 +{\cal O}\left(1/\ln\tilden_d\right).
\eql{8c}
\end{eqnarray}
To see how \Eq~(\ref{8c}) follows from \Eq~(\ref{8ca}) consider the following.
If one of the clock values, say $k_i$,  is larger than the other one, the
corresponding velocity deviation $\vdv_i\sim\tilden_d^{-k_i}$ is very much
larger than $\vdv_j\sim\tilden_d^{-k_j}$.  We will consider the case that the
two clock values differ by less than one in a moment. We say particle $i$ is
the dominant particle. Only the dominant term inside the logarithm in the
denominator of \Eq~(\ref{8ca}) has to be taken into account, as the other one
is much smaller. The free flight times scale, for low densities, inversely with
the density $\tilde n_d$, whereas the terms  $\matr{Q}_{\hat\sigma}(i,j)$ and
$v_0$ do not contain any density dependence.  To see the density dependence
more explicitly, we scale the free flight  time as $s_i=\tilde n_d t_i$, so
that $s_i$ is density independent. The denominator in \Eq~(\ref{8ca}) is then
seen to be the sum of a term proportional to $-\ln\tilden_d$ and a density
independent term which fluctuates from collision to collision:
\[
	\ln |\matr{Q}_{\hat\sigma}(i,j)t_i\delta\vec v_i/\delta v_0|
	= \ln |\matr{Q}_{\hat\sigma}(i,j)s_i (1/\tilden_d)^{k_i+1}\hat e_i|
	= (k_i +1)(-\ln\tilde n_d) +\ln|\matr{Q}_{\hat\sigma}(i,j)s_i\hat
e_i|,
\]
which leads to \Eq~(\ref{8c}). The fluctuating corrections to the addition of
$1$ of the dominant clock value scale with density as $1/\ln \tilde n_d$.
Finally, let us consider what happens when $|k_i-k_j|<1$. In that case, it is
not necessarily true that one of the two terms in the denominator in \eq{8ca}
dominates. To show that \Eq~(\ref{8c}) then still holds, i.e.\
$k_i'=k_j'=\mbox{max}(k_i,k_j)+1$ plus corrections which scale with the density
as $1/\ln\tilden_d$, we consider the extreme case that $k_i=k_j$ exactly:
\begin{eqnarray*}
	\ln |\matr{Q}_{\hat\sigma}(i,j)(t_i\delta\vec
v_i-t_j\delta\vec v_j/\delta v_0|
	&=& \ln |\matr{Q}_{\hat\sigma}(i,j)(1/\tilden_d)^{k_i+1}
(s_i\hat e_i-s_j\hat e_j)|
\\
	&=& (k_i +1)(-\ln\tilde n_d) +\ln|\matr{Q}_{\hat\sigma}(i,j)
(s_i\hat e_i-s_j\hat e_j)|,
\end{eqnarray*}
where $s_j=\tilde n_d t_j$. Again, the second fluctuating term is density
independent, and \Eq~(\ref{8c}) follows.

The clock values have a well-defined dynamics in the limit
$\tilden_d\rightarrow 0$, which gives the leading behavior of $w$. For  this
leading behavior, we can neglect the ${\cal O}(1/\ln\tilden_d)$  terms in
\Eq~(\ref{8c}):
\begin{equation}
        k'_i=k_j'=\max(k_i,k_j)+1.
\eql{kdyn}
\end{equation}
In some collisions the neglected terms may be large, but the lower the
density, the rarer such collisions become. If these terms were kept, they would
give rise to terms of the order of $\ln^{-1}(\tilden_d)$ and higher in $w$, so
we would  get an expansion of the form:
\[
        w = w_0 + w_1 \ln^{-1}(\tilden_d)+w_2\ln^{-2}(\tilden_d)+\ldots.
\]
The coefficients $w_0$, $w_1$, etc., in fact are still density dependent  due
to the neglected terms in \Eq~(\ref{dvdynamics}), and to correlations between
collisions. The former will give rise to powers of the density, the latter may
also give rise to a nonanalytic dependence on the density\cite{llt}. Such terms
however are at least accompanied by one factor of the density. That means that
for $\tilde n_d\rightarrow 0$, the limiting values of the coefficients give the
asymptotic behavior of $w$. Corresponding to the expansion of $w$, the Lyapunov
exponent has an expansion of the form
\begin{equation}
        \lambda_{max} = \nu_d[-w_0\ln\tilden_d- w_1 -w_2\ln^{-1}(\tilden_d)
                        +\ldots].
\eql{lambda}
\end{equation}

We will calculate  the first term in the density expansion of $\lambda_{max}$
(which gives the leading behavior) by calculating the leading clock speed
$w_0$. As we already argued, it is enough to use the dynamics in
\Eq~(\ref{kdyn}), for which we can restrict ourselves to integer clock values.
The calculation will be based on an extended Boltzmann equation for the one
particle distribution function $f(k,\vv,t)$ of clock values and velocities.
Colliding particles are assumed to be uncorrelated before the collision, i.e.,
the probability that a particle with clock value $k_i$ and velocity $\vv_i$ and
a particle with clock value $k_j$ and velocity $\vv_j$ collide is proportional
to $f(k_i,\vv_i,t)$ $f(k_j,\vv_j,t)$; this is the {\em Stosszahlansatz}. For
that collision we still need to specify $\hat\sigma$ and this vector will, in a
collision with given $\vv_{ij}$, be drawn from a probability distribution
proportional to $|\hat\sigma\cdot\vv_{ij}|$. We demand that
$\hat\sigma\cdot\vv_{ij}$ be negative, so that the particles are approaching
each other. The gas will be  uniform in equilibrium, therefore we neglect any
spatial dependence. Considering gains and losses, we can construct the
following extended Boltzmann equation for the time evolution of $f$:
\begin{eqnarray}
        \dd{f(k,\vv_1)}{t} & = & - \nu(\vv_1) f(k,\vv) + \int d\vv_2
\int d\hat\sigma\, \Theta(\hat\sigma\cdot\vv_{12})n a^{d-1}
|\hat\sigma\cdot\vv_{12}|\Bigg[
f(k-1,\vv_1')\sum_{l=-\infty}^{k-2} f(l,\vv_2')
\nonumber
\\ &&+
f(k-1,\vv_2')\sum_{l=-\infty}^{k-2} f(l,\vv_1') +
f(k-1,\vv_1')f(k-1,\vv_2')
\Bigg],
\eql{frate2}
\end{eqnarray}
where in the integral,  primed and unprimed quantities denote values before,
respectively after  collision.  $\nu(\vv_1)$ is the velocity  dependent
collision frequency, given by \Eq~(\ref{141}), where we  take the velocities to
have  their equilibrium  distribution $\varphi_0$. In principle, we could also
allow for   an initially nonequilibrium or nonstationary velocity distribution,
but this will not influence the  asymptotic clock speed, as the velocity
distribution function will be  stationary in due course, whereas the clock
values keep on growing  indefinitely. The extended Boltzmann equation can be
simplified  using cumulatives, defined by
\[
        C(k,\vv) \equiv
\frac{1}{\varphi_0(\vv)}\sum_{l=-\infty}^{k}f(l,\vv).
\]
Because of this definition, $C(k,\vv)$ will tend to $1$ as $k$ tends to
infinity for all $\vv$. In terms of the cumulatives, the Boltzmann equation
reads
\begin{eqnarray}
        \dd{C(k,\vv_1)}{t} &=& - \nu(\vv_1) C(k,\vv_1)
\nonumber\\&&
+ \int d\vv_2
\int d\hat\sigma\,
\Theta(\hat\sigma\cdot\vv_{12})
n a^{d-1} |\hat\sigma\cdot\vv_{12}| \varphi_0(\vv_2)
C(k-1,\vv_1')C(k-1,\vv_2').
\eql{Crate2}
\end{eqnarray}
This is the appropriate equation to calculate $w_0$ from.  The assumptions made
in its derivation were that colliding particles are uncorrelated, that the
clock values increase according to (\ref{kdyn}), that spatial fluctuations do
not play any significant role and that the velocity distribution has reached
its equilibrium  form. The equation is similar to that found in earlier
work\cite{myself,leid}, where  we had not accounted for the velocity
dependence. We will follow the same approach used there, to find a better
estimate for $w_0$, and thus for the leading behavior of the largest Lyapunov
exponent for low density.

The Boltzmann equation is of the same type as equations encountered in  front
propagation\cite{VanSaarloos}. In this analogy, $C=1$ is an  unstable phase,
that lives at the higher clock values, and $C=0$ is a  stable phase, that
propagates from the lower clock values to the higher  ones. Because the
$\vdv_i$ grow exponentially, this propagation  means that the clock values are
increasing linearly in time:
\begin{equation}
        C(k,\vv;t) = F(k-w_0\nu_d t,\vv).
\eql{propagation}
\end{equation}
We assume that the equation is of the pulled front type, meaning that the
equation linearized around the leading edge ($F=1$) sets a critical velocity of
the front. Fronts exist for any velocity above this critical one. If the
initial conditions of $C$ are sufficiently steep, then a front will develop
with this critical velocity.  In our case, as an initial condition, almost any
(single) $\delta\Gamma$ will do to see an exponential growth with the largest
Lyapunov exponent, as almost any $\delta\Gamma$ will have some component along
the  fastest expanding direction in phase space. Because of the finite number
of particles, the initial distribution of clock values corresponding to a
single $\delta\Gamma$, will have a finite support. This is as steep an initial
condition as one can get, so the critical velocity is the clock speed  $w_0$
that determines the Lyapunov exponent.

We will now show how the linearized equation sets a critical velocity,  and
then we will determine it for the  hard-ball gas in two dimensions. We insert
the  {\em Ansatz}~(\ref{propagation})  into the Boltzmann  equation and
concentrate on the leading edge. That is, we write  $F(k,\vv)= 1 -
\Delta(k,\vv)$ and get, to linear order in $\Delta$,
\begin{eqnarray}
        -w_0\nu_d\dd{\Delta(k,\vv_1)}{k} & = & - \nu(\vv_1)
\Delta(k,\vv_1)
\eql{linearized}
\\ &+&
        \int d\vv_2
        \int d\hat\sigma\,
		\Theta(\hat\sigma\cdot\vv_{12})
		n a^{d-1} |\hat\sigma\cdot\vv_{12}|
                \varphi_0(\vv_2)
        \left[\Delta(k-1,\vv_1')+\Delta(k-1,\vv_2')\right].
\nonumber
\end{eqnarray}
We now take  a linear superposition of exponentially decreasing functions in
$k$:
\begin{equation}\eql{superposition}
        \Delta(k,\vv) = \sum_i A_i(\vv)e^{-\gamma_i k}.
\end{equation}
For this to be a solution of \Eq~(\ref{linearized}), only certain
characteristic values $\gamma_i$ and corresponding characteristic functions
$A_i(\vv)$ are allowed, determined by the characteristic equation:
\begin{eqnarray*}
        w_0\gamma\nu_d\,A(\vv_1) & = & - \nu(\vv_1) A(\vv_1)
        +
        e^\gamma\int d\vv_2
        \int d\hat\sigma\,
	\Theta(\hat\sigma\cdot\vv_{12})
	n a^{d-1} |\hat\sigma\cdot\vv_{12}|
                \varphi_0(\vv_2)
        \left[A(\vv_1')+A(\vv_2')\right].
\nonumber
\end{eqnarray*}
which is found by inserting $A(\vv)\exp(-\gamma k)$ into the linearized
equation, \Eq~(\ref{linearized}). The characteristic equation  is non-linear in
$\gamma$. To deal with it, we rewrite this equation first as a generalized
eigenvalue problem:  \newcommand{\mW}{\mbox{\boldmath $W$}_{\!w_0\gamma}}
\newcommand{\mL}{\mbox{\boldmath $L$}}
\begin{equation}
\eql{symbol}
        \Lambda \,\mW \, A =
        \mL \, A ,
\end{equation}
where $\Lambda=e^{-\gamma}$ is the eigenvalue and the operators are defined by
\begin{eqnarray*}
        \left(\mW A\right)(\vv_1) &\
                =&
                         (\nu_d^{\,-1}\nu(\vv_1)+w_0\gamma)
                A(\vv_1), \\
        \left(\mL A\right)(\vv_1) &
                =&
                \bar\nu_d^{-1}
	\int d\vv_2
        \int d\hat\sigma\,
		\Theta(\hat\sigma\cdot\vv_{12})
		n a^{d-1} |\hat\sigma\cdot\vv_{12}|
                \varphi_0(\vv_2)
        \left[A(\vv_1')+A(\vv_2')\right].
\end{eqnarray*}
For solving \Eq~(\ref{linearized}), first the whole spectrum
$\{\Lambda_k(w\gamma)\}$ of the generalized eigenvalue  problem~(\ref{symbol})
should be determined, with the corresponding  eigenfunctions
$\{A_k(\vv;w_0\gamma)\}$. The eigenvalues should also equal  $e^{\gamma}$. So
next, for every $\Lambda_k$, the solutions for $\gamma$ of
\[
        e^{\gamma} = \Lambda_k(w_0\gamma),
\]
for fixed $w_0$, have to be found, and they are characteristic values. For
each term in \Eq~(\ref{superposition}), the $\gamma_i$ is one of  these
solutions $\gamma$ with characteristic function
$A_i(\vv)=A_k(\vv;w_0\gamma_i)$.  The superposition~(\ref{superposition}), with
 arbitrary coefficients, then is the general solution of the linearized
equation~(\ref{linearized}), for a given clock speed $w_0$.

However, we don't need the full solution of \Eq~(\ref{linearized}).  For the
leading edge, $k\rightarrow\infty$, only the term with the slowest decay in
\Eq~(\ref{superposition}) survives. The characteristic value with the smallest
real part, we call this $\gamma_0$, corresponds to the slowest decay. As the
distribution function $f$ should be positive, $F$ should be an increasing
function and $\Delta$ should be monotonically decreasing, meaning that
$\gamma_0$ cannot have a imaginary part. The requirement that $\gamma_0$ be
real is essential: it will be the determining factor for the critical velocity
$w_0$, as follows.  The smallest characteristic value $\gamma_0$ is a function
of the clock speed $w_0$. The inverse of this function, $w_0(\gamma_0)$, has a
minimum. That minimum is the critical value, because, for clock speeds $w_0$
below this minimum, there are no real $\gamma_0$.

The characteristic value with the smallest real part, was required to be  real,
so we only need to consider the eigenvalue problem~(\ref{symbol})  for real
$w_0\gamma$. In that case, the operators are self-adjoint with  respect to the
inner product
\begin{equation}
        \inp{A}{B} = \int d\vv\, \varphi_0(\vv)\, A(\vv)\, B(\vv).
\eql{inner}
\end{equation}
The self-adjointness means that all the eigenvalues $\Lambda_k$ are real.  As
we want the smallest real $\gamma$, we are interested in the largest
eigenvalue, which we will call $\Lambda_0$. The self-adjointness of the
operators, together with the fact that $\mW$ is a positive operator, can  be
used to derive a maximum principle for this eigenvalue,
\begin{equation}
\eql{variational}
        \Lambda_0(w_0\gamma) = \max_{A} \frac{
                                \mel{A}{\mL\,}{A}
                                }{
                                \mel{A}{\mW}{A}
                                }.
\end{equation}
The corresponding eigenvector $A_0(\vv)$ is the $A(\vv)$ for which the
expression takes on its maximum value. Once we have $\Lambda_0(w_0\gamma)$,
$\gamma_0$ is the smallest real solution of
\begin{equation}
        \Lambda_0(w_0\gamma_0) = e^{-\gamma_0}.
\eql{l0}
\end{equation}
Notice that $\gamma_0$ then indeed still depends on the value of $w_0$ in the
{\em Ansatz}~(\ref{propagation}).

As in previous work\cite{myself,leid}, the critical velocity is determined  as
the smallest $w_0$ possible when $\gamma_0$ is real and positive, i.e., the
minimum of $w_0(\gamma_0)$. This minimum is obtained from  \Eq~(\ref{l0})
taking the derivative with respect to $\gamma_0$ and  using $w_0'(\gamma_0)=0$.
This gives
\[
        w_0\Lambda_0'(w_0\gamma_0)=-e^{-\gamma_0}.
\]
This condition together with \Eq~(\ref{l0}) can be captured concisely by saying
that the critical velocity is the minimum of the following function  $\tilde w$
over real positive values of $w_0\gamma_0$:
\begin{equation}
        \tilde w(w_0\gamma_0) \equiv
\frac{w_0\gamma_0}{-\ln\Lambda_0(w_0\gamma_0)}
\eql{tildew}
\end{equation}

Once the critical clock speed $w_0$ has been obtained, the characteristic
function $A_0(\vv)$, determined by \Eq~(\ref{symbol}), gives the velocity
distribution in the head  of the front.  More precisely,
\begin{equation}
        P_{\mbox{\scriptsize head}}(\vv) \equiv A_0(\vv)\varphi_0(\vv),
\eql{headdistr}
\end{equation}
is the velocity distribution of the particles with  clock values $k$    larger
than some $k_0$, for $k_0$ tending to infinity.

We have no exact expression for  the eigenvalue $\Lambda_0$, but the
variational principle will allow us to calculate  approximations  to it, as
follows. $\Lambda_0$ was given in \Eq~(\ref{variational}) as the maximum over
all functions $A(\vv)$. So inserting any function $A(\vv)$ would give a lower
bound on $\Lambda_0$. Using this lower bound in \Eq~(\ref{tildew}) will then
give a lower bound on $w_0$. We take a fixed form of $A(\vv;b_0,b_1,\ldots)$,
depending on variational parameters $\{b_i\}$, and determine the maximum in
\Eq~(\ref{variational}) over these parameters. This gives an approximation to
$\Lambda_0$ and $A_0(\vv)$, which can be systematically improved, and checked
for convergence, by including more variational parameters.

We will now apply this procedure explicitly to the two dimensional case. The
extraction of the factor $\nu_d$ in the definition of $\mL$ and $\mW$  removes
the density from the equations. We can get rid of the temperature dependence by
rescaling the velocities by the typical velocity $v_0$:
$\vv\rightarrow\vv/v_0$. The operator $\mW$ then becomes
\[
        (\mW A)(\vv_1) =
\left(w_0\gamma+
\frac{1}{2\sqrt\pi}\int
d\vv_2\frac{e^{-\frac{1}{2}|\vv_2|^2}}{2\pi}2|\vv_{21}|
\right)A(\vv_1),
\]
and $\mL$ is
\[
        (\mL A)(\vv_1) = \frac{1}{2\sqrt\pi}\int d\vv_2
        \frac{e^{-\frac{1}{2}|\vv_2|^2}}{2\pi}
        \int d\hn |\hn\cdot\vv_{21}|\left[A(\vv_1') + A(\vv_2')\right].
\]
As our variational eigenfunction we choose:  $A(\vv)=a + b_0 |\vv|^2 +
b_1|\vv|^4 + b_2 v_x + b_3v_y$.  $a$ is not a variational parameter, but is set
by the normalization  requirement, $\inp{A}{1}=1$.  The operator $\mL$ contains
no preferred direction, so the eigenfunctions are expected not to contain any
either.  Indeed, it turns out that the maximum in \Eq~(\ref{variational}) is
assumed for $(b_2=0,b_3=0)$. Therefore we will only work with the nontrivial
variational parameters $b_0$ and $b_1$. Because the variational eigenfunction
is linear in the parameters, there is an equivalent method that we will use. We
construct an orthonormal basis out of $1$, $|\vv|^2$ and $|\vv|^4$, and
determine the matrix elements of the operators $\mL$ and $\mW$. The largest
eigenvalue of the truncated matrix (in which all other matrix elements are set
to zero) is an approximation and lower bound to the real eigenvalue
$\Lambda_0$.

The basis vectors are
\[
        |1\rangle = 1,\quad |2\rangle=\sfrac{1}{2}|\vv|^2 -1,\quad
        |3\rangle=\sfrac{1}{8}|\vv|^4-|\vv|^2+1,
\]
which are orthonormal with respect to the inner product defined in
\Eq~(\ref{inner}). A tedious but straightforward calculation gives the
truncated matrices on this basis,
\begin{eqnarray*}
        \mW^{3\times3} =
        w_0\gamma\identity +
        \left(
        \begin{array}{ccc}
                1&\sfrac{1}{4}&-\sfrac{1}{32}\vspace{3pt}\\
                \sfrac{1}{4}&\sfrac{23}{16}&\sfrac{51}{128}\vspace{3pt}\\
                -\sfrac{1}{32}&\sfrac{51}{128}&\sfrac{1809}{1024}
        \end{array}
        \right)&\quad\mbox{and}\quad&
        \mL^{3\times3}=
        \left(
        \begin{array}{ccc}
                2&\sfrac{1}{2}&-\sfrac{1}{16}\vspace{3pt}\\
                \sfrac{1}{2}&\sfrac{11}{8}&\sfrac{27}{64}\vspace{3pt}\\
                -\sfrac{1}{16}&\sfrac{27}{64}&\sfrac{641}{512}
        \end{array}
        \right).
\end{eqnarray*}

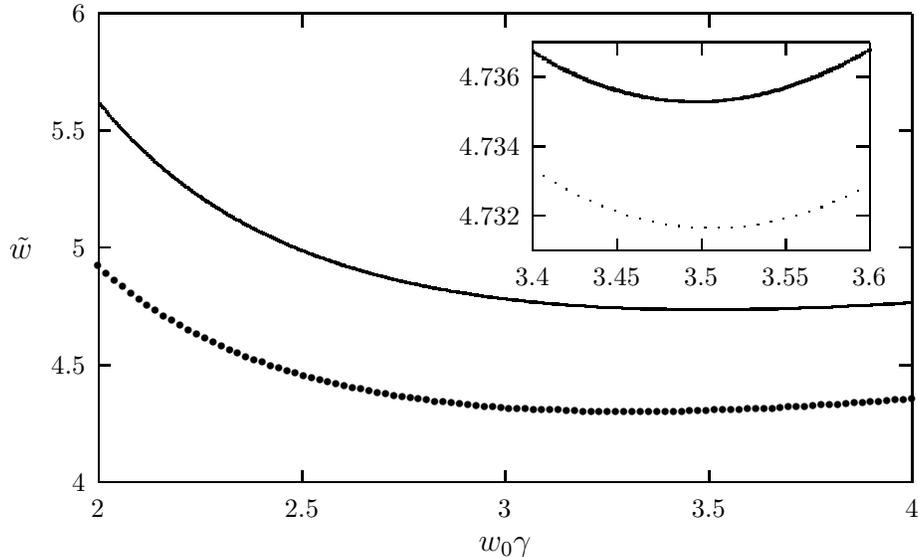
\begin{figure}[t]
\centerline{
\setlength{\unitlength}{0.240900pt}
\ifx\plotpoint\undefined\newsavebox{\plotpoint}\fi
\sbox{\plotpoint}{\rule[-0.200pt]{0.400pt}{0.400pt}}\begin{picture}(1500,900)(0,0)
\put(1050,630){\makebox(0,0){\setlength{\unitlength}{0.240900pt}
\ifx\plotpoint\undefined\newsavebox{\plotpoint}\fi
\sbox{\plotpoint}{\rule[-0.200pt]{0.400pt}{0.400pt}}\begin{picture}(750,450)(0,0)
\font\gnuplot=cmr10 at 10pt
\gnuplot
\sbox{\plotpoint}{\rule[-0.200pt]{0.400pt}{0.400pt}}\put(160.0,137.0){\rule[-0.200pt]{4.818pt}{0.400pt}}
\put(140,137){\makebox(0,0)[r]{4.732}}
\put(669.0,137.0){\rule[-0.200pt]{4.818pt}{0.400pt}}
\put(160.0,246.0){\rule[-0.200pt]{4.818pt}{0.400pt}}
\put(140,246){\makebox(0,0)[r]{4.734}}
\put(669.0,246.0){\rule[-0.200pt]{4.818pt}{0.400pt}}
\put(160.0,355.0){\rule[-0.200pt]{4.818pt}{0.400pt}}
\put(140,355){\makebox(0,0)[r]{4.736}}
\put(669.0,355.0){\rule[-0.200pt]{4.818pt}{0.400pt}}
\put(160.0,82.0){\rule[-0.200pt]{0.400pt}{4.818pt}}
\put(160,41){\makebox(0,0){3.4}}
\put(160.0,390.0){\rule[-0.200pt]{0.400pt}{4.818pt}}
\put(292.0,82.0){\rule[-0.200pt]{0.400pt}{4.818pt}}
\put(292,41){\makebox(0,0){3.45}}
\put(292.0,390.0){\rule[-0.200pt]{0.400pt}{4.818pt}}
\put(425.0,82.0){\rule[-0.200pt]{0.400pt}{4.818pt}}
\put(425,41){\makebox(0,0){3.5}}
\put(425.0,390.0){\rule[-0.200pt]{0.400pt}{4.818pt}}
\put(557.0,82.0){\rule[-0.200pt]{0.400pt}{4.818pt}}
\put(557,41){\makebox(0,0){3.55}}
\put(557.0,390.0){\rule[-0.200pt]{0.400pt}{4.818pt}}
\put(689.0,82.0){\rule[-0.200pt]{0.400pt}{4.818pt}}
\put(689,41){\makebox(0,0){3.6}}
\put(689.0,390.0){\rule[-0.200pt]{0.400pt}{4.818pt}}
\put(160.0,82.0){\rule[-0.200pt]{127.436pt}{0.400pt}}
\put(689.0,82.0){\rule[-0.200pt]{0.400pt}{79.015pt}}
\put(160.0,410.0){\rule[-0.200pt]{127.436pt}{0.400pt}}
\put(160.0,82.0){\rule[-0.200pt]{0.400pt}{79.015pt}}
\put(160,211){\usebox{\plotpoint}}
\put(160.00,211.00){\usebox{\plotpoint}}
\put(177.21,199.41){\usebox{\plotpoint}}
\put(195.06,188.82){\usebox{\plotpoint}}
\put(212.83,178.10){\usebox{\plotpoint}}
\put(231.26,168.57){\usebox{\plotpoint}}
\put(250.24,160.18){\usebox{\plotpoint}}
\put(269.21,151.77){\usebox{\plotpoint}}
\put(288.58,144.32){\usebox{\plotpoint}}
\put(308.09,137.34){\usebox{\plotpoint}}
\put(328.18,132.18){\usebox{\plotpoint}}
\put(348.32,127.23){\usebox{\plotpoint}}
\put(368.72,123.47){\usebox{\plotpoint}}
\put(389.27,120.67){\usebox{\plotpoint}}
\put(409.86,118.17){\usebox{\plotpoint}}
\put(430.61,118.00){\usebox{\plotpoint}}
\put(451.36,118.00){\usebox{\plotpoint}}
\put(472.09,118.62){\usebox{\plotpoint}}
\put(492.78,120.27){\usebox{\plotpoint}}
\put(513.29,123.43){\usebox{\plotpoint}}
\put(533.58,127.77){\usebox{\plotpoint}}
\put(553.84,132.27){\usebox{\plotpoint}}
\put(573.99,137.23){\usebox{\plotpoint}}
\put(593.57,144.07){\usebox{\plotpoint}}
\put(613.26,150.51){\usebox{\plotpoint}}
\put(632.36,158.60){\usebox{\plotpoint}}
\put(651.33,167.00){\usebox{\plotpoint}}
\put(670.10,175.82){\usebox{\plotpoint}}
\put(688.76,184.89){\usebox{\plotpoint}}
\put(689,185){\usebox{\plotpoint}}
\sbox{\plotpoint}{\rule[-0.400pt]{0.800pt}{0.800pt}}\put(160,395){\usebox{\plotpoint}}
\multiput(160.00,393.07)(0.797,-0.536){5}{\rule{1.400pt}{0.129pt}}
\multiput(160.00,393.34)(6.094,-6.000){2}{\rule{0.700pt}{0.800pt}}
\multiput(169.00,387.06)(1.096,-0.560){3}{\rule{1.640pt}{0.135pt}}
\multiput(169.00,387.34)(5.596,-5.000){2}{\rule{0.820pt}{0.800pt}}
\multiput(178.00,382.06)(0.928,-0.560){3}{\rule{1.480pt}{0.135pt}}
\multiput(178.00,382.34)(4.928,-5.000){2}{\rule{0.740pt}{0.800pt}}
\multiput(186.00,377.06)(1.096,-0.560){3}{\rule{1.640pt}{0.135pt}}
\multiput(186.00,377.34)(5.596,-5.000){2}{\rule{0.820pt}{0.800pt}}
\multiput(195.00,372.06)(1.096,-0.560){3}{\rule{1.640pt}{0.135pt}}
\multiput(195.00,372.34)(5.596,-5.000){2}{\rule{0.820pt}{0.800pt}}
\put(204,365.34){\rule{2.000pt}{0.800pt}}
\multiput(204.00,367.34)(4.849,-4.000){2}{\rule{1.000pt}{0.800pt}}
\put(213,361.34){\rule{2.000pt}{0.800pt}}
\multiput(213.00,363.34)(4.849,-4.000){2}{\rule{1.000pt}{0.800pt}}
\put(222,357.34){\rule{2.000pt}{0.800pt}}
\multiput(222.00,359.34)(4.849,-4.000){2}{\rule{1.000pt}{0.800pt}}
\put(231,353.34){\rule{1.800pt}{0.800pt}}
\multiput(231.00,355.34)(4.264,-4.000){2}{\rule{0.900pt}{0.800pt}}
\put(239,349.34){\rule{2.000pt}{0.800pt}}
\multiput(239.00,351.34)(4.849,-4.000){2}{\rule{1.000pt}{0.800pt}}
\put(248,345.84){\rule{2.168pt}{0.800pt}}
\multiput(248.00,347.34)(4.500,-3.000){2}{\rule{1.084pt}{0.800pt}}
\put(257,342.84){\rule{2.168pt}{0.800pt}}
\multiput(257.00,344.34)(4.500,-3.000){2}{\rule{1.084pt}{0.800pt}}
\put(266,339.34){\rule{2.000pt}{0.800pt}}
\multiput(266.00,341.34)(4.849,-4.000){2}{\rule{1.000pt}{0.800pt}}
\put(275,336.34){\rule{1.927pt}{0.800pt}}
\multiput(275.00,337.34)(4.000,-2.000){2}{\rule{0.964pt}{0.800pt}}
\put(283,333.84){\rule{2.168pt}{0.800pt}}
\multiput(283.00,335.34)(4.500,-3.000){2}{\rule{1.084pt}{0.800pt}}
\put(292,330.84){\rule{2.168pt}{0.800pt}}
\multiput(292.00,332.34)(4.500,-3.000){2}{\rule{1.084pt}{0.800pt}}
\put(301,328.34){\rule{2.168pt}{0.800pt}}
\multiput(301.00,329.34)(4.500,-2.000){2}{\rule{1.084pt}{0.800pt}}
\put(310,326.34){\rule{2.168pt}{0.800pt}}
\multiput(310.00,327.34)(4.500,-2.000){2}{\rule{1.084pt}{0.800pt}}
\put(319,324.34){\rule{2.168pt}{0.800pt}}
\multiput(319.00,325.34)(4.500,-2.000){2}{\rule{1.084pt}{0.800pt}}
\put(328,322.34){\rule{1.927pt}{0.800pt}}
\multiput(328.00,323.34)(4.000,-2.000){2}{\rule{0.964pt}{0.800pt}}
\put(336,320.84){\rule{2.168pt}{0.800pt}}
\multiput(336.00,321.34)(4.500,-1.000){2}{\rule{1.084pt}{0.800pt}}
\put(345,319.34){\rule{2.168pt}{0.800pt}}
\multiput(345.00,320.34)(4.500,-2.000){2}{\rule{1.084pt}{0.800pt}}
\put(354,317.84){\rule{2.168pt}{0.800pt}}
\multiput(354.00,318.34)(4.500,-1.000){2}{\rule{1.084pt}{0.800pt}}
\put(363,316.84){\rule{2.168pt}{0.800pt}}
\multiput(363.00,317.34)(4.500,-1.000){2}{\rule{1.084pt}{0.800pt}}
\put(372,315.84){\rule{1.927pt}{0.800pt}}
\multiput(372.00,316.34)(4.000,-1.000){2}{\rule{0.964pt}{0.800pt}}
\put(389,314.84){\rule{2.168pt}{0.800pt}}
\multiput(389.00,315.34)(4.500,-1.000){2}{\rule{1.084pt}{0.800pt}}
\put(380.0,317.0){\rule[-0.400pt]{2.168pt}{0.800pt}}
\put(442,314.84){\rule{2.168pt}{0.800pt}}
\multiput(442.00,314.34)(4.500,1.000){2}{\rule{1.084pt}{0.800pt}}
\put(451,315.84){\rule{2.168pt}{0.800pt}}
\multiput(451.00,315.34)(4.500,1.000){2}{\rule{1.084pt}{0.800pt}}
\put(460,316.84){\rule{2.168pt}{0.800pt}}
\multiput(460.00,316.34)(4.500,1.000){2}{\rule{1.084pt}{0.800pt}}
\put(469,317.84){\rule{1.927pt}{0.800pt}}
\multiput(469.00,317.34)(4.000,1.000){2}{\rule{0.964pt}{0.800pt}}
\put(477,318.84){\rule{2.168pt}{0.800pt}}
\multiput(477.00,318.34)(4.500,1.000){2}{\rule{1.084pt}{0.800pt}}
\put(486,320.34){\rule{2.168pt}{0.800pt}}
\multiput(486.00,319.34)(4.500,2.000){2}{\rule{1.084pt}{0.800pt}}
\put(495,321.84){\rule{2.168pt}{0.800pt}}
\multiput(495.00,321.34)(4.500,1.000){2}{\rule{1.084pt}{0.800pt}}
\put(504,323.34){\rule{2.168pt}{0.800pt}}
\multiput(504.00,322.34)(4.500,2.000){2}{\rule{1.084pt}{0.800pt}}
\put(513,325.34){\rule{1.927pt}{0.800pt}}
\multiput(513.00,324.34)(4.000,2.000){2}{\rule{0.964pt}{0.800pt}}
\put(521,327.34){\rule{2.168pt}{0.800pt}}
\multiput(521.00,326.34)(4.500,2.000){2}{\rule{1.084pt}{0.800pt}}
\put(530,329.84){\rule{2.168pt}{0.800pt}}
\multiput(530.00,328.34)(4.500,3.000){2}{\rule{1.084pt}{0.800pt}}
\put(539,332.34){\rule{2.168pt}{0.800pt}}
\multiput(539.00,331.34)(4.500,2.000){2}{\rule{1.084pt}{0.800pt}}
\put(548,334.84){\rule{2.168pt}{0.800pt}}
\multiput(548.00,333.34)(4.500,3.000){2}{\rule{1.084pt}{0.800pt}}
\put(557,337.84){\rule{2.168pt}{0.800pt}}
\multiput(557.00,336.34)(4.500,3.000){2}{\rule{1.084pt}{0.800pt}}
\put(566,340.84){\rule{1.927pt}{0.800pt}}
\multiput(566.00,339.34)(4.000,3.000){2}{\rule{0.964pt}{0.800pt}}
\put(574,343.84){\rule{2.168pt}{0.800pt}}
\multiput(574.00,342.34)(4.500,3.000){2}{\rule{1.084pt}{0.800pt}}
\put(583,346.84){\rule{2.168pt}{0.800pt}}
\multiput(583.00,345.34)(4.500,3.000){2}{\rule{1.084pt}{0.800pt}}
\put(592,350.34){\rule{2.000pt}{0.800pt}}
\multiput(592.00,348.34)(4.849,4.000){2}{\rule{1.000pt}{0.800pt}}
\put(601,353.84){\rule{2.168pt}{0.800pt}}
\multiput(601.00,352.34)(4.500,3.000){2}{\rule{1.084pt}{0.800pt}}
\put(610,357.34){\rule{1.800pt}{0.800pt}}
\multiput(610.00,355.34)(4.264,4.000){2}{\rule{0.900pt}{0.800pt}}
\put(618,361.34){\rule{2.000pt}{0.800pt}}
\multiput(618.00,359.34)(4.849,4.000){2}{\rule{1.000pt}{0.800pt}}
\put(627,365.34){\rule{2.000pt}{0.800pt}}
\multiput(627.00,363.34)(4.849,4.000){2}{\rule{1.000pt}{0.800pt}}
\put(636,369.34){\rule{2.000pt}{0.800pt}}
\multiput(636.00,367.34)(4.849,4.000){2}{\rule{1.000pt}{0.800pt}}
\multiput(645.00,374.38)(1.096,0.560){3}{\rule{1.640pt}{0.135pt}}
\multiput(645.00,371.34)(5.596,5.000){2}{\rule{0.820pt}{0.800pt}}
\multiput(654.00,379.38)(1.096,0.560){3}{\rule{1.640pt}{0.135pt}}
\multiput(654.00,376.34)(5.596,5.000){2}{\rule{0.820pt}{0.800pt}}
\put(663,383.34){\rule{1.800pt}{0.800pt}}
\multiput(663.00,381.34)(4.264,4.000){2}{\rule{0.900pt}{0.800pt}}
\multiput(671.00,388.38)(1.096,0.560){3}{\rule{1.640pt}{0.135pt}}
\multiput(671.00,385.34)(5.596,5.000){2}{\rule{0.820pt}{0.800pt}}
\multiput(680.00,393.38)(1.096,0.560){3}{\rule{1.640pt}{0.135pt}}
\multiput(680.00,390.34)(5.596,5.000){2}{\rule{0.820pt}{0.800pt}}
\put(398.0,316.0){\rule[-0.400pt]{10.600pt}{0.800pt}}
\put(689,397){\usebox{\plotpoint}}
\end{picture}}}
\font\gnuplot=cmr10 at 10pt
\gnuplot
\sbox{\plotpoint}{\rule[-0.200pt]{0.400pt}{0.400pt}}\put(161.0,123.0){\rule[-0.200pt]{4.818pt}{0.400pt}}
\put(141,123){\makebox(0,0)[r]{4}}
\put(1419.0,123.0){\rule[-0.200pt]{4.818pt}{0.400pt}}
\put(161.0,307.0){\rule[-0.200pt]{4.818pt}{0.400pt}}
\put(141,307){\makebox(0,0)[r]{4.5}}
\put(1419.0,307.0){\rule[-0.200pt]{4.818pt}{0.400pt}}
\put(161.0,492.0){\rule[-0.200pt]{4.818pt}{0.400pt}}
\put(141,492){\makebox(0,0)[r]{5}}
\put(1419.0,492.0){\rule[-0.200pt]{4.818pt}{0.400pt}}
\put(161.0,676.0){\rule[-0.200pt]{4.818pt}{0.400pt}}
\put(141,676){\makebox(0,0)[r]{5.5}}
\put(1419.0,676.0){\rule[-0.200pt]{4.818pt}{0.400pt}}
\put(161.0,860.0){\rule[-0.200pt]{4.818pt}{0.400pt}}
\put(141,860){\makebox(0,0)[r]{6}}
\put(1419.0,860.0){\rule[-0.200pt]{4.818pt}{0.400pt}}
\put(161.0,123.0){\rule[-0.200pt]{0.400pt}{4.818pt}}
\put(161,82){\makebox(0,0){2}}
\put(161.0,840.0){\rule[-0.200pt]{0.400pt}{4.818pt}}
\put(481.0,123.0){\rule[-0.200pt]{0.400pt}{4.818pt}}
\put(481,82){\makebox(0,0){2.5}}
\put(481.0,840.0){\rule[-0.200pt]{0.400pt}{4.818pt}}
\put(800.0,123.0){\rule[-0.200pt]{0.400pt}{4.818pt}}
\put(800,82){\makebox(0,0){3}}
\put(800.0,840.0){\rule[-0.200pt]{0.400pt}{4.818pt}}
\put(1120.0,123.0){\rule[-0.200pt]{0.400pt}{4.818pt}}
\put(1120,82){\makebox(0,0){3.5}}
\put(1120.0,840.0){\rule[-0.200pt]{0.400pt}{4.818pt}}
\put(1439.0,123.0){\rule[-0.200pt]{0.400pt}{4.818pt}}
\put(1439,82){\makebox(0,0){4}}
\put(1439.0,840.0){\rule[-0.200pt]{0.400pt}{4.818pt}}
\put(161.0,123.0){\rule[-0.200pt]{307.870pt}{0.400pt}}
\put(1439.0,123.0){\rule[-0.200pt]{0.400pt}{177.543pt}}
\put(161.0,860.0){\rule[-0.200pt]{307.870pt}{0.400pt}}
\put(40,491){\makebox(0,0){$\tilde w$}}
\put(800,21){\makebox(0,0){$w_0\gamma$}}
\put(161.0,123.0){\rule[-0.200pt]{0.400pt}{177.543pt}}
\sbox{\plotpoint}{\rule[-0.400pt]{0.800pt}{0.800pt}}\put(163,718){\usebox{\plotpoint}}
\put(163,715.34){\rule{0.482pt}{0.800pt}}
\multiput(163.00,716.34)(1.000,-2.000){2}{\rule{0.241pt}{0.800pt}}
\put(164.34,713){\rule{0.800pt}{0.723pt}}
\multiput(163.34,714.50)(2.000,-1.500){2}{\rule{0.800pt}{0.361pt}}
\put(167,710.34){\rule{0.723pt}{0.800pt}}
\multiput(167.00,711.34)(1.500,-2.000){2}{\rule{0.361pt}{0.800pt}}
\put(169.34,708){\rule{0.800pt}{0.723pt}}
\multiput(168.34,709.50)(2.000,-1.500){2}{\rule{0.800pt}{0.361pt}}
\put(172,705.34){\rule{0.482pt}{0.800pt}}
\multiput(172.00,706.34)(1.000,-2.000){2}{\rule{0.241pt}{0.800pt}}
\put(173.34,703){\rule{0.800pt}{0.723pt}}
\multiput(172.34,704.50)(2.000,-1.500){2}{\rule{0.800pt}{0.361pt}}
\put(176,700.34){\rule{0.482pt}{0.800pt}}
\multiput(176.00,701.34)(1.000,-2.000){2}{\rule{0.241pt}{0.800pt}}
\put(177.34,698){\rule{0.800pt}{0.723pt}}
\multiput(176.34,699.50)(2.000,-1.500){2}{\rule{0.800pt}{0.361pt}}
\put(180,695.34){\rule{0.482pt}{0.800pt}}
\multiput(180.00,696.34)(1.000,-2.000){2}{\rule{0.241pt}{0.800pt}}
\put(181.34,693){\rule{0.800pt}{0.723pt}}
\multiput(180.34,694.50)(2.000,-1.500){2}{\rule{0.800pt}{0.361pt}}
\put(184,690.34){\rule{0.723pt}{0.800pt}}
\multiput(184.00,691.34)(1.500,-2.000){2}{\rule{0.361pt}{0.800pt}}
\put(186.34,688){\rule{0.800pt}{0.723pt}}
\multiput(185.34,689.50)(2.000,-1.500){2}{\rule{0.800pt}{0.361pt}}
\put(189,685.34){\rule{0.482pt}{0.800pt}}
\multiput(189.00,686.34)(1.000,-2.000){2}{\rule{0.241pt}{0.800pt}}
\put(191,683.34){\rule{0.482pt}{0.800pt}}
\multiput(191.00,684.34)(1.000,-2.000){2}{\rule{0.241pt}{0.800pt}}
\put(192.34,681){\rule{0.800pt}{0.723pt}}
\multiput(191.34,682.50)(2.000,-1.500){2}{\rule{0.800pt}{0.361pt}}
\put(195,678.34){\rule{0.482pt}{0.800pt}}
\multiput(195.00,679.34)(1.000,-2.000){2}{\rule{0.241pt}{0.800pt}}
\put(197,676.34){\rule{0.482pt}{0.800pt}}
\multiput(197.00,677.34)(1.000,-2.000){2}{\rule{0.241pt}{0.800pt}}
\put(199,674.34){\rule{0.482pt}{0.800pt}}
\multiput(199.00,675.34)(1.000,-2.000){2}{\rule{0.241pt}{0.800pt}}
\put(201,671.84){\rule{0.723pt}{0.800pt}}
\multiput(201.00,673.34)(1.500,-3.000){2}{\rule{0.361pt}{0.800pt}}
\put(204,669.34){\rule{0.482pt}{0.800pt}}
\multiput(204.00,670.34)(1.000,-2.000){2}{\rule{0.241pt}{0.800pt}}
\put(206,667.34){\rule{0.482pt}{0.800pt}}
\multiput(206.00,668.34)(1.000,-2.000){2}{\rule{0.241pt}{0.800pt}}
\put(208,665.34){\rule{0.482pt}{0.800pt}}
\multiput(208.00,666.34)(1.000,-2.000){2}{\rule{0.241pt}{0.800pt}}
\put(209.34,663){\rule{0.800pt}{0.723pt}}
\multiput(208.34,664.50)(2.000,-1.500){2}{\rule{0.800pt}{0.361pt}}
\put(212,660.34){\rule{0.482pt}{0.800pt}}
\multiput(212.00,661.34)(1.000,-2.000){2}{\rule{0.241pt}{0.800pt}}
\put(214,658.34){\rule{0.482pt}{0.800pt}}
\multiput(214.00,659.34)(1.000,-2.000){2}{\rule{0.241pt}{0.800pt}}
\put(216,656.34){\rule{0.723pt}{0.800pt}}
\multiput(216.00,657.34)(1.500,-2.000){2}{\rule{0.361pt}{0.800pt}}
\put(219,654.34){\rule{0.482pt}{0.800pt}}
\multiput(219.00,655.34)(1.000,-2.000){2}{\rule{0.241pt}{0.800pt}}
\put(221,652.34){\rule{0.482pt}{0.800pt}}
\multiput(221.00,653.34)(1.000,-2.000){2}{\rule{0.241pt}{0.800pt}}
\put(223,650.34){\rule{0.482pt}{0.800pt}}
\multiput(223.00,651.34)(1.000,-2.000){2}{\rule{0.241pt}{0.800pt}}
\put(225,648.34){\rule{0.482pt}{0.800pt}}
\multiput(225.00,649.34)(1.000,-2.000){2}{\rule{0.241pt}{0.800pt}}
\put(227,646.34){\rule{0.482pt}{0.800pt}}
\multiput(227.00,647.34)(1.000,-2.000){2}{\rule{0.241pt}{0.800pt}}
\put(229,644.34){\rule{0.482pt}{0.800pt}}
\multiput(229.00,645.34)(1.000,-2.000){2}{\rule{0.241pt}{0.800pt}}
\put(231,642.34){\rule{0.482pt}{0.800pt}}
\multiput(231.00,643.34)(1.000,-2.000){2}{\rule{0.241pt}{0.800pt}}
\put(233,640.34){\rule{0.723pt}{0.800pt}}
\multiput(233.00,641.34)(1.500,-2.000){2}{\rule{0.361pt}{0.800pt}}
\put(236,638.34){\rule{0.482pt}{0.800pt}}
\multiput(236.00,639.34)(1.000,-2.000){2}{\rule{0.241pt}{0.800pt}}
\put(238,636.34){\rule{0.482pt}{0.800pt}}
\multiput(238.00,637.34)(1.000,-2.000){2}{\rule{0.241pt}{0.800pt}}
\put(240,634.34){\rule{0.482pt}{0.800pt}}
\multiput(240.00,635.34)(1.000,-2.000){2}{\rule{0.241pt}{0.800pt}}
\put(242,632.34){\rule{0.482pt}{0.800pt}}
\multiput(242.00,633.34)(1.000,-2.000){2}{\rule{0.241pt}{0.800pt}}
\put(244,630.34){\rule{0.482pt}{0.800pt}}
\multiput(244.00,631.34)(1.000,-2.000){2}{\rule{0.241pt}{0.800pt}}
\put(246,628.34){\rule{0.482pt}{0.800pt}}
\multiput(246.00,629.34)(1.000,-2.000){2}{\rule{0.241pt}{0.800pt}}
\put(248,626.34){\rule{0.482pt}{0.800pt}}
\multiput(248.00,627.34)(1.000,-2.000){2}{\rule{0.241pt}{0.800pt}}
\put(250,624.34){\rule{0.723pt}{0.800pt}}
\multiput(250.00,625.34)(1.500,-2.000){2}{\rule{0.361pt}{0.800pt}}
\put(253,622.34){\rule{0.482pt}{0.800pt}}
\multiput(253.00,623.34)(1.000,-2.000){2}{\rule{0.241pt}{0.800pt}}
\put(255,620.34){\rule{0.482pt}{0.800pt}}
\multiput(255.00,621.34)(1.000,-2.000){2}{\rule{0.241pt}{0.800pt}}
\put(257,618.84){\rule{0.482pt}{0.800pt}}
\multiput(257.00,619.34)(1.000,-1.000){2}{\rule{0.241pt}{0.800pt}}
\put(259,617.34){\rule{0.482pt}{0.800pt}}
\multiput(259.00,618.34)(1.000,-2.000){2}{\rule{0.241pt}{0.800pt}}
\put(261,615.34){\rule{0.482pt}{0.800pt}}
\multiput(261.00,616.34)(1.000,-2.000){2}{\rule{0.241pt}{0.800pt}}
\put(263,613.34){\rule{0.482pt}{0.800pt}}
\multiput(263.00,614.34)(1.000,-2.000){2}{\rule{0.241pt}{0.800pt}}
\put(265,611.34){\rule{0.723pt}{0.800pt}}
\multiput(265.00,612.34)(1.500,-2.000){2}{\rule{0.361pt}{0.800pt}}
\put(268,609.84){\rule{0.482pt}{0.800pt}}
\multiput(268.00,610.34)(1.000,-1.000){2}{\rule{0.241pt}{0.800pt}}
\put(270,608.34){\rule{0.482pt}{0.800pt}}
\multiput(270.00,609.34)(1.000,-2.000){2}{\rule{0.241pt}{0.800pt}}
\put(272,606.34){\rule{0.482pt}{0.800pt}}
\multiput(272.00,607.34)(1.000,-2.000){2}{\rule{0.241pt}{0.800pt}}
\put(274,604.34){\rule{0.482pt}{0.800pt}}
\multiput(274.00,605.34)(1.000,-2.000){2}{\rule{0.241pt}{0.800pt}}
\put(276,602.84){\rule{0.482pt}{0.800pt}}
\multiput(276.00,603.34)(1.000,-1.000){2}{\rule{0.241pt}{0.800pt}}
\put(278,601.34){\rule{0.482pt}{0.800pt}}
\multiput(278.00,602.34)(1.000,-2.000){2}{\rule{0.241pt}{0.800pt}}
\put(280,599.34){\rule{0.482pt}{0.800pt}}
\multiput(280.00,600.34)(1.000,-2.000){2}{\rule{0.241pt}{0.800pt}}
\put(282,597.84){\rule{0.723pt}{0.800pt}}
\multiput(282.00,598.34)(1.500,-1.000){2}{\rule{0.361pt}{0.800pt}}
\put(285,596.34){\rule{0.482pt}{0.800pt}}
\multiput(285.00,597.34)(1.000,-2.000){2}{\rule{0.241pt}{0.800pt}}
\put(287,594.34){\rule{0.482pt}{0.800pt}}
\multiput(287.00,595.34)(1.000,-2.000){2}{\rule{0.241pt}{0.800pt}}
\put(289,592.84){\rule{0.482pt}{0.800pt}}
\multiput(289.00,593.34)(1.000,-1.000){2}{\rule{0.241pt}{0.800pt}}
\put(291,591.34){\rule{0.482pt}{0.800pt}}
\multiput(291.00,592.34)(1.000,-2.000){2}{\rule{0.241pt}{0.800pt}}
\put(293,589.34){\rule{0.482pt}{0.800pt}}
\multiput(293.00,590.34)(1.000,-2.000){2}{\rule{0.241pt}{0.800pt}}
\put(295,587.84){\rule{0.482pt}{0.800pt}}
\multiput(295.00,588.34)(1.000,-1.000){2}{\rule{0.241pt}{0.800pt}}
\put(297,586.34){\rule{0.482pt}{0.800pt}}
\multiput(297.00,587.34)(1.000,-2.000){2}{\rule{0.241pt}{0.800pt}}
\put(299,584.84){\rule{0.723pt}{0.800pt}}
\multiput(299.00,585.34)(1.500,-1.000){2}{\rule{0.361pt}{0.800pt}}
\put(302,583.34){\rule{0.482pt}{0.800pt}}
\multiput(302.00,584.34)(1.000,-2.000){2}{\rule{0.241pt}{0.800pt}}
\put(304,581.84){\rule{0.482pt}{0.800pt}}
\multiput(304.00,582.34)(1.000,-1.000){2}{\rule{0.241pt}{0.800pt}}
\put(306,580.34){\rule{0.482pt}{0.800pt}}
\multiput(306.00,581.34)(1.000,-2.000){2}{\rule{0.241pt}{0.800pt}}
\put(308,578.34){\rule{0.482pt}{0.800pt}}
\multiput(308.00,579.34)(1.000,-2.000){2}{\rule{0.241pt}{0.800pt}}
\put(310,576.84){\rule{0.482pt}{0.800pt}}
\multiput(310.00,577.34)(1.000,-1.000){2}{\rule{0.241pt}{0.800pt}}
\put(312,575.34){\rule{0.482pt}{0.800pt}}
\multiput(312.00,576.34)(1.000,-2.000){2}{\rule{0.241pt}{0.800pt}}
\put(314,573.84){\rule{0.482pt}{0.800pt}}
\multiput(314.00,574.34)(1.000,-1.000){2}{\rule{0.241pt}{0.800pt}}
\put(316,572.34){\rule{0.723pt}{0.800pt}}
\multiput(316.00,573.34)(1.500,-2.000){2}{\rule{0.361pt}{0.800pt}}
\put(319,570.84){\rule{0.482pt}{0.800pt}}
\multiput(319.00,571.34)(1.000,-1.000){2}{\rule{0.241pt}{0.800pt}}
\put(321,569.84){\rule{0.482pt}{0.800pt}}
\multiput(321.00,570.34)(1.000,-1.000){2}{\rule{0.241pt}{0.800pt}}
\put(323,568.34){\rule{0.482pt}{0.800pt}}
\multiput(323.00,569.34)(1.000,-2.000){2}{\rule{0.241pt}{0.800pt}}
\put(325,566.84){\rule{0.482pt}{0.800pt}}
\multiput(325.00,567.34)(1.000,-1.000){2}{\rule{0.241pt}{0.800pt}}
\put(327,565.34){\rule{0.482pt}{0.800pt}}
\multiput(327.00,566.34)(1.000,-2.000){2}{\rule{0.241pt}{0.800pt}}
\put(329,563.84){\rule{0.482pt}{0.800pt}}
\multiput(329.00,564.34)(1.000,-1.000){2}{\rule{0.241pt}{0.800pt}}
\put(331,562.34){\rule{0.723pt}{0.800pt}}
\multiput(331.00,563.34)(1.500,-2.000){2}{\rule{0.361pt}{0.800pt}}
\put(334,560.84){\rule{0.482pt}{0.800pt}}
\multiput(334.00,561.34)(1.000,-1.000){2}{\rule{0.241pt}{0.800pt}}
\put(336,559.84){\rule{0.482pt}{0.800pt}}
\multiput(336.00,560.34)(1.000,-1.000){2}{\rule{0.241pt}{0.800pt}}
\put(338,558.34){\rule{0.482pt}{0.800pt}}
\multiput(338.00,559.34)(1.000,-2.000){2}{\rule{0.241pt}{0.800pt}}
\put(340,556.84){\rule{0.482pt}{0.800pt}}
\multiput(340.00,557.34)(1.000,-1.000){2}{\rule{0.241pt}{0.800pt}}
\put(342,555.84){\rule{0.482pt}{0.800pt}}
\multiput(342.00,556.34)(1.000,-1.000){2}{\rule{0.241pt}{0.800pt}}
\put(344,554.34){\rule{0.482pt}{0.800pt}}
\multiput(344.00,555.34)(1.000,-2.000){2}{\rule{0.241pt}{0.800pt}}
\put(346,552.84){\rule{0.482pt}{0.800pt}}
\multiput(346.00,553.34)(1.000,-1.000){2}{\rule{0.241pt}{0.800pt}}
\put(348,551.84){\rule{0.723pt}{0.800pt}}
\multiput(348.00,552.34)(1.500,-1.000){2}{\rule{0.361pt}{0.800pt}}
\put(351,550.34){\rule{0.482pt}{0.800pt}}
\multiput(351.00,551.34)(1.000,-2.000){2}{\rule{0.241pt}{0.800pt}}
\put(353,548.84){\rule{0.482pt}{0.800pt}}
\multiput(353.00,549.34)(1.000,-1.000){2}{\rule{0.241pt}{0.800pt}}
\put(355,547.84){\rule{0.482pt}{0.800pt}}
\multiput(355.00,548.34)(1.000,-1.000){2}{\rule{0.241pt}{0.800pt}}
\put(357,546.34){\rule{0.482pt}{0.800pt}}
\multiput(357.00,547.34)(1.000,-2.000){2}{\rule{0.241pt}{0.800pt}}
\put(359,544.84){\rule{0.482pt}{0.800pt}}
\multiput(359.00,545.34)(1.000,-1.000){2}{\rule{0.241pt}{0.800pt}}
\put(361,543.84){\rule{0.482pt}{0.800pt}}
\multiput(361.00,544.34)(1.000,-1.000){2}{\rule{0.241pt}{0.800pt}}
\put(363,542.34){\rule{0.482pt}{0.800pt}}
\multiput(363.00,543.34)(1.000,-2.000){2}{\rule{0.241pt}{0.800pt}}
\put(365,540.84){\rule{0.723pt}{0.800pt}}
\multiput(365.00,541.34)(1.500,-1.000){2}{\rule{0.361pt}{0.800pt}}
\put(368,539.84){\rule{0.482pt}{0.800pt}}
\multiput(368.00,540.34)(1.000,-1.000){2}{\rule{0.241pt}{0.800pt}}
\put(370,538.84){\rule{0.482pt}{0.800pt}}
\multiput(370.00,539.34)(1.000,-1.000){2}{\rule{0.241pt}{0.800pt}}
\put(372,537.84){\rule{0.482pt}{0.800pt}}
\multiput(372.00,538.34)(1.000,-1.000){2}{\rule{0.241pt}{0.800pt}}
\put(374,536.34){\rule{0.482pt}{0.800pt}}
\multiput(374.00,537.34)(1.000,-2.000){2}{\rule{0.241pt}{0.800pt}}
\put(376,534.84){\rule{0.482pt}{0.800pt}}
\multiput(376.00,535.34)(1.000,-1.000){2}{\rule{0.241pt}{0.800pt}}
\put(378,533.84){\rule{0.482pt}{0.800pt}}
\multiput(378.00,534.34)(1.000,-1.000){2}{\rule{0.241pt}{0.800pt}}
\put(380,532.84){\rule{0.723pt}{0.800pt}}
\multiput(380.00,533.34)(1.500,-1.000){2}{\rule{0.361pt}{0.800pt}}
\put(383,531.84){\rule{0.482pt}{0.800pt}}
\multiput(383.00,532.34)(1.000,-1.000){2}{\rule{0.241pt}{0.800pt}}
\put(385,530.34){\rule{0.482pt}{0.800pt}}
\multiput(385.00,531.34)(1.000,-2.000){2}{\rule{0.241pt}{0.800pt}}
\put(387,528.84){\rule{0.482pt}{0.800pt}}
\multiput(387.00,529.34)(1.000,-1.000){2}{\rule{0.241pt}{0.800pt}}
\put(389,527.84){\rule{0.482pt}{0.800pt}}
\multiput(389.00,528.34)(1.000,-1.000){2}{\rule{0.241pt}{0.800pt}}
\put(391,526.84){\rule{0.482pt}{0.800pt}}
\multiput(391.00,527.34)(1.000,-1.000){2}{\rule{0.241pt}{0.800pt}}
\put(393,525.84){\rule{0.482pt}{0.800pt}}
\multiput(393.00,526.34)(1.000,-1.000){2}{\rule{0.241pt}{0.800pt}}
\put(395,524.84){\rule{0.482pt}{0.800pt}}
\multiput(395.00,525.34)(1.000,-1.000){2}{\rule{0.241pt}{0.800pt}}
\put(397,523.84){\rule{0.723pt}{0.800pt}}
\multiput(397.00,524.34)(1.500,-1.000){2}{\rule{0.361pt}{0.800pt}}
\put(400,522.34){\rule{0.482pt}{0.800pt}}
\multiput(400.00,523.34)(1.000,-2.000){2}{\rule{0.241pt}{0.800pt}}
\put(402,520.84){\rule{0.482pt}{0.800pt}}
\multiput(402.00,521.34)(1.000,-1.000){2}{\rule{0.241pt}{0.800pt}}
\put(404,519.84){\rule{0.482pt}{0.800pt}}
\multiput(404.00,520.34)(1.000,-1.000){2}{\rule{0.241pt}{0.800pt}}
\put(406,518.84){\rule{0.482pt}{0.800pt}}
\multiput(406.00,519.34)(1.000,-1.000){2}{\rule{0.241pt}{0.800pt}}
\put(408,517.84){\rule{0.482pt}{0.800pt}}
\multiput(408.00,518.34)(1.000,-1.000){2}{\rule{0.241pt}{0.800pt}}
\put(410,516.84){\rule{0.482pt}{0.800pt}}
\multiput(410.00,517.34)(1.000,-1.000){2}{\rule{0.241pt}{0.800pt}}
\put(412,515.84){\rule{0.482pt}{0.800pt}}
\multiput(412.00,516.34)(1.000,-1.000){2}{\rule{0.241pt}{0.800pt}}
\put(414,514.84){\rule{0.723pt}{0.800pt}}
\multiput(414.00,515.34)(1.500,-1.000){2}{\rule{0.361pt}{0.800pt}}
\put(417,513.84){\rule{0.482pt}{0.800pt}}
\multiput(417.00,514.34)(1.000,-1.000){2}{\rule{0.241pt}{0.800pt}}
\put(419,512.84){\rule{0.482pt}{0.800pt}}
\multiput(419.00,513.34)(1.000,-1.000){2}{\rule{0.241pt}{0.800pt}}
\put(421,511.84){\rule{0.482pt}{0.800pt}}
\multiput(421.00,512.34)(1.000,-1.000){2}{\rule{0.241pt}{0.800pt}}
\put(423,510.84){\rule{0.482pt}{0.800pt}}
\multiput(423.00,511.34)(1.000,-1.000){2}{\rule{0.241pt}{0.800pt}}
\put(425,509.84){\rule{0.482pt}{0.800pt}}
\multiput(425.00,510.34)(1.000,-1.000){2}{\rule{0.241pt}{0.800pt}}
\put(427,508.84){\rule{0.482pt}{0.800pt}}
\multiput(427.00,509.34)(1.000,-1.000){2}{\rule{0.241pt}{0.800pt}}
\put(429,507.84){\rule{0.723pt}{0.800pt}}
\multiput(429.00,508.34)(1.500,-1.000){2}{\rule{0.361pt}{0.800pt}}
\put(432,506.84){\rule{0.482pt}{0.800pt}}
\multiput(432.00,507.34)(1.000,-1.000){2}{\rule{0.241pt}{0.800pt}}
\put(434,505.84){\rule{0.482pt}{0.800pt}}
\multiput(434.00,506.34)(1.000,-1.000){2}{\rule{0.241pt}{0.800pt}}
\put(436,504.84){\rule{0.482pt}{0.800pt}}
\multiput(436.00,505.34)(1.000,-1.000){2}{\rule{0.241pt}{0.800pt}}
\put(438,503.84){\rule{0.482pt}{0.800pt}}
\multiput(438.00,504.34)(1.000,-1.000){2}{\rule{0.241pt}{0.800pt}}
\put(440,502.84){\rule{0.482pt}{0.800pt}}
\multiput(440.00,503.34)(1.000,-1.000){2}{\rule{0.241pt}{0.800pt}}
\put(442,501.84){\rule{0.482pt}{0.800pt}}
\multiput(442.00,502.34)(1.000,-1.000){2}{\rule{0.241pt}{0.800pt}}
\put(444,500.84){\rule{0.482pt}{0.800pt}}
\multiput(444.00,501.34)(1.000,-1.000){2}{\rule{0.241pt}{0.800pt}}
\put(446,499.84){\rule{0.723pt}{0.800pt}}
\multiput(446.00,500.34)(1.500,-1.000){2}{\rule{0.361pt}{0.800pt}}
\put(449,498.84){\rule{0.482pt}{0.800pt}}
\multiput(449.00,499.34)(1.000,-1.000){2}{\rule{0.241pt}{0.800pt}}
\put(451,497.84){\rule{0.482pt}{0.800pt}}
\multiput(451.00,498.34)(1.000,-1.000){2}{\rule{0.241pt}{0.800pt}}
\put(453,496.84){\rule{0.482pt}{0.800pt}}
\multiput(453.00,497.34)(1.000,-1.000){2}{\rule{0.241pt}{0.800pt}}
\put(455,495.84){\rule{0.482pt}{0.800pt}}
\multiput(455.00,496.34)(1.000,-1.000){2}{\rule{0.241pt}{0.800pt}}
\put(457,494.84){\rule{0.482pt}{0.800pt}}
\multiput(457.00,495.34)(1.000,-1.000){2}{\rule{0.241pt}{0.800pt}}
\put(459,493.84){\rule{0.482pt}{0.800pt}}
\multiput(459.00,494.34)(1.000,-1.000){2}{\rule{0.241pt}{0.800pt}}
\put(461,492.84){\rule{0.482pt}{0.800pt}}
\multiput(461.00,493.34)(1.000,-1.000){2}{\rule{0.241pt}{0.800pt}}
\put(463,491.84){\rule{0.723pt}{0.800pt}}
\multiput(463.00,492.34)(1.500,-1.000){2}{\rule{0.361pt}{0.800pt}}
\put(468,490.84){\rule{0.482pt}{0.800pt}}
\multiput(468.00,491.34)(1.000,-1.000){2}{\rule{0.241pt}{0.800pt}}
\put(470,489.84){\rule{0.482pt}{0.800pt}}
\multiput(470.00,490.34)(1.000,-1.000){2}{\rule{0.241pt}{0.800pt}}
\put(472,488.84){\rule{0.482pt}{0.800pt}}
\multiput(472.00,489.34)(1.000,-1.000){2}{\rule{0.241pt}{0.800pt}}
\put(474,487.84){\rule{0.482pt}{0.800pt}}
\multiput(474.00,488.34)(1.000,-1.000){2}{\rule{0.241pt}{0.800pt}}
\put(476,486.84){\rule{0.482pt}{0.800pt}}
\multiput(476.00,487.34)(1.000,-1.000){2}{\rule{0.241pt}{0.800pt}}
\put(478,485.84){\rule{0.723pt}{0.800pt}}
\multiput(478.00,486.34)(1.500,-1.000){2}{\rule{0.361pt}{0.800pt}}
\put(466.0,493.0){\usebox{\plotpoint}}
\put(483,484.84){\rule{0.482pt}{0.800pt}}
\multiput(483.00,485.34)(1.000,-1.000){2}{\rule{0.241pt}{0.800pt}}
\put(485,483.84){\rule{0.482pt}{0.800pt}}
\multiput(485.00,484.34)(1.000,-1.000){2}{\rule{0.241pt}{0.800pt}}
\put(487,482.84){\rule{0.482pt}{0.800pt}}
\multiput(487.00,483.34)(1.000,-1.000){2}{\rule{0.241pt}{0.800pt}}
\put(489,481.84){\rule{0.482pt}{0.800pt}}
\multiput(489.00,482.34)(1.000,-1.000){2}{\rule{0.241pt}{0.800pt}}
\put(491,480.84){\rule{0.482pt}{0.800pt}}
\multiput(491.00,481.34)(1.000,-1.000){2}{\rule{0.241pt}{0.800pt}}
\put(481.0,487.0){\usebox{\plotpoint}}
\put(495,479.84){\rule{0.723pt}{0.800pt}}
\multiput(495.00,480.34)(1.500,-1.000){2}{\rule{0.361pt}{0.800pt}}
\put(498,478.84){\rule{0.482pt}{0.800pt}}
\multiput(498.00,479.34)(1.000,-1.000){2}{\rule{0.241pt}{0.800pt}}
\put(500,477.84){\rule{0.482pt}{0.800pt}}
\multiput(500.00,478.34)(1.000,-1.000){2}{\rule{0.241pt}{0.800pt}}
\put(502,476.84){\rule{0.482pt}{0.800pt}}
\multiput(502.00,477.34)(1.000,-1.000){2}{\rule{0.241pt}{0.800pt}}
\put(493.0,482.0){\usebox{\plotpoint}}
\put(506,475.84){\rule{0.482pt}{0.800pt}}
\multiput(506.00,476.34)(1.000,-1.000){2}{\rule{0.241pt}{0.800pt}}
\put(508,474.84){\rule{0.482pt}{0.800pt}}
\multiput(508.00,475.34)(1.000,-1.000){2}{\rule{0.241pt}{0.800pt}}
\put(510,473.84){\rule{0.482pt}{0.800pt}}
\multiput(510.00,474.34)(1.000,-1.000){2}{\rule{0.241pt}{0.800pt}}
\put(504.0,478.0){\usebox{\plotpoint}}
\put(515,472.84){\rule{0.482pt}{0.800pt}}
\multiput(515.00,473.34)(1.000,-1.000){2}{\rule{0.241pt}{0.800pt}}
\put(517,471.84){\rule{0.482pt}{0.800pt}}
\multiput(517.00,472.34)(1.000,-1.000){2}{\rule{0.241pt}{0.800pt}}
\put(519,470.84){\rule{0.482pt}{0.800pt}}
\multiput(519.00,471.34)(1.000,-1.000){2}{\rule{0.241pt}{0.800pt}}
\put(512.0,475.0){\usebox{\plotpoint}}
\put(523,469.84){\rule{0.482pt}{0.800pt}}
\multiput(523.00,470.34)(1.000,-1.000){2}{\rule{0.241pt}{0.800pt}}
\put(525,468.84){\rule{0.482pt}{0.800pt}}
\multiput(525.00,469.34)(1.000,-1.000){2}{\rule{0.241pt}{0.800pt}}
\put(527,467.84){\rule{0.482pt}{0.800pt}}
\multiput(527.00,468.34)(1.000,-1.000){2}{\rule{0.241pt}{0.800pt}}
\put(521.0,472.0){\usebox{\plotpoint}}
\put(532,466.84){\rule{0.482pt}{0.800pt}}
\multiput(532.00,467.34)(1.000,-1.000){2}{\rule{0.241pt}{0.800pt}}
\put(534,465.84){\rule{0.482pt}{0.800pt}}
\multiput(534.00,466.34)(1.000,-1.000){2}{\rule{0.241pt}{0.800pt}}
\put(529.0,469.0){\usebox{\plotpoint}}
\put(538,464.84){\rule{0.482pt}{0.800pt}}
\multiput(538.00,465.34)(1.000,-1.000){2}{\rule{0.241pt}{0.800pt}}
\put(540,463.84){\rule{0.482pt}{0.800pt}}
\multiput(540.00,464.34)(1.000,-1.000){2}{\rule{0.241pt}{0.800pt}}
\put(542,462.84){\rule{0.482pt}{0.800pt}}
\multiput(542.00,463.34)(1.000,-1.000){2}{\rule{0.241pt}{0.800pt}}
\put(536.0,467.0){\usebox{\plotpoint}}
\put(547,461.84){\rule{0.482pt}{0.800pt}}
\multiput(547.00,462.34)(1.000,-1.000){2}{\rule{0.241pt}{0.800pt}}
\put(549,460.84){\rule{0.482pt}{0.800pt}}
\multiput(549.00,461.34)(1.000,-1.000){2}{\rule{0.241pt}{0.800pt}}
\put(544.0,464.0){\usebox{\plotpoint}}
\put(553,459.84){\rule{0.482pt}{0.800pt}}
\multiput(553.00,460.34)(1.000,-1.000){2}{\rule{0.241pt}{0.800pt}}
\put(555,458.84){\rule{0.482pt}{0.800pt}}
\multiput(555.00,459.34)(1.000,-1.000){2}{\rule{0.241pt}{0.800pt}}
\put(551.0,462.0){\usebox{\plotpoint}}
\put(559,457.84){\rule{0.482pt}{0.800pt}}
\multiput(559.00,458.34)(1.000,-1.000){2}{\rule{0.241pt}{0.800pt}}
\put(557.0,460.0){\usebox{\plotpoint}}
\put(564,456.84){\rule{0.482pt}{0.800pt}}
\multiput(564.00,457.34)(1.000,-1.000){2}{\rule{0.241pt}{0.800pt}}
\put(566,455.84){\rule{0.482pt}{0.800pt}}
\multiput(566.00,456.34)(1.000,-1.000){2}{\rule{0.241pt}{0.800pt}}
\put(561.0,459.0){\usebox{\plotpoint}}
\put(570,454.84){\rule{0.482pt}{0.800pt}}
\multiput(570.00,455.34)(1.000,-1.000){2}{\rule{0.241pt}{0.800pt}}
\put(572,453.84){\rule{0.482pt}{0.800pt}}
\multiput(572.00,454.34)(1.000,-1.000){2}{\rule{0.241pt}{0.800pt}}
\put(568.0,457.0){\usebox{\plotpoint}}
\put(576,452.84){\rule{0.482pt}{0.800pt}}
\multiput(576.00,453.34)(1.000,-1.000){2}{\rule{0.241pt}{0.800pt}}
\put(574.0,455.0){\usebox{\plotpoint}}
\put(581,451.84){\rule{0.482pt}{0.800pt}}
\multiput(581.00,452.34)(1.000,-1.000){2}{\rule{0.241pt}{0.800pt}}
\put(583,450.84){\rule{0.482pt}{0.800pt}}
\multiput(583.00,451.34)(1.000,-1.000){2}{\rule{0.241pt}{0.800pt}}
\put(578.0,454.0){\usebox{\plotpoint}}
\put(587,449.84){\rule{0.482pt}{0.800pt}}
\multiput(587.00,450.34)(1.000,-1.000){2}{\rule{0.241pt}{0.800pt}}
\put(585.0,452.0){\usebox{\plotpoint}}
\put(591,448.84){\rule{0.482pt}{0.800pt}}
\multiput(591.00,449.34)(1.000,-1.000){2}{\rule{0.241pt}{0.800pt}}
\put(593,447.84){\rule{0.723pt}{0.800pt}}
\multiput(593.00,448.34)(1.500,-1.000){2}{\rule{0.361pt}{0.800pt}}
\put(589.0,451.0){\usebox{\plotpoint}}
\put(598,446.84){\rule{0.482pt}{0.800pt}}
\multiput(598.00,447.34)(1.000,-1.000){2}{\rule{0.241pt}{0.800pt}}
\put(596.0,449.0){\usebox{\plotpoint}}
\put(602,445.84){\rule{0.482pt}{0.800pt}}
\multiput(602.00,446.34)(1.000,-1.000){2}{\rule{0.241pt}{0.800pt}}
\put(600.0,448.0){\usebox{\plotpoint}}
\put(606,444.84){\rule{0.482pt}{0.800pt}}
\multiput(606.00,445.34)(1.000,-1.000){2}{\rule{0.241pt}{0.800pt}}
\put(604.0,447.0){\usebox{\plotpoint}}
\put(610,443.84){\rule{0.723pt}{0.800pt}}
\multiput(610.00,444.34)(1.500,-1.000){2}{\rule{0.361pt}{0.800pt}}
\put(613,442.84){\rule{0.482pt}{0.800pt}}
\multiput(613.00,443.34)(1.000,-1.000){2}{\rule{0.241pt}{0.800pt}}
\put(608.0,446.0){\usebox{\plotpoint}}
\put(617,441.84){\rule{0.482pt}{0.800pt}}
\multiput(617.00,442.34)(1.000,-1.000){2}{\rule{0.241pt}{0.800pt}}
\put(615.0,444.0){\usebox{\plotpoint}}
\put(621,440.84){\rule{0.482pt}{0.800pt}}
\multiput(621.00,441.34)(1.000,-1.000){2}{\rule{0.241pt}{0.800pt}}
\put(619.0,443.0){\usebox{\plotpoint}}
\put(625,439.84){\rule{0.482pt}{0.800pt}}
\multiput(625.00,440.34)(1.000,-1.000){2}{\rule{0.241pt}{0.800pt}}
\put(623.0,442.0){\usebox{\plotpoint}}
\put(630,438.84){\rule{0.482pt}{0.800pt}}
\multiput(630.00,439.34)(1.000,-1.000){2}{\rule{0.241pt}{0.800pt}}
\put(627.0,441.0){\usebox{\plotpoint}}
\put(634,437.84){\rule{0.482pt}{0.800pt}}
\multiput(634.00,438.34)(1.000,-1.000){2}{\rule{0.241pt}{0.800pt}}
\put(632.0,440.0){\usebox{\plotpoint}}
\put(638,436.84){\rule{0.482pt}{0.800pt}}
\multiput(638.00,437.34)(1.000,-1.000){2}{\rule{0.241pt}{0.800pt}}
\put(636.0,439.0){\usebox{\plotpoint}}
\put(642,435.84){\rule{0.723pt}{0.800pt}}
\multiput(642.00,436.34)(1.500,-1.000){2}{\rule{0.361pt}{0.800pt}}
\put(640.0,438.0){\usebox{\plotpoint}}
\put(647,434.84){\rule{0.482pt}{0.800pt}}
\multiput(647.00,435.34)(1.000,-1.000){2}{\rule{0.241pt}{0.800pt}}
\put(645.0,437.0){\usebox{\plotpoint}}
\put(651,433.84){\rule{0.482pt}{0.800pt}}
\multiput(651.00,434.34)(1.000,-1.000){2}{\rule{0.241pt}{0.800pt}}
\put(649.0,436.0){\usebox{\plotpoint}}
\put(657,432.84){\rule{0.482pt}{0.800pt}}
\multiput(657.00,433.34)(1.000,-1.000){2}{\rule{0.241pt}{0.800pt}}
\put(653.0,435.0){\rule[-0.400pt]{0.964pt}{0.800pt}}
\put(662,431.84){\rule{0.482pt}{0.800pt}}
\multiput(662.00,432.34)(1.000,-1.000){2}{\rule{0.241pt}{0.800pt}}
\put(659.0,434.0){\usebox{\plotpoint}}
\put(666,430.84){\rule{0.482pt}{0.800pt}}
\multiput(666.00,431.34)(1.000,-1.000){2}{\rule{0.241pt}{0.800pt}}
\put(664.0,433.0){\usebox{\plotpoint}}
\put(670,429.84){\rule{0.482pt}{0.800pt}}
\multiput(670.00,430.34)(1.000,-1.000){2}{\rule{0.241pt}{0.800pt}}
\put(668.0,432.0){\usebox{\plotpoint}}
\put(676,428.84){\rule{0.723pt}{0.800pt}}
\multiput(676.00,429.34)(1.500,-1.000){2}{\rule{0.361pt}{0.800pt}}
\put(672.0,431.0){\rule[-0.400pt]{0.964pt}{0.800pt}}
\put(681,427.84){\rule{0.482pt}{0.800pt}}
\multiput(681.00,428.34)(1.000,-1.000){2}{\rule{0.241pt}{0.800pt}}
\put(679.0,430.0){\usebox{\plotpoint}}
\put(685,426.84){\rule{0.482pt}{0.800pt}}
\multiput(685.00,427.34)(1.000,-1.000){2}{\rule{0.241pt}{0.800pt}}
\put(683.0,429.0){\usebox{\plotpoint}}
\put(691,425.84){\rule{0.482pt}{0.800pt}}
\multiput(691.00,426.34)(1.000,-1.000){2}{\rule{0.241pt}{0.800pt}}
\put(687.0,428.0){\rule[-0.400pt]{0.964pt}{0.800pt}}
\put(696,424.84){\rule{0.482pt}{0.800pt}}
\multiput(696.00,425.34)(1.000,-1.000){2}{\rule{0.241pt}{0.800pt}}
\put(693.0,427.0){\usebox{\plotpoint}}
\put(702,423.84){\rule{0.482pt}{0.800pt}}
\multiput(702.00,424.34)(1.000,-1.000){2}{\rule{0.241pt}{0.800pt}}
\put(698.0,426.0){\rule[-0.400pt]{0.964pt}{0.800pt}}
\put(708,422.84){\rule{0.723pt}{0.800pt}}
\multiput(708.00,423.34)(1.500,-1.000){2}{\rule{0.361pt}{0.800pt}}
\put(704.0,425.0){\rule[-0.400pt]{0.964pt}{0.800pt}}
\put(713,421.84){\rule{0.482pt}{0.800pt}}
\multiput(713.00,422.34)(1.000,-1.000){2}{\rule{0.241pt}{0.800pt}}
\put(711.0,424.0){\usebox{\plotpoint}}
\put(719,420.84){\rule{0.482pt}{0.800pt}}
\multiput(719.00,421.34)(1.000,-1.000){2}{\rule{0.241pt}{0.800pt}}
\put(715.0,423.0){\rule[-0.400pt]{0.964pt}{0.800pt}}
\put(725,419.84){\rule{0.723pt}{0.800pt}}
\multiput(725.00,420.34)(1.500,-1.000){2}{\rule{0.361pt}{0.800pt}}
\put(721.0,422.0){\rule[-0.400pt]{0.964pt}{0.800pt}}
\put(732,418.84){\rule{0.482pt}{0.800pt}}
\multiput(732.00,419.34)(1.000,-1.000){2}{\rule{0.241pt}{0.800pt}}
\put(728.0,421.0){\rule[-0.400pt]{0.964pt}{0.800pt}}
\put(738,417.84){\rule{0.482pt}{0.800pt}}
\multiput(738.00,418.34)(1.000,-1.000){2}{\rule{0.241pt}{0.800pt}}
\put(734.0,420.0){\rule[-0.400pt]{0.964pt}{0.800pt}}
\put(745,416.84){\rule{0.482pt}{0.800pt}}
\multiput(745.00,417.34)(1.000,-1.000){2}{\rule{0.241pt}{0.800pt}}
\put(740.0,419.0){\rule[-0.400pt]{1.204pt}{0.800pt}}
\put(751,415.84){\rule{0.482pt}{0.800pt}}
\multiput(751.00,416.34)(1.000,-1.000){2}{\rule{0.241pt}{0.800pt}}
\put(747.0,418.0){\rule[-0.400pt]{0.964pt}{0.800pt}}
\put(757,414.84){\rule{0.723pt}{0.800pt}}
\multiput(757.00,415.34)(1.500,-1.000){2}{\rule{0.361pt}{0.800pt}}
\put(753.0,417.0){\rule[-0.400pt]{0.964pt}{0.800pt}}
\put(764,413.84){\rule{0.482pt}{0.800pt}}
\multiput(764.00,414.34)(1.000,-1.000){2}{\rule{0.241pt}{0.800pt}}
\put(760.0,416.0){\rule[-0.400pt]{0.964pt}{0.800pt}}
\put(772,412.84){\rule{0.482pt}{0.800pt}}
\multiput(772.00,413.34)(1.000,-1.000){2}{\rule{0.241pt}{0.800pt}}
\put(766.0,415.0){\rule[-0.400pt]{1.445pt}{0.800pt}}
\put(779,411.84){\rule{0.482pt}{0.800pt}}
\multiput(779.00,412.34)(1.000,-1.000){2}{\rule{0.241pt}{0.800pt}}
\put(774.0,414.0){\rule[-0.400pt]{1.204pt}{0.800pt}}
\put(787,410.84){\rule{0.482pt}{0.800pt}}
\multiput(787.00,411.34)(1.000,-1.000){2}{\rule{0.241pt}{0.800pt}}
\put(781.0,413.0){\rule[-0.400pt]{1.445pt}{0.800pt}}
\put(794,409.84){\rule{0.482pt}{0.800pt}}
\multiput(794.00,410.34)(1.000,-1.000){2}{\rule{0.241pt}{0.800pt}}
\put(789.0,412.0){\rule[-0.400pt]{1.204pt}{0.800pt}}
\put(802,408.84){\rule{0.482pt}{0.800pt}}
\multiput(802.00,409.34)(1.000,-1.000){2}{\rule{0.241pt}{0.800pt}}
\put(796.0,411.0){\rule[-0.400pt]{1.445pt}{0.800pt}}
\put(811,407.84){\rule{0.482pt}{0.800pt}}
\multiput(811.00,408.34)(1.000,-1.000){2}{\rule{0.241pt}{0.800pt}}
\put(804.0,410.0){\rule[-0.400pt]{1.686pt}{0.800pt}}
\put(819,406.84){\rule{0.482pt}{0.800pt}}
\multiput(819.00,407.34)(1.000,-1.000){2}{\rule{0.241pt}{0.800pt}}
\put(813.0,409.0){\rule[-0.400pt]{1.445pt}{0.800pt}}
\put(830,405.84){\rule{0.482pt}{0.800pt}}
\multiput(830.00,406.34)(1.000,-1.000){2}{\rule{0.241pt}{0.800pt}}
\put(821.0,408.0){\rule[-0.400pt]{2.168pt}{0.800pt}}
\put(838,404.84){\rule{0.482pt}{0.800pt}}
\multiput(838.00,405.34)(1.000,-1.000){2}{\rule{0.241pt}{0.800pt}}
\put(832.0,407.0){\rule[-0.400pt]{1.445pt}{0.800pt}}
\put(849,403.84){\rule{0.482pt}{0.800pt}}
\multiput(849.00,404.34)(1.000,-1.000){2}{\rule{0.241pt}{0.800pt}}
\put(840.0,406.0){\rule[-0.400pt]{2.168pt}{0.800pt}}
\put(860,402.84){\rule{0.482pt}{0.800pt}}
\multiput(860.00,403.34)(1.000,-1.000){2}{\rule{0.241pt}{0.800pt}}
\put(851.0,405.0){\rule[-0.400pt]{2.168pt}{0.800pt}}
\put(872,401.84){\rule{0.723pt}{0.800pt}}
\multiput(872.00,402.34)(1.500,-1.000){2}{\rule{0.361pt}{0.800pt}}
\put(862.0,404.0){\rule[-0.400pt]{2.409pt}{0.800pt}}
\put(883,400.84){\rule{0.482pt}{0.800pt}}
\multiput(883.00,401.34)(1.000,-1.000){2}{\rule{0.241pt}{0.800pt}}
\put(875.0,403.0){\rule[-0.400pt]{1.927pt}{0.800pt}}
\put(896,399.84){\rule{0.482pt}{0.800pt}}
\multiput(896.00,400.34)(1.000,-1.000){2}{\rule{0.241pt}{0.800pt}}
\put(885.0,402.0){\rule[-0.400pt]{2.650pt}{0.800pt}}
\put(911,398.84){\rule{0.482pt}{0.800pt}}
\multiput(911.00,399.34)(1.000,-1.000){2}{\rule{0.241pt}{0.800pt}}
\put(898.0,401.0){\rule[-0.400pt]{3.132pt}{0.800pt}}
\put(926,397.84){\rule{0.482pt}{0.800pt}}
\multiput(926.00,398.34)(1.000,-1.000){2}{\rule{0.241pt}{0.800pt}}
\put(913.0,400.0){\rule[-0.400pt]{3.132pt}{0.800pt}}
\put(943,396.84){\rule{0.482pt}{0.800pt}}
\multiput(943.00,397.34)(1.000,-1.000){2}{\rule{0.241pt}{0.800pt}}
\put(928.0,399.0){\rule[-0.400pt]{3.613pt}{0.800pt}}
\put(962,395.84){\rule{0.482pt}{0.800pt}}
\multiput(962.00,396.34)(1.000,-1.000){2}{\rule{0.241pt}{0.800pt}}
\put(945.0,398.0){\rule[-0.400pt]{4.095pt}{0.800pt}}
\put(983,394.84){\rule{0.482pt}{0.800pt}}
\multiput(983.00,395.34)(1.000,-1.000){2}{\rule{0.241pt}{0.800pt}}
\put(964.0,397.0){\rule[-0.400pt]{4.577pt}{0.800pt}}
\put(1013,393.84){\rule{0.482pt}{0.800pt}}
\multiput(1013.00,394.34)(1.000,-1.000){2}{\rule{0.241pt}{0.800pt}}
\put(985.0,396.0){\rule[-0.400pt]{6.745pt}{0.800pt}}
\put(1053,392.84){\rule{0.723pt}{0.800pt}}
\multiput(1053.00,393.34)(1.500,-1.000){2}{\rule{0.361pt}{0.800pt}}
\put(1015.0,395.0){\rule[-0.400pt]{9.154pt}{0.800pt}}
\put(1183,392.84){\rule{0.723pt}{0.800pt}}
\multiput(1183.00,392.34)(1.500,1.000){2}{\rule{0.361pt}{0.800pt}}
\put(1056.0,394.0){\rule[-0.400pt]{30.594pt}{0.800pt}}
\put(1228,393.84){\rule{0.482pt}{0.800pt}}
\multiput(1228.00,393.34)(1.000,1.000){2}{\rule{0.241pt}{0.800pt}}
\put(1186.0,395.0){\rule[-0.400pt]{10.118pt}{0.800pt}}
\put(1262,394.84){\rule{0.482pt}{0.800pt}}
\multiput(1262.00,394.34)(1.000,1.000){2}{\rule{0.241pt}{0.800pt}}
\put(1230.0,396.0){\rule[-0.400pt]{7.709pt}{0.800pt}}
\put(1290,395.84){\rule{0.482pt}{0.800pt}}
\multiput(1290.00,395.34)(1.000,1.000){2}{\rule{0.241pt}{0.800pt}}
\put(1264.0,397.0){\rule[-0.400pt]{6.263pt}{0.800pt}}
\put(1313,396.84){\rule{0.482pt}{0.800pt}}
\multiput(1313.00,396.34)(1.000,1.000){2}{\rule{0.241pt}{0.800pt}}
\put(1292.0,398.0){\rule[-0.400pt]{5.059pt}{0.800pt}}
\put(1337,397.84){\rule{0.482pt}{0.800pt}}
\multiput(1337.00,397.34)(1.000,1.000){2}{\rule{0.241pt}{0.800pt}}
\put(1315.0,399.0){\rule[-0.400pt]{5.300pt}{0.800pt}}
\put(1356,398.84){\rule{0.482pt}{0.800pt}}
\multiput(1356.00,398.34)(1.000,1.000){2}{\rule{0.241pt}{0.800pt}}
\put(1339.0,400.0){\rule[-0.400pt]{4.095pt}{0.800pt}}
\put(1375,399.84){\rule{0.482pt}{0.800pt}}
\multiput(1375.00,399.34)(1.000,1.000){2}{\rule{0.241pt}{0.800pt}}
\put(1358.0,401.0){\rule[-0.400pt]{4.095pt}{0.800pt}}
\put(1394,400.84){\rule{0.482pt}{0.800pt}}
\multiput(1394.00,400.34)(1.000,1.000){2}{\rule{0.241pt}{0.800pt}}
\put(1377.0,402.0){\rule[-0.400pt]{4.095pt}{0.800pt}}
\put(1411,401.84){\rule{0.482pt}{0.800pt}}
\multiput(1411.00,401.34)(1.000,1.000){2}{\rule{0.241pt}{0.800pt}}
\put(1396.0,403.0){\rule[-0.400pt]{3.613pt}{0.800pt}}
\put(1426,402.84){\rule{0.482pt}{0.800pt}}
\multiput(1426.00,402.34)(1.000,1.000){2}{\rule{0.241pt}{0.800pt}}
\put(1413.0,404.0){\rule[-0.400pt]{3.132pt}{0.800pt}}
\put(1428.0,405.0){\rule[-0.400pt]{2.650pt}{0.800pt}}
\sbox{\plotpoint}{\rule[-0.500pt]{1.000pt}{1.000pt}}\let\oldDiamond=\Diamond
\renewcommand{\Diamond}{\mbox{\tiny$\bullet$}}
\put(161,467){\raisebox{-.8pt}{\makebox(0,0){$\Diamond$}}}
\put(174,455){\raisebox{-.8pt}{\makebox(0,0){$\Diamond$}}}
\put(187,444){\raisebox{-.8pt}{\makebox(0,0){$\Diamond$}}}
\put(200,434){\raisebox{-.8pt}{\makebox(0,0){$\Diamond$}}}
\put(213,424){\raisebox{-.8pt}{\makebox(0,0){$\Diamond$}}}
\put(226,414){\raisebox{-.8pt}{\makebox(0,0){$\Diamond$}}}
\put(238,405){\raisebox{-.8pt}{\makebox(0,0){$\Diamond$}}}
\put(251,397){\raisebox{-.8pt}{\makebox(0,0){$\Diamond$}}}
\put(264,388){\raisebox{-.8pt}{\makebox(0,0){$\Diamond$}}}
\put(277,381){\raisebox{-.8pt}{\makebox(0,0){$\Diamond$}}}
\put(290,373){\raisebox{-.8pt}{\makebox(0,0){$\Diamond$}}}
\put(303,366){\raisebox{-.8pt}{\makebox(0,0){$\Diamond$}}}
\put(316,359){\raisebox{-.8pt}{\makebox(0,0){$\Diamond$}}}
\put(329,353){\raisebox{-.8pt}{\makebox(0,0){$\Diamond$}}}
\put(342,347){\raisebox{-.8pt}{\makebox(0,0){$\Diamond$}}}
\put(355,341){\raisebox{-.8pt}{\makebox(0,0){$\Diamond$}}}
\put(368,335){\raisebox{-.8pt}{\makebox(0,0){$\Diamond$}}}
\put(380,330){\raisebox{-.8pt}{\makebox(0,0){$\Diamond$}}}
\put(393,324){\raisebox{-.8pt}{\makebox(0,0){$\Diamond$}}}
\put(406,319){\raisebox{-.8pt}{\makebox(0,0){$\Diamond$}}}
\put(419,315){\raisebox{-.8pt}{\makebox(0,0){$\Diamond$}}}
\put(432,310){\raisebox{-.8pt}{\makebox(0,0){$\Diamond$}}}
\put(445,306){\raisebox{-.8pt}{\makebox(0,0){$\Diamond$}}}
\put(458,302){\raisebox{-.8pt}{\makebox(0,0){$\Diamond$}}}
\put(471,298){\raisebox{-.8pt}{\makebox(0,0){$\Diamond$}}}
\put(484,294){\raisebox{-.8pt}{\makebox(0,0){$\Diamond$}}}
\put(497,291){\raisebox{-.8pt}{\makebox(0,0){$\Diamond$}}}
\put(510,287){\raisebox{-.8pt}{\makebox(0,0){$\Diamond$}}}
\put(522,284){\raisebox{-.8pt}{\makebox(0,0){$\Diamond$}}}
\put(535,281){\raisebox{-.8pt}{\makebox(0,0){$\Diamond$}}}
\put(548,278){\raisebox{-.8pt}{\makebox(0,0){$\Diamond$}}}
\put(561,275){\raisebox{-.8pt}{\makebox(0,0){$\Diamond$}}}
\put(574,273){\raisebox{-.8pt}{\makebox(0,0){$\Diamond$}}}
\put(587,270){\raisebox{-.8pt}{\makebox(0,0){$\Diamond$}}}
\put(600,268){\raisebox{-.8pt}{\makebox(0,0){$\Diamond$}}}
\put(613,266){\raisebox{-.8pt}{\makebox(0,0){$\Diamond$}}}
\put(626,263){\raisebox{-.8pt}{\makebox(0,0){$\Diamond$}}}
\put(639,261){\raisebox{-.8pt}{\makebox(0,0){$\Diamond$}}}
\put(652,259){\raisebox{-.8pt}{\makebox(0,0){$\Diamond$}}}
\put(664,258){\raisebox{-.8pt}{\makebox(0,0){$\Diamond$}}}
\put(677,256){\raisebox{-.8pt}{\makebox(0,0){$\Diamond$}}}
\put(690,254){\raisebox{-.8pt}{\makebox(0,0){$\Diamond$}}}
\put(703,253){\raisebox{-.8pt}{\makebox(0,0){$\Diamond$}}}
\put(716,251){\raisebox{-.8pt}{\makebox(0,0){$\Diamond$}}}
\put(729,250){\raisebox{-.8pt}{\makebox(0,0){$\Diamond$}}}
\put(742,249){\raisebox{-.8pt}{\makebox(0,0){$\Diamond$}}}
\put(755,247){\raisebox{-.8pt}{\makebox(0,0){$\Diamond$}}}
\put(768,246){\raisebox{-.8pt}{\makebox(0,0){$\Diamond$}}}
\put(781,245){\raisebox{-.8pt}{\makebox(0,0){$\Diamond$}}}
\put(794,244){\raisebox{-.8pt}{\makebox(0,0){$\Diamond$}}}
\put(806,243){\raisebox{-.8pt}{\makebox(0,0){$\Diamond$}}}
\put(819,243){\raisebox{-.8pt}{\makebox(0,0){$\Diamond$}}}
\put(832,242){\raisebox{-.8pt}{\makebox(0,0){$\Diamond$}}}
\put(845,241){\raisebox{-.8pt}{\makebox(0,0){$\Diamond$}}}
\put(858,241){\raisebox{-.8pt}{\makebox(0,0){$\Diamond$}}}
\put(871,240){\raisebox{-.8pt}{\makebox(0,0){$\Diamond$}}}
\put(884,240){\raisebox{-.8pt}{\makebox(0,0){$\Diamond$}}}
\put(897,239){\raisebox{-.8pt}{\makebox(0,0){$\Diamond$}}}
\put(910,239){\raisebox{-.8pt}{\makebox(0,0){$\Diamond$}}}
\put(923,238){\raisebox{-.8pt}{\makebox(0,0){$\Diamond$}}}
\put(936,238){\raisebox{-.8pt}{\makebox(0,0){$\Diamond$}}}
\put(948,238){\raisebox{-.8pt}{\makebox(0,0){$\Diamond$}}}
\put(961,238){\raisebox{-.8pt}{\makebox(0,0){$\Diamond$}}}
\put(974,238){\raisebox{-.8pt}{\makebox(0,0){$\Diamond$}}}
\put(987,238){\raisebox{-.8pt}{\makebox(0,0){$\Diamond$}}}
\put(1000,238){\raisebox{-.8pt}{\makebox(0,0){$\Diamond$}}}
\put(1013,238){\raisebox{-.8pt}{\makebox(0,0){$\Diamond$}}}
\put(1026,238){\raisebox{-.8pt}{\makebox(0,0){$\Diamond$}}}
\put(1039,238){\raisebox{-.8pt}{\makebox(0,0){$\Diamond$}}}
\put(1052,238){\raisebox{-.8pt}{\makebox(0,0){$\Diamond$}}}
\put(1065,238){\raisebox{-.8pt}{\makebox(0,0){$\Diamond$}}}
\put(1078,238){\raisebox{-.8pt}{\makebox(0,0){$\Diamond$}}}
\put(1090,239){\raisebox{-.8pt}{\makebox(0,0){$\Diamond$}}}
\put(1103,239){\raisebox{-.8pt}{\makebox(0,0){$\Diamond$}}}
\put(1116,239){\raisebox{-.8pt}{\makebox(0,0){$\Diamond$}}}
\put(1129,240){\raisebox{-.8pt}{\makebox(0,0){$\Diamond$}}}
\put(1142,240){\raisebox{-.8pt}{\makebox(0,0){$\Diamond$}}}
\put(1155,241){\raisebox{-.8pt}{\makebox(0,0){$\Diamond$}}}
\put(1168,241){\raisebox{-.8pt}{\makebox(0,0){$\Diamond$}}}
\put(1181,242){\raisebox{-.8pt}{\makebox(0,0){$\Diamond$}}}
\put(1194,242){\raisebox{-.8pt}{\makebox(0,0){$\Diamond$}}}
\put(1207,243){\raisebox{-.8pt}{\makebox(0,0){$\Diamond$}}}
\put(1220,243){\raisebox{-.8pt}{\makebox(0,0){$\Diamond$}}}
\put(1232,244){\raisebox{-.8pt}{\makebox(0,0){$\Diamond$}}}
\put(1245,245){\raisebox{-.8pt}{\makebox(0,0){$\Diamond$}}}
\put(1258,245){\raisebox{-.8pt}{\makebox(0,0){$\Diamond$}}}
\put(1271,246){\raisebox{-.8pt}{\makebox(0,0){$\Diamond$}}}
\put(1284,247){\raisebox{-.8pt}{\makebox(0,0){$\Diamond$}}}
\put(1297,248){\raisebox{-.8pt}{\makebox(0,0){$\Diamond$}}}
\put(1310,248){\raisebox{-.8pt}{\makebox(0,0){$\Diamond$}}}
\put(1323,249){\raisebox{-.8pt}{\makebox(0,0){$\Diamond$}}}
\put(1336,250){\raisebox{-.8pt}{\makebox(0,0){$\Diamond$}}}
\put(1349,251){\raisebox{-.8pt}{\makebox(0,0){$\Diamond$}}}
\put(1362,252){\raisebox{-.8pt}{\makebox(0,0){$\Diamond$}}}
\put(1374,253){\raisebox{-.8pt}{\makebox(0,0){$\Diamond$}}}
\put(1387,254){\raisebox{-.8pt}{\makebox(0,0){$\Diamond$}}}
\put(1400,255){\raisebox{-.8pt}{\makebox(0,0){$\Diamond$}}}
\put(1413,256){\raisebox{-.8pt}{\makebox(0,0){$\Diamond$}}}
\put(1426,257){\raisebox{-.8pt}{\makebox(0,0){$\Diamond$}}}
\put(1439,258){\raisebox{-.8pt}{\makebox(0,0){$\Diamond$}}}
\let\Diamond=\oldDiamond
\end{picture}}
\caption{Several approximations for the function $\tilde
w$.
A $1\times1$ truncation of the matrices yields the dotted curve. The
solid curve results from the $3\times3$ matrices.
As the difference is so small, only the blow-up in the inset
can show the $2\times2$ results
separately as the curve of points.
}
\label{fig:tildew}
\end{figure}

For many values of $w_0\gamma$ the eigenvalue problem
\[
        \Lambda \mW^{3\times3} A = \mL^{3\times3} A,
\]
was solved numerically using ``Maple V'', and the largest  eigenvalue,
$\Lambda_0(w_0\gamma)$ was taken to determine $\tilde w$, according to
\Eq~(\ref{tildew}). The minimum $w_0$ is assumed at about  $w_0\gamma_0=3.5$,
therefore, in \Fig~\ref{fig:tildew}, the values of  $\tilde w$ were plotted for
$2<w_0\gamma<4$.

The minimum of $\tilde w$ can be determined numerically. It is given by
\begin{eqnarray}
\eql{approx}
        w_0&
                \approx&
                        4.735\ldots ,
\nonumber\\
        w_0\gamma_0 &
                \approx&
                        3.497\ldots,
\end{eqnarray}
and the  normalized eigenfunction is approximately
\begin{equation}
        A_0(\vv) = 0.63943 + 0.16289 |\vv|^2 + 0.004351 |\vv|^4.
\eql{A}
\end{equation}
In \Fig~\ref{fig:tildew}, also the result is plotted when the matrices are
truncated further.   Keeping only $|1\rangle$ means no variational parameters
and no velocity dependence. Hence we recover the result
$w_0\approx4.31107\ldots$of previous work\cite{myself,leid}, in which the
velocity dependence of the collision frequency was neglected. The enhancement
of taking two basis vectors, $|1\rangle$ and $|2\rangle$,  is rather large, but
taking one more  gives only a small difference, of about 0.1\percent, as the
inset of \Fig~\ref{fig:tildew} shows. Therefore we assume that these values
have converged up to  at most  0.1\percent.

The results can be qualitatively understood as follows.  In the first place,
particles with a higher velocity have a higher collision frequency, and so
their clock values increase faster than average. Thus, one expects higher
velocities to be more prominent in the head than in the rest of the
distribution. Indeed, this is seen in \Eq~(\ref{A}), where according to
formula~(\ref{headdistr}), the positive value of the coefficient $0.16289$
signifies a shift to higher velocities in $P_{head}$ as compared to
$\varphi_0$. Secondly, collisions of other particles with the head just
synchronize this particle to the one in the head. So the collision frequency of
the particle in the bulk is irrelevant and unable to compensate the increase in
collision frequency in the head. This increase is what really causes the
increase of $w$ over the velocity independent estimate of $4.31107\ldots$ in
Refs.\cite{myself,leid}, as can be seen from   the relative increase in this
frequency,
\begin{eqnarray*}
	\frac{\nu_{head}}{\nu_d} &=& \frac{1}{2\sqrt{\pi}}
	\int d\vv_1 d\vv_2\, 2|\vv_{12}| A_0(\vv_2)
		e^{-\sfrac{1}{2}(|\vv_1|^2+|\vv_2|^2)} \\
	&=& \mel{A_0}{\nu_d^{-1}\nu}{1} = 1+\frac{1}{2}b_0+\frac{15}{4}b_1
	= 1.0977\ldots,
\end{eqnarray*}
which matches closely the relative increase in clock speed,
$w_0/4.31107=1.0983\ldots$.

\subsection{Comparison with Simulations}

The results for the largest Lyapunov exponent of a system of $N$ hard  balls
will  now be checked against simulations. We want to check the following
points: the validity of the clock model, the relation between the Lyapunov
exponent and the clock speed, and the velocity distribution in the head.

The simulation method we shall use is a variant of the Direct Simulation Monte
Carlo Method (DSMC), that was also used by Dellago and Posch\cite{Dellago3} for
the calculation of Lyapunov exponents in a ``spatially homogeneous system''.
These authors also checked the equivalence for low densities of DSMC and
molecular dynamics simulations. The low densities that we are interested in,
are not accessible to the molecular dynamics simulations. The DSMC method
simulates the Boltzmann equation, and consists of the following. In each time
step, a pair of particles is picked at random to collide. The probability of
accepting a pair is proportional to the relative velocity so that the collision
frequency of two particles with given velocity is also linear with the relative
velocity.  In the original method due to Bird\cite{Bird}, the system is divided
into cells, and only particles within one cell can collide, but as we are
simulating a homogeneous system, the whole system is in one cell. Once a pair
is picked, a collision normal $\hat\sigma$ is drawn from a distribution
proportional to  $|\vv_{21}\cdot\hat\sigma|$, like in the Boltzmann equation.
The velocities of the pair are transformed according to Eq.  (\ref{vdynamics})
and their deviations are subjected to the  transformations in Eq. (\ref{ks9}),
to a account for free flight, and,  subsequently Eq. (\ref{eq:dvd}). Even
though in the DSMC method the positions themselves are discarded, we still keep
track of the deviations in position by integrating the deviations in velocity
by means of \Eq~(\ref{ks9}).

In a parallel simulation, for the same collision sequence,  the particles are
given  separate clock values, which are updated in a collision according to
\Eq~(\ref{kdyn}). This enables us to check how well the clock values from the
dynamics of \Eq~(\ref{kdyn}) represent the  actual dynamics of the velocity
deviations.

Initially, the velocities of the particles are picked from a Maxwellian
distribution, scaled by $v_0$, i.e., with  $k_BT/m$  set to one. The initial
velocity and position deviations are unit vectors drawn from an isotropic
distribution, $\delta v_0$ is set to one (this can be done because the
deviations follow a linear equation),  and correspondingly all clock values are
zero.

\newcommand{\densA}{\mbox{$10^{-4}$}}
\newcommand{\densB}{\mbox{$10^{-12}$}}
\newcommand{\simtime}{$3,887,196$}

The first check is to see whether the clock values are accurate in  describing
the behavior of the velocity deviations. As mentioned  above, in the simulation
the particles have their real  deviations in position and  velocity as well as
a clock value,  evolving independently of the deviations, according to
\Eq~(\ref{kdyn}). If that equation were exact,   for each particle the clock
value $k_i$ would equal  $\ln|\vdv_i/\delta v_0|/|\ln\tilden|$ for all time, if
 it did so initially.  But \Eq~(\ref{8c}) shows that there are density
corrections of  order $1/\ln\tilden$. If the difference between $k_i$ and
$\ln|\vdv_i/\delta v_0|/|\ln\tilden|$ became very big too often, the  maximum
of the  clock values might not correspond to the maximum of the  velocity
deviations, and the dynamics from Eq.~(\ref{kdyn}) would not  even be
approximately correct.  If things go as we expected, the clock values will give
the right $w$  in the limit of zero density, i.e., $w_0$, with deviations  that
scale as $1/\ln\tilden$.

Two simulations for $N=128$ particles, and $d=2$ dimensions, were performed
for, in total, \simtime\ collisions, one simulation at a density
$\tilden_2=10^{-4}$  and one at a density  $\tilden_2=10^{-12}$.  In
\Fig~\ref{fig:kdv} results are plotted for both  simulations. For each particle
a symbol is plotted, a plus for the simulation at density $\tilden_2=10^{-12}$
and a diamond for the simulation at density $\tilden_2=10^{-4}$.  The
horizontal position of each symbol is given by the final clock value $k_i$ of
particle $i$ (of which 226885 is subtracted), the vertical position by  the
final value of $\ln|\vdv_i/\delta v_0|/|\ln\tilden_2|$.  The scale of the
velocity deviations is a bit different for the two simulations, therefore
different scales were used on the vertical axis, as indicated in the  figure.

{}From the plot, we can see the following. There is a linear relation between
$\ln|\vdv_i/\delta v_0|/|\ln\tilden_2|$  and $k_i$, around which there are
fluctuations. The fluctuations get smaller, and the slope gets closer to one,
as the density decreases. Small fluctuations indicate that the maximum in
\Eq~(\ref{kdyn}) is correct, i.e., the particle with the largest clock value
has the largest velocity deviation. The value of $\ln|\vdv_i/\delta
v_0|/|\ln\tilden_2|$  is not exactly equal to the clock value $k_i$.
\Eq~(\ref{8c}) shows that the correction in the increase of the clock value in
a collision is of the order of  $1/\ln\tilden_2$.  Because  such a correction
is present in each collision, the correction to the relation
$\ln|\vdv_i/\delta v_0|/|\ln\tilden_2|=k_i$  also grows. The relative
correction is, for density  $\tilden_2=10^{-4}$  of the order of 24\percent,
while for the density  $\tilden_2=10^{-12}$  it is about 8\percent. The
difference by a factor of three shows that the correction is of the order of
$1/|\ln\tilden_2|$,  as $\ln10^{-12}/\ln10^{-4}=3$ also.  We conclude that the
dynamics of the clock value in  \Eq~(\ref{kdyn}) indeed describes the dynamics
of the  $\ln|\vdv_i|$ accurately for low densities.

\begin{figure}[t]
\centerline{\setlength{\unitlength}{0.240900pt}
\ifx\plotpoint\undefined\newsavebox{\plotpoint}\fi
\sbox{\plotpoint}{\rule[-0.200pt]{0.400pt}{0.400pt}}\begin{picture}(1500,900)(0,0)
\font\gnuplot=cmr10 at 10pt
\gnuplot
\sbox{\plotpoint}{\rule[-0.200pt]{0.400pt}{0.400pt}}\put(161.0,123.0){\rule[-0.200pt]{4.818pt}{0.400pt}}
\put(141,123){\makebox(0,0)[r]{-10}}
\put(1419.0,123.0){\rule[-0.200pt]{4.818pt}{0.400pt}}
\put(161.0,215.0){\rule[-0.200pt]{4.818pt}{0.400pt}}
\put(141,215){\makebox(0,0)[r]{-5}}
\put(1419.0,215.0){\rule[-0.200pt]{4.818pt}{0.400pt}}
\put(161.0,307.0){\rule[-0.200pt]{4.818pt}{0.400pt}}
\put(141,307){\makebox(0,0)[r]{0}}
\put(1419.0,307.0){\rule[-0.200pt]{4.818pt}{0.400pt}}
\put(161.0,399.0){\rule[-0.200pt]{4.818pt}{0.400pt}}
\put(141,399){\makebox(0,0)[r]{5}}
\put(1419.0,399.0){\rule[-0.200pt]{4.818pt}{0.400pt}}
\put(161.0,492.0){\rule[-0.200pt]{4.818pt}{0.400pt}}
\put(141,492){\makebox(0,0)[r]{10}}
\put(1419.0,492.0){\rule[-0.200pt]{4.818pt}{0.400pt}}
\put(161.0,584.0){\rule[-0.200pt]{4.818pt}{0.400pt}}
\put(141,584){\makebox(0,0)[r]{15}}
\put(1419.0,584.0){\rule[-0.200pt]{4.818pt}{0.400pt}}
\put(161.0,676.0){\rule[-0.200pt]{4.818pt}{0.400pt}}
\put(141,676){\makebox(0,0)[r]{20}}
\put(1419.0,676.0){\rule[-0.200pt]{4.818pt}{0.400pt}}
\put(161.0,768.0){\rule[-0.200pt]{4.818pt}{0.400pt}}
\put(141,768){\makebox(0,0)[r]{25}}
\put(1419.0,768.0){\rule[-0.200pt]{4.818pt}{0.400pt}}
\put(161.0,860.0){\rule[-0.200pt]{4.818pt}{0.400pt}}
\put(141,860){\makebox(0,0)[r]{30}}
\put(1419.0,860.0){\rule[-0.200pt]{4.818pt}{0.400pt}}
\put(161.0,123.0){\rule[-0.200pt]{0.400pt}{4.818pt}}
\put(161,82){\makebox(0,0){5}}
\put(161.0,840.0){\rule[-0.200pt]{0.400pt}{4.818pt}}
\put(417.0,123.0){\rule[-0.200pt]{0.400pt}{4.818pt}}
\put(417,82){\makebox(0,0){10}}
\put(417.0,840.0){\rule[-0.200pt]{0.400pt}{4.818pt}}
\put(672.0,123.0){\rule[-0.200pt]{0.400pt}{4.818pt}}
\put(672,82){\makebox(0,0){15}}
\put(672.0,840.0){\rule[-0.200pt]{0.400pt}{4.818pt}}
\put(928.0,123.0){\rule[-0.200pt]{0.400pt}{4.818pt}}
\put(928,82){\makebox(0,0){20}}
\put(928.0,840.0){\rule[-0.200pt]{0.400pt}{4.818pt}}
\put(1183.0,123.0){\rule[-0.200pt]{0.400pt}{4.818pt}}
\put(1183,82){\makebox(0,0){25}}
\put(1183.0,840.0){\rule[-0.200pt]{0.400pt}{4.818pt}}
\put(1439.0,123.0){\rule[-0.200pt]{0.400pt}{4.818pt}}
\put(1439,82){\makebox(0,0){30}}
\put(1439.0,840.0){\rule[-0.200pt]{0.400pt}{4.818pt}}
\put(161.0,123.0){\rule[-0.200pt]{307.870pt}{0.400pt}}
\put(1439.0,123.0){\rule[-0.200pt]{0.400pt}{177.543pt}}
\put(161.0,860.0){\rule[-0.200pt]{307.870pt}{0.400pt}}
\put(40,491){\makebox(0,0){\rotatebox{90}{\small\parbox{2.8in}{
$+: (\ln|\delta\vec v_i|/|\ln\tilde n_2|)-171690$($\tilde n_2=10^{-4}$)
\\
$\Diamond:(\ln|\delta\vec v_i|/|\ln\tilde n_2|)-208153$($\tilde
n_2=10^{-12}$)\\\ }}}}
\put(800,21){\makebox(0,0){$k_i-226885$}}
\put(161.0,123.0){\rule[-0.200pt]{0.400pt}{177.543pt}}
\put(928,379){\raisebox{-.8pt}{\makebox(0,0){$\Diamond$}}}
\put(979,387){\raisebox{-.8pt}{\makebox(0,0){$\Diamond$}}}
\put(723,337){\raisebox{-.8pt}{\makebox(0,0){$\Diamond$}}}
\put(1132,437){\raisebox{-.8pt}{\makebox(0,0){$\Diamond$}}}
\put(1132,441){\raisebox{-.8pt}{\makebox(0,0){$\Diamond$}}}
\put(1132,443){\raisebox{-.8pt}{\makebox(0,0){$\Diamond$}}}
\put(1030,415){\raisebox{-.8pt}{\makebox(0,0){$\Diamond$}}}
\put(1132,437){\raisebox{-.8pt}{\makebox(0,0){$\Diamond$}}}
\put(1388,503){\raisebox{-.8pt}{\makebox(0,0){$\Diamond$}}}
\put(826,371){\raisebox{-.8pt}{\makebox(0,0){$\Diamond$}}}
\put(1132,433){\raisebox{-.8pt}{\makebox(0,0){$\Diamond$}}}
\put(1235,463){\raisebox{-.8pt}{\makebox(0,0){$\Diamond$}}}
\put(1132,437){\raisebox{-.8pt}{\makebox(0,0){$\Diamond$}}}
\put(826,355){\raisebox{-.8pt}{\makebox(0,0){$\Diamond$}}}
\put(1235,465){\raisebox{-.8pt}{\makebox(0,0){$\Diamond$}}}
\put(1081,413){\raisebox{-.8pt}{\makebox(0,0){$\Diamond$}}}
\put(1235,471){\raisebox{-.8pt}{\makebox(0,0){$\Diamond$}}}
\put(621,305){\raisebox{-.8pt}{\makebox(0,0){$\Diamond$}}}
\put(1132,443){\raisebox{-.8pt}{\makebox(0,0){$\Diamond$}}}
\put(979,391){\raisebox{-.8pt}{\makebox(0,0){$\Diamond$}}}
\put(774,337){\raisebox{-.8pt}{\makebox(0,0){$\Diamond$}}}
\put(723,331){\raisebox{-.8pt}{\makebox(0,0){$\Diamond$}}}
\put(263,197){\raisebox{-.8pt}{\makebox(0,0){$\Diamond$}}}
\put(928,399){\raisebox{-.8pt}{\makebox(0,0){$\Diamond$}}}
\put(1388,503){\raisebox{-.8pt}{\makebox(0,0){$\Diamond$}}}
\put(979,391){\raisebox{-.8pt}{\makebox(0,0){$\Diamond$}}}
\put(1337,483){\raisebox{-.8pt}{\makebox(0,0){$\Diamond$}}}
\put(1235,465){\raisebox{-.8pt}{\makebox(0,0){$\Diamond$}}}
\put(1030,411){\raisebox{-.8pt}{\makebox(0,0){$\Diamond$}}}
\put(1081,429){\raisebox{-.8pt}{\makebox(0,0){$\Diamond$}}}
\put(928,371){\raisebox{-.8pt}{\makebox(0,0){$\Diamond$}}}
\put(1235,477){\raisebox{-.8pt}{\makebox(0,0){$\Diamond$}}}
\put(1030,419){\raisebox{-.8pt}{\makebox(0,0){$\Diamond$}}}
\put(1388,495){\raisebox{-.8pt}{\makebox(0,0){$\Diamond$}}}
\put(1286,483){\raisebox{-.8pt}{\makebox(0,0){$\Diamond$}}}
\put(1081,413){\raisebox{-.8pt}{\makebox(0,0){$\Diamond$}}}
\put(1286,479){\raisebox{-.8pt}{\makebox(0,0){$\Diamond$}}}
\put(1235,475){\raisebox{-.8pt}{\makebox(0,0){$\Diamond$}}}
\put(1132,433){\raisebox{-.8pt}{\makebox(0,0){$\Diamond$}}}
\put(1030,417){\raisebox{-.8pt}{\makebox(0,0){$\Diamond$}}}
\put(1183,455){\raisebox{-.8pt}{\makebox(0,0){$\Diamond$}}}
\put(1132,433){\raisebox{-.8pt}{\makebox(0,0){$\Diamond$}}}
\put(877,375){\raisebox{-.8pt}{\makebox(0,0){$\Diamond$}}}
\put(1235,457){\raisebox{-.8pt}{\makebox(0,0){$\Diamond$}}}
\put(1183,459){\raisebox{-.8pt}{\makebox(0,0){$\Diamond$}}}
\put(417,247){\raisebox{-.8pt}{\makebox(0,0){$\Diamond$}}}
\put(979,389){\raisebox{-.8pt}{\makebox(0,0){$\Diamond$}}}
\put(826,351){\raisebox{-.8pt}{\makebox(0,0){$\Diamond$}}}
\put(774,339){\raisebox{-.8pt}{\makebox(0,0){$\Diamond$}}}
\put(468,269){\raisebox{-.8pt}{\makebox(0,0){$\Diamond$}}}
\put(1030,413){\raisebox{-.8pt}{\makebox(0,0){$\Diamond$}}}
\put(774,363){\raisebox{-.8pt}{\makebox(0,0){$\Diamond$}}}
\put(1030,411){\raisebox{-.8pt}{\makebox(0,0){$\Diamond$}}}
\put(1183,449){\raisebox{-.8pt}{\makebox(0,0){$\Diamond$}}}
\put(979,403){\raisebox{-.8pt}{\makebox(0,0){$\Diamond$}}}
\put(1183,451){\raisebox{-.8pt}{\makebox(0,0){$\Diamond$}}}
\put(1030,413){\raisebox{-.8pt}{\makebox(0,0){$\Diamond$}}}
\put(979,395){\raisebox{-.8pt}{\makebox(0,0){$\Diamond$}}}
\put(1081,423){\raisebox{-.8pt}{\makebox(0,0){$\Diamond$}}}
\put(1081,427){\raisebox{-.8pt}{\makebox(0,0){$\Diamond$}}}
\put(468,271){\raisebox{-.8pt}{\makebox(0,0){$\Diamond$}}}
\put(1132,437){\raisebox{-.8pt}{\makebox(0,0){$\Diamond$}}}
\put(1183,465){\raisebox{-.8pt}{\makebox(0,0){$\Diamond$}}}
\put(1030,413){\raisebox{-.8pt}{\makebox(0,0){$\Diamond$}}}
\put(979,391){\raisebox{-.8pt}{\makebox(0,0){$\Diamond$}}}
\put(1081,427){\raisebox{-.8pt}{\makebox(0,0){$\Diamond$}}}
\put(1081,417){\raisebox{-.8pt}{\makebox(0,0){$\Diamond$}}}
\put(1337,489){\raisebox{-.8pt}{\makebox(0,0){$\Diamond$}}}
\put(877,369){\raisebox{-.8pt}{\makebox(0,0){$\Diamond$}}}
\put(1286,475){\raisebox{-.8pt}{\makebox(0,0){$\Diamond$}}}
\put(979,401){\raisebox{-.8pt}{\makebox(0,0){$\Diamond$}}}
\put(928,381){\raisebox{-.8pt}{\makebox(0,0){$\Diamond$}}}
\put(979,395){\raisebox{-.8pt}{\makebox(0,0){$\Diamond$}}}
\put(928,381){\raisebox{-.8pt}{\makebox(0,0){$\Diamond$}}}
\put(979,387){\raisebox{-.8pt}{\makebox(0,0){$\Diamond$}}}
\put(1388,495){\raisebox{-.8pt}{\makebox(0,0){$\Diamond$}}}
\put(774,343){\raisebox{-.8pt}{\makebox(0,0){$\Diamond$}}}
\put(263,191){\raisebox{-.8pt}{\makebox(0,0){$\Diamond$}}}
\put(979,397){\raisebox{-.8pt}{\makebox(0,0){$\Diamond$}}}
\put(1081,429){\raisebox{-.8pt}{\makebox(0,0){$\Diamond$}}}
\put(519,281){\raisebox{-.8pt}{\makebox(0,0){$\Diamond$}}}
\put(1132,437){\raisebox{-.8pt}{\makebox(0,0){$\Diamond$}}}
\put(1132,437){\raisebox{-.8pt}{\makebox(0,0){$\Diamond$}}}
\put(1030,411){\raisebox{-.8pt}{\makebox(0,0){$\Diamond$}}}
\put(928,377){\raisebox{-.8pt}{\makebox(0,0){$\Diamond$}}}
\put(723,321){\raisebox{-.8pt}{\makebox(0,0){$\Diamond$}}}
\put(723,331){\raisebox{-.8pt}{\makebox(0,0){$\Diamond$}}}
\put(1183,455){\raisebox{-.8pt}{\makebox(0,0){$\Diamond$}}}
\put(928,379){\raisebox{-.8pt}{\makebox(0,0){$\Diamond$}}}
\put(672,319){\raisebox{-.8pt}{\makebox(0,0){$\Diamond$}}}
\put(979,395){\raisebox{-.8pt}{\makebox(0,0){$\Diamond$}}}
\put(314,207){\raisebox{-.8pt}{\makebox(0,0){$\Diamond$}}}
\put(1286,479){\raisebox{-.8pt}{\makebox(0,0){$\Diamond$}}}
\put(1030,417){\raisebox{-.8pt}{\makebox(0,0){$\Diamond$}}}
\put(774,349){\raisebox{-.8pt}{\makebox(0,0){$\Diamond$}}}
\put(1132,443){\raisebox{-.8pt}{\makebox(0,0){$\Diamond$}}}
\put(1183,455){\raisebox{-.8pt}{\makebox(0,0){$\Diamond$}}}
\put(979,415){\raisebox{-.8pt}{\makebox(0,0){$\Diamond$}}}
\put(1132,437){\raisebox{-.8pt}{\makebox(0,0){$\Diamond$}}}
\put(1081,417){\raisebox{-.8pt}{\makebox(0,0){$\Diamond$}}}
\put(570,303){\raisebox{-.8pt}{\makebox(0,0){$\Diamond$}}}
\put(1286,483){\raisebox{-.8pt}{\makebox(0,0){$\Diamond$}}}
\put(1235,475){\raisebox{-.8pt}{\makebox(0,0){$\Diamond$}}}
\put(877,369){\raisebox{-.8pt}{\makebox(0,0){$\Diamond$}}}
\put(1183,449){\raisebox{-.8pt}{\makebox(0,0){$\Diamond$}}}
\put(1081,427){\raisebox{-.8pt}{\makebox(0,0){$\Diamond$}}}
\put(621,305){\raisebox{-.8pt}{\makebox(0,0){$\Diamond$}}}
\put(365,241){\raisebox{-.8pt}{\makebox(0,0){$\Diamond$}}}
\put(979,415){\raisebox{-.8pt}{\makebox(0,0){$\Diamond$}}}
\put(979,403){\raisebox{-.8pt}{\makebox(0,0){$\Diamond$}}}
\put(979,387){\raisebox{-.8pt}{\makebox(0,0){$\Diamond$}}}
\put(1081,429){\raisebox{-.8pt}{\makebox(0,0){$\Diamond$}}}
\put(1030,419){\raisebox{-.8pt}{\makebox(0,0){$\Diamond$}}}
\put(519,271){\raisebox{-.8pt}{\makebox(0,0){$\Diamond$}}}
\put(621,305){\raisebox{-.8pt}{\makebox(0,0){$\Diamond$}}}
\put(979,397){\raisebox{-.8pt}{\makebox(0,0){$\Diamond$}}}
\put(1235,457){\raisebox{-.8pt}{\makebox(0,0){$\Diamond$}}}
\put(1183,459){\raisebox{-.8pt}{\makebox(0,0){$\Diamond$}}}
\put(1286,473){\raisebox{-.8pt}{\makebox(0,0){$\Diamond$}}}
\put(1081,429){\raisebox{-.8pt}{\makebox(0,0){$\Diamond$}}}
\put(1132,433){\raisebox{-.8pt}{\makebox(0,0){$\Diamond$}}}
\put(1132,441){\raisebox{-.8pt}{\makebox(0,0){$\Diamond$}}}
\put(1183,455){\raisebox{-.8pt}{\makebox(0,0){$\Diamond$}}}
\put(1235,477){\raisebox{-.8pt}{\makebox(0,0){$\Diamond$}}}
\put(1183,449){\raisebox{-.8pt}{\makebox(0,0){$\Diamond$}}}
\put(1030,403){\raisebox{-.8pt}{\makebox(0,0){$\Diamond$}}}
\put(774,337){\raisebox{-.8pt}{\makebox(0,0){$\Diamond$}}}
\put(1183,451){\raisebox{-.8pt}{\makebox(0,0){$\Diamond$}}}
\put(928,545){\makebox(0,0){$+$}}
\put(979,560){\makebox(0,0){$+$}}
\put(723,481){\makebox(0,0){$+$}}
\put(1132,613){\makebox(0,0){$+$}}
\put(1132,614){\makebox(0,0){$+$}}
\put(1132,615){\makebox(0,0){$+$}}
\put(1030,581){\makebox(0,0){$+$}}
\put(1132,613){\makebox(0,0){$+$}}
\put(1388,697){\makebox(0,0){$+$}}
\put(826,517){\makebox(0,0){$+$}}
\put(1132,612){\makebox(0,0){$+$}}
\put(1235,647){\makebox(0,0){$+$}}
\put(1132,613){\makebox(0,0){$+$}}
\put(826,512){\makebox(0,0){$+$}}
\put(1235,647){\makebox(0,0){$+$}}
\put(1081,593){\makebox(0,0){$+$}}
\put(1235,649){\makebox(0,0){$+$}}
\put(621,446){\makebox(0,0){$+$}}
\put(1132,615){\makebox(0,0){$+$}}
\put(979,561){\makebox(0,0){$+$}}
\put(774,494){\makebox(0,0){$+$}}
\put(723,480){\makebox(0,0){$+$}}
\put(263,325){\makebox(0,0){$+$}}
\put(928,551){\makebox(0,0){$+$}}
\put(1388,697){\makebox(0,0){$+$}}
\put(979,561){\makebox(0,0){$+$}}
\put(1337,678){\makebox(0,0){$+$}}
\put(1235,647){\makebox(0,0){$+$}}
\put(1030,580){\makebox(0,0){$+$}}
\put(1081,598){\makebox(0,0){$+$}}
\put(928,542){\makebox(0,0){$+$}}
\put(1235,651){\makebox(0,0){$+$}}
\put(1030,583){\makebox(0,0){$+$}}
\put(1388,694){\makebox(0,0){$+$}}
\put(1286,665){\makebox(0,0){$+$}}
\put(1081,593){\makebox(0,0){$+$}}
\put(1286,664){\makebox(0,0){$+$}}
\put(1235,651){\makebox(0,0){$+$}}
\put(1132,611){\makebox(0,0){$+$}}
\put(1030,581){\makebox(0,0){$+$}}
\put(1183,631){\makebox(0,0){$+$}}
\put(1132,612){\makebox(0,0){$+$}}
\put(877,531){\makebox(0,0){$+$}}
\put(1235,645){\makebox(0,0){$+$}}
\put(1183,633){\makebox(0,0){$+$}}
\put(417,378){\makebox(0,0){$+$}}
\put(979,561){\makebox(0,0){$+$}}
\put(826,511){\makebox(0,0){$+$}}
\put(774,493){\makebox(0,0){$+$}}
\put(468,398){\makebox(0,0){$+$}}
\put(1030,581){\makebox(0,0){$+$}}
\put(774,503){\makebox(0,0){$+$}}
\put(1030,580){\makebox(0,0){$+$}}
\put(1183,629){\makebox(0,0){$+$}}
\put(979,565){\makebox(0,0){$+$}}
\put(1183,630){\makebox(0,0){$+$}}
\put(1030,580){\makebox(0,0){$+$}}
\put(979,563){\makebox(0,0){$+$}}
\put(1081,596){\makebox(0,0){$+$}}
\put(1081,598){\makebox(0,0){$+$}}
\put(468,398){\makebox(0,0){$+$}}
\put(1132,613){\makebox(0,0){$+$}}
\put(1183,635){\makebox(0,0){$+$}}
\put(1030,581){\makebox(0,0){$+$}}
\put(979,561){\makebox(0,0){$+$}}
\put(1081,598){\makebox(0,0){$+$}}
\put(1081,594){\makebox(0,0){$+$}}
\put(1337,679){\makebox(0,0){$+$}}
\put(877,529){\makebox(0,0){$+$}}
\put(1286,663){\makebox(0,0){$+$}}
\put(979,565){\makebox(0,0){$+$}}
\put(928,545){\makebox(0,0){$+$}}
\put(979,562){\makebox(0,0){$+$}}
\put(928,545){\makebox(0,0){$+$}}
\put(979,559){\makebox(0,0){$+$}}
\put(1388,694){\makebox(0,0){$+$}}
\put(774,495){\makebox(0,0){$+$}}
\put(263,322){\makebox(0,0){$+$}}
\put(979,563){\makebox(0,0){$+$}}
\put(1081,598){\makebox(0,0){$+$}}
\put(519,414){\makebox(0,0){$+$}}
\put(1132,613){\makebox(0,0){$+$}}
\put(1132,613){\makebox(0,0){$+$}}
\put(1030,580){\makebox(0,0){$+$}}
\put(928,544){\makebox(0,0){$+$}}
\put(723,476){\makebox(0,0){$+$}}
\put(723,479){\makebox(0,0){$+$}}
\put(1183,631){\makebox(0,0){$+$}}
\put(928,545){\makebox(0,0){$+$}}
\put(672,463){\makebox(0,0){$+$}}
\put(979,563){\makebox(0,0){$+$}}
\put(314,340){\makebox(0,0){$+$}}
\put(1286,664){\makebox(0,0){$+$}}
\put(1030,581){\makebox(0,0){$+$}}
\put(774,497){\makebox(0,0){$+$}}
\put(1132,615){\makebox(0,0){$+$}}
\put(1183,631){\makebox(0,0){$+$}}
\put(979,569){\makebox(0,0){$+$}}
\put(1132,613){\makebox(0,0){$+$}}
\put(1081,594){\makebox(0,0){$+$}}
\put(570,433){\makebox(0,0){$+$}}
\put(1286,665){\makebox(0,0){$+$}}
\put(1235,651){\makebox(0,0){$+$}}
\put(877,529){\makebox(0,0){$+$}}
\put(1183,629){\makebox(0,0){$+$}}
\put(1081,598){\makebox(0,0){$+$}}
\put(621,446){\makebox(0,0){$+$}}
\put(365,364){\makebox(0,0){$+$}}
\put(979,569){\makebox(0,0){$+$}}
\put(979,565){\makebox(0,0){$+$}}
\put(979,559){\makebox(0,0){$+$}}
\put(1081,599){\makebox(0,0){$+$}}
\put(1030,583){\makebox(0,0){$+$}}
\put(519,410){\makebox(0,0){$+$}}
\put(621,447){\makebox(0,0){$+$}}
\put(979,563){\makebox(0,0){$+$}}
\put(1235,645){\makebox(0,0){$+$}}
\put(1183,633){\makebox(0,0){$+$}}
\put(1286,662){\makebox(0,0){$+$}}
\put(1081,599){\makebox(0,0){$+$}}
\put(1132,612){\makebox(0,0){$+$}}
\put(1132,615){\makebox(0,0){$+$}}
\put(1183,631){\makebox(0,0){$+$}}
\put(1235,651){\makebox(0,0){$+$}}
\put(1183,629){\makebox(0,0){$+$}}
\put(1030,577){\makebox(0,0){$+$}}
\put(774,494){\makebox(0,0){$+$}}
\put(1183,630){\makebox(0,0){$+$}}
\sbox{\plotpoint}{\rule[-0.400pt]{0.800pt}{0.800pt}}\put(263,436){\usebox{\plotpoint}}
\put(263,436.34){\rule{2.600pt}{0.800pt}}
\multiput(263.00,434.34)(6.604,4.000){2}{\rule{1.300pt}{0.800pt}}
\put(275,440.34){\rule{2.400pt}{0.800pt}}
\multiput(275.00,438.34)(6.019,4.000){2}{\rule{1.200pt}{0.800pt}}
\multiput(286.00,445.38)(1.432,0.560){3}{\rule{1.960pt}{0.135pt}}
\multiput(286.00,442.34)(6.932,5.000){2}{\rule{0.980pt}{0.800pt}}
\put(297,449.34){\rule{2.600pt}{0.800pt}}
\multiput(297.00,447.34)(6.604,4.000){2}{\rule{1.300pt}{0.800pt}}
\put(309,453.34){\rule{2.400pt}{0.800pt}}
\multiput(309.00,451.34)(6.019,4.000){2}{\rule{1.200pt}{0.800pt}}
\put(320,457.34){\rule{2.400pt}{0.800pt}}
\multiput(320.00,455.34)(6.019,4.000){2}{\rule{1.200pt}{0.800pt}}
\put(331,461.34){\rule{2.600pt}{0.800pt}}
\multiput(331.00,459.34)(6.604,4.000){2}{\rule{1.300pt}{0.800pt}}
\put(343,465.34){\rule{2.400pt}{0.800pt}}
\multiput(343.00,463.34)(6.019,4.000){2}{\rule{1.200pt}{0.800pt}}
\put(354,469.34){\rule{2.400pt}{0.800pt}}
\multiput(354.00,467.34)(6.019,4.000){2}{\rule{1.200pt}{0.800pt}}
\put(365,473.34){\rule{2.600pt}{0.800pt}}
\multiput(365.00,471.34)(6.604,4.000){2}{\rule{1.300pt}{0.800pt}}
\put(377,477.34){\rule{2.400pt}{0.800pt}}
\multiput(377.00,475.34)(6.019,4.000){2}{\rule{1.200pt}{0.800pt}}
\put(388,481.34){\rule{2.600pt}{0.800pt}}
\multiput(388.00,479.34)(6.604,4.000){2}{\rule{1.300pt}{0.800pt}}
\put(400,485.34){\rule{2.400pt}{0.800pt}}
\multiput(400.00,483.34)(6.019,4.000){2}{\rule{1.200pt}{0.800pt}}
\multiput(411.00,490.38)(1.432,0.560){3}{\rule{1.960pt}{0.135pt}}
\multiput(411.00,487.34)(6.932,5.000){2}{\rule{0.980pt}{0.800pt}}
\put(422,494.34){\rule{2.600pt}{0.800pt}}
\multiput(422.00,492.34)(6.604,4.000){2}{\rule{1.300pt}{0.800pt}}
\put(434,498.34){\rule{2.400pt}{0.800pt}}
\multiput(434.00,496.34)(6.019,4.000){2}{\rule{1.200pt}{0.800pt}}
\put(445,502.34){\rule{2.400pt}{0.800pt}}
\multiput(445.00,500.34)(6.019,4.000){2}{\rule{1.200pt}{0.800pt}}
\put(456,506.34){\rule{2.600pt}{0.800pt}}
\multiput(456.00,504.34)(6.604,4.000){2}{\rule{1.300pt}{0.800pt}}
\put(468,510.34){\rule{2.400pt}{0.800pt}}
\multiput(468.00,508.34)(6.019,4.000){2}{\rule{1.200pt}{0.800pt}}
\put(479,514.34){\rule{2.400pt}{0.800pt}}
\multiput(479.00,512.34)(6.019,4.000){2}{\rule{1.200pt}{0.800pt}}
\put(490,518.34){\rule{2.600pt}{0.800pt}}
\multiput(490.00,516.34)(6.604,4.000){2}{\rule{1.300pt}{0.800pt}}
\put(502,522.34){\rule{2.400pt}{0.800pt}}
\multiput(502.00,520.34)(6.019,4.000){2}{\rule{1.200pt}{0.800pt}}
\put(513,526.34){\rule{2.600pt}{0.800pt}}
\multiput(513.00,524.34)(6.604,4.000){2}{\rule{1.300pt}{0.800pt}}
\put(525,530.34){\rule{2.400pt}{0.800pt}}
\multiput(525.00,528.34)(6.019,4.000){2}{\rule{1.200pt}{0.800pt}}
\multiput(536.00,535.38)(1.432,0.560){3}{\rule{1.960pt}{0.135pt}}
\multiput(536.00,532.34)(6.932,5.000){2}{\rule{0.980pt}{0.800pt}}
\put(547,539.34){\rule{2.600pt}{0.800pt}}
\multiput(547.00,537.34)(6.604,4.000){2}{\rule{1.300pt}{0.800pt}}
\put(559,543.34){\rule{2.400pt}{0.800pt}}
\multiput(559.00,541.34)(6.019,4.000){2}{\rule{1.200pt}{0.800pt}}
\put(570,547.34){\rule{2.400pt}{0.800pt}}
\multiput(570.00,545.34)(6.019,4.000){2}{\rule{1.200pt}{0.800pt}}
\put(581,551.34){\rule{2.600pt}{0.800pt}}
\multiput(581.00,549.34)(6.604,4.000){2}{\rule{1.300pt}{0.800pt}}
\put(593,555.34){\rule{2.400pt}{0.800pt}}
\multiput(593.00,553.34)(6.019,4.000){2}{\rule{1.200pt}{0.800pt}}
\put(604,559.34){\rule{2.400pt}{0.800pt}}
\multiput(604.00,557.34)(6.019,4.000){2}{\rule{1.200pt}{0.800pt}}
\put(615,563.34){\rule{2.600pt}{0.800pt}}
\multiput(615.00,561.34)(6.604,4.000){2}{\rule{1.300pt}{0.800pt}}
\put(627,567.34){\rule{2.400pt}{0.800pt}}
\multiput(627.00,565.34)(6.019,4.000){2}{\rule{1.200pt}{0.800pt}}
\put(638,571.34){\rule{2.400pt}{0.800pt}}
\multiput(638.00,569.34)(6.019,4.000){2}{\rule{1.200pt}{0.800pt}}
\multiput(649.00,576.38)(1.600,0.560){3}{\rule{2.120pt}{0.135pt}}
\multiput(649.00,573.34)(7.600,5.000){2}{\rule{1.060pt}{0.800pt}}
\put(661,580.34){\rule{2.400pt}{0.800pt}}
\multiput(661.00,578.34)(6.019,4.000){2}{\rule{1.200pt}{0.800pt}}
\put(672,584.34){\rule{2.600pt}{0.800pt}}
\multiput(672.00,582.34)(6.604,4.000){2}{\rule{1.300pt}{0.800pt}}
\put(684,588.34){\rule{2.400pt}{0.800pt}}
\multiput(684.00,586.34)(6.019,4.000){2}{\rule{1.200pt}{0.800pt}}
\put(695,592.34){\rule{2.400pt}{0.800pt}}
\multiput(695.00,590.34)(6.019,4.000){2}{\rule{1.200pt}{0.800pt}}
\put(706,596.34){\rule{2.600pt}{0.800pt}}
\multiput(706.00,594.34)(6.604,4.000){2}{\rule{1.300pt}{0.800pt}}
\put(718,600.34){\rule{2.400pt}{0.800pt}}
\multiput(718.00,598.34)(6.019,4.000){2}{\rule{1.200pt}{0.800pt}}
\put(729,604.34){\rule{2.400pt}{0.800pt}}
\multiput(729.00,602.34)(6.019,4.000){2}{\rule{1.200pt}{0.800pt}}
\put(740,608.34){\rule{2.600pt}{0.800pt}}
\multiput(740.00,606.34)(6.604,4.000){2}{\rule{1.300pt}{0.800pt}}
\put(752,612.34){\rule{2.400pt}{0.800pt}}
\multiput(752.00,610.34)(6.019,4.000){2}{\rule{1.200pt}{0.800pt}}
\put(763,616.34){\rule{2.400pt}{0.800pt}}
\multiput(763.00,614.34)(6.019,4.000){2}{\rule{1.200pt}{0.800pt}}
\multiput(774.00,621.38)(1.600,0.560){3}{\rule{2.120pt}{0.135pt}}
\multiput(774.00,618.34)(7.600,5.000){2}{\rule{1.060pt}{0.800pt}}
\put(786,625.34){\rule{2.400pt}{0.800pt}}
\multiput(786.00,623.34)(6.019,4.000){2}{\rule{1.200pt}{0.800pt}}
\put(797,629.34){\rule{2.600pt}{0.800pt}}
\multiput(797.00,627.34)(6.604,4.000){2}{\rule{1.300pt}{0.800pt}}
\put(809,633.34){\rule{2.400pt}{0.800pt}}
\multiput(809.00,631.34)(6.019,4.000){2}{\rule{1.200pt}{0.800pt}}
\put(820,637.34){\rule{2.400pt}{0.800pt}}
\multiput(820.00,635.34)(6.019,4.000){2}{\rule{1.200pt}{0.800pt}}
\put(831,641.34){\rule{2.600pt}{0.800pt}}
\multiput(831.00,639.34)(6.604,4.000){2}{\rule{1.300pt}{0.800pt}}
\put(843,645.34){\rule{2.400pt}{0.800pt}}
\multiput(843.00,643.34)(6.019,4.000){2}{\rule{1.200pt}{0.800pt}}
\put(854,649.34){\rule{2.400pt}{0.800pt}}
\multiput(854.00,647.34)(6.019,4.000){2}{\rule{1.200pt}{0.800pt}}
\put(865,653.34){\rule{2.600pt}{0.800pt}}
\multiput(865.00,651.34)(6.604,4.000){2}{\rule{1.300pt}{0.800pt}}
\put(877,657.34){\rule{2.400pt}{0.800pt}}
\multiput(877.00,655.34)(6.019,4.000){2}{\rule{1.200pt}{0.800pt}}
\multiput(888.00,662.38)(1.432,0.560){3}{\rule{1.960pt}{0.135pt}}
\multiput(888.00,659.34)(6.932,5.000){2}{\rule{0.980pt}{0.800pt}}
\put(899,666.34){\rule{2.600pt}{0.800pt}}
\multiput(899.00,664.34)(6.604,4.000){2}{\rule{1.300pt}{0.800pt}}
\put(911,670.34){\rule{2.400pt}{0.800pt}}
\multiput(911.00,668.34)(6.019,4.000){2}{\rule{1.200pt}{0.800pt}}
\put(922,674.34){\rule{2.400pt}{0.800pt}}
\multiput(922.00,672.34)(6.019,4.000){2}{\rule{1.200pt}{0.800pt}}
\put(933,678.34){\rule{2.600pt}{0.800pt}}
\multiput(933.00,676.34)(6.604,4.000){2}{\rule{1.300pt}{0.800pt}}
\put(945,682.34){\rule{2.400pt}{0.800pt}}
\multiput(945.00,680.34)(6.019,4.000){2}{\rule{1.200pt}{0.800pt}}
\put(956,686.34){\rule{2.600pt}{0.800pt}}
\multiput(956.00,684.34)(6.604,4.000){2}{\rule{1.300pt}{0.800pt}}
\put(968,690.34){\rule{2.400pt}{0.800pt}}
\multiput(968.00,688.34)(6.019,4.000){2}{\rule{1.200pt}{0.800pt}}
\put(979,694.34){\rule{2.400pt}{0.800pt}}
\multiput(979.00,692.34)(6.019,4.000){2}{\rule{1.200pt}{0.800pt}}
\put(990,698.34){\rule{2.600pt}{0.800pt}}
\multiput(990.00,696.34)(6.604,4.000){2}{\rule{1.300pt}{0.800pt}}
\put(1002,702.34){\rule{2.400pt}{0.800pt}}
\multiput(1002.00,700.34)(6.019,4.000){2}{\rule{1.200pt}{0.800pt}}
\multiput(1013.00,707.38)(1.432,0.560){3}{\rule{1.960pt}{0.135pt}}
\multiput(1013.00,704.34)(6.932,5.000){2}{\rule{0.980pt}{0.800pt}}
\put(1024,711.34){\rule{2.600pt}{0.800pt}}
\multiput(1024.00,709.34)(6.604,4.000){2}{\rule{1.300pt}{0.800pt}}
\put(1036,715.34){\rule{2.400pt}{0.800pt}}
\multiput(1036.00,713.34)(6.019,4.000){2}{\rule{1.200pt}{0.800pt}}
\put(1047,719.34){\rule{2.400pt}{0.800pt}}
\multiput(1047.00,717.34)(6.019,4.000){2}{\rule{1.200pt}{0.800pt}}
\put(1058,723.34){\rule{2.600pt}{0.800pt}}
\multiput(1058.00,721.34)(6.604,4.000){2}{\rule{1.300pt}{0.800pt}}
\put(1070,727.34){\rule{2.400pt}{0.800pt}}
\multiput(1070.00,725.34)(6.019,4.000){2}{\rule{1.200pt}{0.800pt}}
\put(1081,731.34){\rule{2.600pt}{0.800pt}}
\multiput(1081.00,729.34)(6.604,4.000){2}{\rule{1.300pt}{0.800pt}}
\put(1093,735.34){\rule{2.400pt}{0.800pt}}
\multiput(1093.00,733.34)(6.019,4.000){2}{\rule{1.200pt}{0.800pt}}
\put(1104,739.34){\rule{2.400pt}{0.800pt}}
\multiput(1104.00,737.34)(6.019,4.000){2}{\rule{1.200pt}{0.800pt}}
\put(1115,743.34){\rule{2.600pt}{0.800pt}}
\multiput(1115.00,741.34)(6.604,4.000){2}{\rule{1.300pt}{0.800pt}}
\put(1127,747.34){\rule{2.400pt}{0.800pt}}
\multiput(1127.00,745.34)(6.019,4.000){2}{\rule{1.200pt}{0.800pt}}
\multiput(1138.00,752.38)(1.432,0.560){3}{\rule{1.960pt}{0.135pt}}
\multiput(1138.00,749.34)(6.932,5.000){2}{\rule{0.980pt}{0.800pt}}
\put(1149,756.34){\rule{2.600pt}{0.800pt}}
\multiput(1149.00,754.34)(6.604,4.000){2}{\rule{1.300pt}{0.800pt}}
\put(1161,760.34){\rule{2.400pt}{0.800pt}}
\multiput(1161.00,758.34)(6.019,4.000){2}{\rule{1.200pt}{0.800pt}}
\put(1172,764.34){\rule{2.400pt}{0.800pt}}
\multiput(1172.00,762.34)(6.019,4.000){2}{\rule{1.200pt}{0.800pt}}
\put(1183,768.34){\rule{2.600pt}{0.800pt}}
\multiput(1183.00,766.34)(6.604,4.000){2}{\rule{1.300pt}{0.800pt}}
\put(1195,772.34){\rule{2.400pt}{0.800pt}}
\multiput(1195.00,770.34)(6.019,4.000){2}{\rule{1.200pt}{0.800pt}}
\put(1206,776.34){\rule{2.400pt}{0.800pt}}
\multiput(1206.00,774.34)(6.019,4.000){2}{\rule{1.200pt}{0.800pt}}
\put(1217,780.34){\rule{2.600pt}{0.800pt}}
\multiput(1217.00,778.34)(6.604,4.000){2}{\rule{1.300pt}{0.800pt}}
\put(1229,784.34){\rule{2.400pt}{0.800pt}}
\multiput(1229.00,782.34)(6.019,4.000){2}{\rule{1.200pt}{0.800pt}}
\put(1240,788.34){\rule{2.600pt}{0.800pt}}
\multiput(1240.00,786.34)(6.604,4.000){2}{\rule{1.300pt}{0.800pt}}
\multiput(1252.00,793.38)(1.432,0.560){3}{\rule{1.960pt}{0.135pt}}
\multiput(1252.00,790.34)(6.932,5.000){2}{\rule{0.980pt}{0.800pt}}
\put(1263,797.34){\rule{2.400pt}{0.800pt}}
\multiput(1263.00,795.34)(6.019,4.000){2}{\rule{1.200pt}{0.800pt}}
\put(1274,801.34){\rule{2.600pt}{0.800pt}}
\multiput(1274.00,799.34)(6.604,4.000){2}{\rule{1.300pt}{0.800pt}}
\put(1286,805.34){\rule{2.400pt}{0.800pt}}
\multiput(1286.00,803.34)(6.019,4.000){2}{\rule{1.200pt}{0.800pt}}
\put(1297,809.34){\rule{2.400pt}{0.800pt}}
\multiput(1297.00,807.34)(6.019,4.000){2}{\rule{1.200pt}{0.800pt}}
\put(1308,813.34){\rule{2.600pt}{0.800pt}}
\multiput(1308.00,811.34)(6.604,4.000){2}{\rule{1.300pt}{0.800pt}}
\put(1320,817.34){\rule{2.400pt}{0.800pt}}
\multiput(1320.00,815.34)(6.019,4.000){2}{\rule{1.200pt}{0.800pt}}
\put(1331,821.34){\rule{2.400pt}{0.800pt}}
\multiput(1331.00,819.34)(6.019,4.000){2}{\rule{1.200pt}{0.800pt}}
\put(1342,825.34){\rule{2.600pt}{0.800pt}}
\multiput(1342.00,823.34)(6.604,4.000){2}{\rule{1.300pt}{0.800pt}}
\put(1354,829.34){\rule{2.400pt}{0.800pt}}
\multiput(1354.00,827.34)(6.019,4.000){2}{\rule{1.200pt}{0.800pt}}
\put(1365,833.34){\rule{2.600pt}{0.800pt}}
\multiput(1365.00,831.34)(6.604,4.000){2}{\rule{1.300pt}{0.800pt}}
\multiput(1377.00,838.38)(1.432,0.560){3}{\rule{1.960pt}{0.135pt}}
\multiput(1377.00,835.34)(6.932,5.000){2}{\rule{0.980pt}{0.800pt}}
\end{picture}}
\vspace{3pt}
\caption{Check on the clock model, for $N=128$ and $d=2$.
For each particle,
$|\vdv_i|/|\ln\tilden_2|$
is plotted against $k_i$. The
diamonds are from a simulation with
$\tilden_2=\densA$,
after \simtime\
collisions. The plusses are from a simulation with
$\tilden_2=\densB$.
Also
plotted is a line with slope $1$.}
\label{fig:kdv}
\end{figure}

The second check is on the asymptotic form~(\ref{lambda}) of the Lyapunov
exponent, and its relation to the leading clock speed. To check the  leading
behavior of the largest Lyapunov exponent as a function of density, we need to
perform a number of simulations for different densities, all very low, because
the coefficient in the next term in  \Eq~(\ref{lambda}), although it could be
calculated in principle, is  not yet known and thus we cannot assume that we
can neglect it.  The leading clock speed $w_0$, which can be  determined from
the  parallel calculation of the clock values, \eq{kdyn}, does not depend on
density and can be  determined from just one simulation at an arbitrary
density.

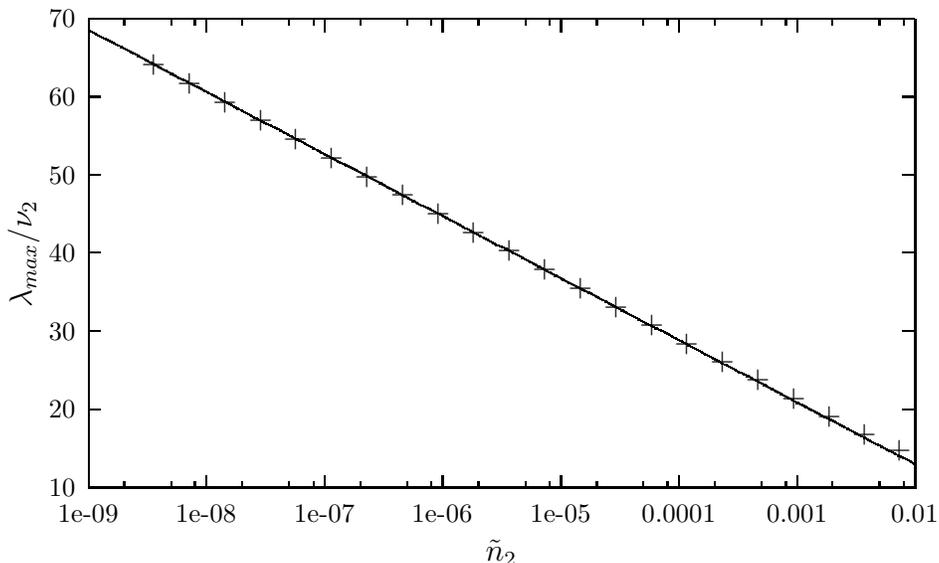
\begin{figure}
\centerline{\setlength{\unitlength}{0.240900pt}
\ifx\plotpoint\undefined\newsavebox{\plotpoint}\fi
\sbox{\plotpoint}{\rule[-0.200pt]{0.400pt}{0.400pt}}\begin{picture}(1500,900)(0,0)
\font\gnuplot=cmr10 at 10pt
\gnuplot
\sbox{\plotpoint}{\rule[-0.200pt]{0.400pt}{0.400pt}}\put(141.0,123.0){\rule[-0.200pt]{4.818pt}{0.400pt}}
\put(121,123){\makebox(0,0)[r]{10}}
\put(1419.0,123.0){\rule[-0.200pt]{4.818pt}{0.400pt}}
\put(141.0,246.0){\rule[-0.200pt]{4.818pt}{0.400pt}}
\put(121,246){\makebox(0,0)[r]{20}}
\put(1419.0,246.0){\rule[-0.200pt]{4.818pt}{0.400pt}}
\put(141.0,369.0){\rule[-0.200pt]{4.818pt}{0.400pt}}
\put(121,369){\makebox(0,0)[r]{30}}
\put(1419.0,369.0){\rule[-0.200pt]{4.818pt}{0.400pt}}
\put(141.0,492.0){\rule[-0.200pt]{4.818pt}{0.400pt}}
\put(121,492){\makebox(0,0)[r]{40}}
\put(1419.0,492.0){\rule[-0.200pt]{4.818pt}{0.400pt}}
\put(141.0,614.0){\rule[-0.200pt]{4.818pt}{0.400pt}}
\put(121,614){\makebox(0,0)[r]{50}}
\put(1419.0,614.0){\rule[-0.200pt]{4.818pt}{0.400pt}}
\put(141.0,737.0){\rule[-0.200pt]{4.818pt}{0.400pt}}
\put(121,737){\makebox(0,0)[r]{60}}
\put(1419.0,737.0){\rule[-0.200pt]{4.818pt}{0.400pt}}
\put(141.0,860.0){\rule[-0.200pt]{4.818pt}{0.400pt}}
\put(121,860){\makebox(0,0)[r]{70}}
\put(1419.0,860.0){\rule[-0.200pt]{4.818pt}{0.400pt}}
\put(141.0,123.0){\rule[-0.200pt]{0.400pt}{4.818pt}}
\put(141,82){\makebox(0,0){1e-09}}
\put(141.0,840.0){\rule[-0.200pt]{0.400pt}{4.818pt}}
\put(197.0,123.0){\rule[-0.200pt]{0.400pt}{2.409pt}}
\put(197.0,850.0){\rule[-0.200pt]{0.400pt}{2.409pt}}
\put(271.0,123.0){\rule[-0.200pt]{0.400pt}{2.409pt}}
\put(271.0,850.0){\rule[-0.200pt]{0.400pt}{2.409pt}}
\put(308.0,123.0){\rule[-0.200pt]{0.400pt}{2.409pt}}
\put(308.0,850.0){\rule[-0.200pt]{0.400pt}{2.409pt}}
\put(326.0,123.0){\rule[-0.200pt]{0.400pt}{4.818pt}}
\put(326,82){\makebox(0,0){1e-08}}
\put(326.0,840.0){\rule[-0.200pt]{0.400pt}{4.818pt}}
\put(382.0,123.0){\rule[-0.200pt]{0.400pt}{2.409pt}}
\put(382.0,850.0){\rule[-0.200pt]{0.400pt}{2.409pt}}
\put(456.0,123.0){\rule[-0.200pt]{0.400pt}{2.409pt}}
\put(456.0,850.0){\rule[-0.200pt]{0.400pt}{2.409pt}}
\put(494.0,123.0){\rule[-0.200pt]{0.400pt}{2.409pt}}
\put(494.0,850.0){\rule[-0.200pt]{0.400pt}{2.409pt}}
\put(512.0,123.0){\rule[-0.200pt]{0.400pt}{4.818pt}}
\put(512,82){\makebox(0,0){1e-07}}
\put(512.0,840.0){\rule[-0.200pt]{0.400pt}{4.818pt}}
\put(568.0,123.0){\rule[-0.200pt]{0.400pt}{2.409pt}}
\put(568.0,850.0){\rule[-0.200pt]{0.400pt}{2.409pt}}
\put(641.0,123.0){\rule[-0.200pt]{0.400pt}{2.409pt}}
\put(641.0,850.0){\rule[-0.200pt]{0.400pt}{2.409pt}}
\put(679.0,123.0){\rule[-0.200pt]{0.400pt}{2.409pt}}
\put(679.0,850.0){\rule[-0.200pt]{0.400pt}{2.409pt}}
\put(697.0,123.0){\rule[-0.200pt]{0.400pt}{4.818pt}}
\put(697,82){\makebox(0,0){1e-06}}
\put(697.0,840.0){\rule[-0.200pt]{0.400pt}{4.818pt}}
\put(753.0,123.0){\rule[-0.200pt]{0.400pt}{2.409pt}}
\put(753.0,850.0){\rule[-0.200pt]{0.400pt}{2.409pt}}
\put(827.0,123.0){\rule[-0.200pt]{0.400pt}{2.409pt}}
\put(827.0,850.0){\rule[-0.200pt]{0.400pt}{2.409pt}}
\put(865.0,123.0){\rule[-0.200pt]{0.400pt}{2.409pt}}
\put(865.0,850.0){\rule[-0.200pt]{0.400pt}{2.409pt}}
\put(883.0,123.0){\rule[-0.200pt]{0.400pt}{4.818pt}}
\put(883,82){\makebox(0,0){1e-05}}
\put(883.0,840.0){\rule[-0.200pt]{0.400pt}{4.818pt}}
\put(939.0,123.0){\rule[-0.200pt]{0.400pt}{2.409pt}}
\put(939.0,850.0){\rule[-0.200pt]{0.400pt}{2.409pt}}
\put(1012.0,123.0){\rule[-0.200pt]{0.400pt}{2.409pt}}
\put(1012.0,850.0){\rule[-0.200pt]{0.400pt}{2.409pt}}
\put(1050.0,123.0){\rule[-0.200pt]{0.400pt}{2.409pt}}
\put(1050.0,850.0){\rule[-0.200pt]{0.400pt}{2.409pt}}
\put(1068.0,123.0){\rule[-0.200pt]{0.400pt}{4.818pt}}
\put(1068,82){\makebox(0,0){0.0001}}
\put(1068.0,840.0){\rule[-0.200pt]{0.400pt}{4.818pt}}
\put(1124.0,123.0){\rule[-0.200pt]{0.400pt}{2.409pt}}
\put(1124.0,850.0){\rule[-0.200pt]{0.400pt}{2.409pt}}
\put(1198.0,123.0){\rule[-0.200pt]{0.400pt}{2.409pt}}
\put(1198.0,850.0){\rule[-0.200pt]{0.400pt}{2.409pt}}
\put(1236.0,123.0){\rule[-0.200pt]{0.400pt}{2.409pt}}
\put(1236.0,850.0){\rule[-0.200pt]{0.400pt}{2.409pt}}
\put(1254.0,123.0){\rule[-0.200pt]{0.400pt}{4.818pt}}
\put(1254,82){\makebox(0,0){0.001}}
\put(1254.0,840.0){\rule[-0.200pt]{0.400pt}{4.818pt}}
\put(1309.0,123.0){\rule[-0.200pt]{0.400pt}{2.409pt}}
\put(1309.0,850.0){\rule[-0.200pt]{0.400pt}{2.409pt}}
\put(1383.0,123.0){\rule[-0.200pt]{0.400pt}{2.409pt}}
\put(1383.0,850.0){\rule[-0.200pt]{0.400pt}{2.409pt}}
\put(1421.0,123.0){\rule[-0.200pt]{0.400pt}{2.409pt}}
\put(1421.0,850.0){\rule[-0.200pt]{0.400pt}{2.409pt}}
\put(1439.0,123.0){\rule[-0.200pt]{0.400pt}{4.818pt}}
\put(1439,82){\makebox(0,0){0.01}}
\put(1439.0,840.0){\rule[-0.200pt]{0.400pt}{4.818pt}}
\put(141.0,123.0){\rule[-0.200pt]{312.688pt}{0.400pt}}
\put(1439.0,123.0){\rule[-0.200pt]{0.400pt}{177.543pt}}
\put(141.0,860.0){\rule[-0.200pt]{312.688pt}{0.400pt}}
\put(40,491){\makebox(0,0){\rotatebox{90}{$\lambda_{max}/\nu_2$}}}
\put(790,21){\makebox(0,0){$\tilde n_2$}}
\put(141.0,123.0){\rule[-0.200pt]{0.400pt}{177.543pt}}
\put(141,842){\usebox{\plotpoint}}
\multiput(141.00,840.93)(0.950,-0.485){11}{\rule{0.843pt}{0.117pt}}
\multiput(141.00,841.17)(11.251,-7.000){2}{\rule{0.421pt}{0.400pt}}
\multiput(154.00,833.93)(0.950,-0.485){11}{\rule{0.843pt}{0.117pt}}
\multiput(154.00,834.17)(11.251,-7.000){2}{\rule{0.421pt}{0.400pt}}
\multiput(167.00,826.93)(1.123,-0.482){9}{\rule{0.967pt}{0.116pt}}
\multiput(167.00,827.17)(10.994,-6.000){2}{\rule{0.483pt}{0.400pt}}
\multiput(180.00,820.93)(0.950,-0.485){11}{\rule{0.843pt}{0.117pt}}
\multiput(180.00,821.17)(11.251,-7.000){2}{\rule{0.421pt}{0.400pt}}
\multiput(193.00,813.93)(1.026,-0.485){11}{\rule{0.900pt}{0.117pt}}
\multiput(193.00,814.17)(12.132,-7.000){2}{\rule{0.450pt}{0.400pt}}
\multiput(207.00,806.93)(0.950,-0.485){11}{\rule{0.843pt}{0.117pt}}
\multiput(207.00,807.17)(11.251,-7.000){2}{\rule{0.421pt}{0.400pt}}
\multiput(220.00,799.93)(0.950,-0.485){11}{\rule{0.843pt}{0.117pt}}
\multiput(220.00,800.17)(11.251,-7.000){2}{\rule{0.421pt}{0.400pt}}
\multiput(233.00,792.93)(0.950,-0.485){11}{\rule{0.843pt}{0.117pt}}
\multiput(233.00,793.17)(11.251,-7.000){2}{\rule{0.421pt}{0.400pt}}
\multiput(246.00,785.93)(0.950,-0.485){11}{\rule{0.843pt}{0.117pt}}
\multiput(246.00,786.17)(11.251,-7.000){2}{\rule{0.421pt}{0.400pt}}
\multiput(259.00,778.93)(0.950,-0.485){11}{\rule{0.843pt}{0.117pt}}
\multiput(259.00,779.17)(11.251,-7.000){2}{\rule{0.421pt}{0.400pt}}
\multiput(272.00,771.93)(0.950,-0.485){11}{\rule{0.843pt}{0.117pt}}
\multiput(272.00,772.17)(11.251,-7.000){2}{\rule{0.421pt}{0.400pt}}
\multiput(285.00,764.93)(1.123,-0.482){9}{\rule{0.967pt}{0.116pt}}
\multiput(285.00,765.17)(10.994,-6.000){2}{\rule{0.483pt}{0.400pt}}
\multiput(298.00,758.93)(0.950,-0.485){11}{\rule{0.843pt}{0.117pt}}
\multiput(298.00,759.17)(11.251,-7.000){2}{\rule{0.421pt}{0.400pt}}
\multiput(311.00,751.93)(1.026,-0.485){11}{\rule{0.900pt}{0.117pt}}
\multiput(311.00,752.17)(12.132,-7.000){2}{\rule{0.450pt}{0.400pt}}
\multiput(325.00,744.93)(0.950,-0.485){11}{\rule{0.843pt}{0.117pt}}
\multiput(325.00,745.17)(11.251,-7.000){2}{\rule{0.421pt}{0.400pt}}
\multiput(338.00,737.93)(0.950,-0.485){11}{\rule{0.843pt}{0.117pt}}
\multiput(338.00,738.17)(11.251,-7.000){2}{\rule{0.421pt}{0.400pt}}
\multiput(351.00,730.93)(0.950,-0.485){11}{\rule{0.843pt}{0.117pt}}
\multiput(351.00,731.17)(11.251,-7.000){2}{\rule{0.421pt}{0.400pt}}
\multiput(364.00,723.93)(0.950,-0.485){11}{\rule{0.843pt}{0.117pt}}
\multiput(364.00,724.17)(11.251,-7.000){2}{\rule{0.421pt}{0.400pt}}
\multiput(377.00,716.93)(0.950,-0.485){11}{\rule{0.843pt}{0.117pt}}
\multiput(377.00,717.17)(11.251,-7.000){2}{\rule{0.421pt}{0.400pt}}
\multiput(390.00,709.93)(0.950,-0.485){11}{\rule{0.843pt}{0.117pt}}
\multiput(390.00,710.17)(11.251,-7.000){2}{\rule{0.421pt}{0.400pt}}
\multiput(403.00,702.93)(0.950,-0.485){11}{\rule{0.843pt}{0.117pt}}
\multiput(403.00,703.17)(11.251,-7.000){2}{\rule{0.421pt}{0.400pt}}
\multiput(416.00,695.93)(1.123,-0.482){9}{\rule{0.967pt}{0.116pt}}
\multiput(416.00,696.17)(10.994,-6.000){2}{\rule{0.483pt}{0.400pt}}
\multiput(429.00,689.93)(1.026,-0.485){11}{\rule{0.900pt}{0.117pt}}
\multiput(429.00,690.17)(12.132,-7.000){2}{\rule{0.450pt}{0.400pt}}
\multiput(443.00,682.93)(0.950,-0.485){11}{\rule{0.843pt}{0.117pt}}
\multiput(443.00,683.17)(11.251,-7.000){2}{\rule{0.421pt}{0.400pt}}
\multiput(456.00,675.93)(0.950,-0.485){11}{\rule{0.843pt}{0.117pt}}
\multiput(456.00,676.17)(11.251,-7.000){2}{\rule{0.421pt}{0.400pt}}
\multiput(469.00,668.93)(0.950,-0.485){11}{\rule{0.843pt}{0.117pt}}
\multiput(469.00,669.17)(11.251,-7.000){2}{\rule{0.421pt}{0.400pt}}
\multiput(482.00,661.93)(0.950,-0.485){11}{\rule{0.843pt}{0.117pt}}
\multiput(482.00,662.17)(11.251,-7.000){2}{\rule{0.421pt}{0.400pt}}
\multiput(495.00,654.93)(0.950,-0.485){11}{\rule{0.843pt}{0.117pt}}
\multiput(495.00,655.17)(11.251,-7.000){2}{\rule{0.421pt}{0.400pt}}
\multiput(508.00,647.93)(0.950,-0.485){11}{\rule{0.843pt}{0.117pt}}
\multiput(508.00,648.17)(11.251,-7.000){2}{\rule{0.421pt}{0.400pt}}
\multiput(521.00,640.93)(0.950,-0.485){11}{\rule{0.843pt}{0.117pt}}
\multiput(521.00,641.17)(11.251,-7.000){2}{\rule{0.421pt}{0.400pt}}
\multiput(534.00,633.93)(1.123,-0.482){9}{\rule{0.967pt}{0.116pt}}
\multiput(534.00,634.17)(10.994,-6.000){2}{\rule{0.483pt}{0.400pt}}
\multiput(547.00,627.93)(1.026,-0.485){11}{\rule{0.900pt}{0.117pt}}
\multiput(547.00,628.17)(12.132,-7.000){2}{\rule{0.450pt}{0.400pt}}
\multiput(561.00,620.93)(0.950,-0.485){11}{\rule{0.843pt}{0.117pt}}
\multiput(561.00,621.17)(11.251,-7.000){2}{\rule{0.421pt}{0.400pt}}
\multiput(574.00,613.93)(0.950,-0.485){11}{\rule{0.843pt}{0.117pt}}
\multiput(574.00,614.17)(11.251,-7.000){2}{\rule{0.421pt}{0.400pt}}
\multiput(587.00,606.93)(0.950,-0.485){11}{\rule{0.843pt}{0.117pt}}
\multiput(587.00,607.17)(11.251,-7.000){2}{\rule{0.421pt}{0.400pt}}
\multiput(600.00,599.93)(0.950,-0.485){11}{\rule{0.843pt}{0.117pt}}
\multiput(600.00,600.17)(11.251,-7.000){2}{\rule{0.421pt}{0.400pt}}
\multiput(613.00,592.93)(0.950,-0.485){11}{\rule{0.843pt}{0.117pt}}
\multiput(613.00,593.17)(11.251,-7.000){2}{\rule{0.421pt}{0.400pt}}
\multiput(626.00,585.93)(0.950,-0.485){11}{\rule{0.843pt}{0.117pt}}
\multiput(626.00,586.17)(11.251,-7.000){2}{\rule{0.421pt}{0.400pt}}
\multiput(639.00,578.93)(0.950,-0.485){11}{\rule{0.843pt}{0.117pt}}
\multiput(639.00,579.17)(11.251,-7.000){2}{\rule{0.421pt}{0.400pt}}
\multiput(652.00,571.93)(0.950,-0.485){11}{\rule{0.843pt}{0.117pt}}
\multiput(652.00,572.17)(11.251,-7.000){2}{\rule{0.421pt}{0.400pt}}
\multiput(665.00,564.93)(1.214,-0.482){9}{\rule{1.033pt}{0.116pt}}
\multiput(665.00,565.17)(11.855,-6.000){2}{\rule{0.517pt}{0.400pt}}
\multiput(679.00,558.93)(0.950,-0.485){11}{\rule{0.843pt}{0.117pt}}
\multiput(679.00,559.17)(11.251,-7.000){2}{\rule{0.421pt}{0.400pt}}
\multiput(692.00,551.93)(0.950,-0.485){11}{\rule{0.843pt}{0.117pt}}
\multiput(692.00,552.17)(11.251,-7.000){2}{\rule{0.421pt}{0.400pt}}
\multiput(705.00,544.93)(0.950,-0.485){11}{\rule{0.843pt}{0.117pt}}
\multiput(705.00,545.17)(11.251,-7.000){2}{\rule{0.421pt}{0.400pt}}
\multiput(718.00,537.93)(0.950,-0.485){11}{\rule{0.843pt}{0.117pt}}
\multiput(718.00,538.17)(11.251,-7.000){2}{\rule{0.421pt}{0.400pt}}
\multiput(731.00,530.93)(0.950,-0.485){11}{\rule{0.843pt}{0.117pt}}
\multiput(731.00,531.17)(11.251,-7.000){2}{\rule{0.421pt}{0.400pt}}
\multiput(744.00,523.93)(0.950,-0.485){11}{\rule{0.843pt}{0.117pt}}
\multiput(744.00,524.17)(11.251,-7.000){2}{\rule{0.421pt}{0.400pt}}
\multiput(757.00,516.93)(0.950,-0.485){11}{\rule{0.843pt}{0.117pt}}
\multiput(757.00,517.17)(11.251,-7.000){2}{\rule{0.421pt}{0.400pt}}
\multiput(770.00,509.93)(0.950,-0.485){11}{\rule{0.843pt}{0.117pt}}
\multiput(770.00,510.17)(11.251,-7.000){2}{\rule{0.421pt}{0.400pt}}
\multiput(783.00,502.93)(1.214,-0.482){9}{\rule{1.033pt}{0.116pt}}
\multiput(783.00,503.17)(11.855,-6.000){2}{\rule{0.517pt}{0.400pt}}
\multiput(797.00,496.93)(0.950,-0.485){11}{\rule{0.843pt}{0.117pt}}
\multiput(797.00,497.17)(11.251,-7.000){2}{\rule{0.421pt}{0.400pt}}
\multiput(810.00,489.93)(0.950,-0.485){11}{\rule{0.843pt}{0.117pt}}
\multiput(810.00,490.17)(11.251,-7.000){2}{\rule{0.421pt}{0.400pt}}
\multiput(823.00,482.93)(0.950,-0.485){11}{\rule{0.843pt}{0.117pt}}
\multiput(823.00,483.17)(11.251,-7.000){2}{\rule{0.421pt}{0.400pt}}
\multiput(836.00,475.93)(0.950,-0.485){11}{\rule{0.843pt}{0.117pt}}
\multiput(836.00,476.17)(11.251,-7.000){2}{\rule{0.421pt}{0.400pt}}
\multiput(849.00,468.93)(0.950,-0.485){11}{\rule{0.843pt}{0.117pt}}
\multiput(849.00,469.17)(11.251,-7.000){2}{\rule{0.421pt}{0.400pt}}
\multiput(862.00,461.93)(0.950,-0.485){11}{\rule{0.843pt}{0.117pt}}
\multiput(862.00,462.17)(11.251,-7.000){2}{\rule{0.421pt}{0.400pt}}
\multiput(875.00,454.93)(0.950,-0.485){11}{\rule{0.843pt}{0.117pt}}
\multiput(875.00,455.17)(11.251,-7.000){2}{\rule{0.421pt}{0.400pt}}
\multiput(888.00,447.93)(0.950,-0.485){11}{\rule{0.843pt}{0.117pt}}
\multiput(888.00,448.17)(11.251,-7.000){2}{\rule{0.421pt}{0.400pt}}
\multiput(901.00,440.93)(1.026,-0.485){11}{\rule{0.900pt}{0.117pt}}
\multiput(901.00,441.17)(12.132,-7.000){2}{\rule{0.450pt}{0.400pt}}
\multiput(915.00,433.93)(1.123,-0.482){9}{\rule{0.967pt}{0.116pt}}
\multiput(915.00,434.17)(10.994,-6.000){2}{\rule{0.483pt}{0.400pt}}
\multiput(928.00,427.93)(0.950,-0.485){11}{\rule{0.843pt}{0.117pt}}
\multiput(928.00,428.17)(11.251,-7.000){2}{\rule{0.421pt}{0.400pt}}
\multiput(941.00,420.93)(0.950,-0.485){11}{\rule{0.843pt}{0.117pt}}
\multiput(941.00,421.17)(11.251,-7.000){2}{\rule{0.421pt}{0.400pt}}
\multiput(954.00,413.93)(0.950,-0.485){11}{\rule{0.843pt}{0.117pt}}
\multiput(954.00,414.17)(11.251,-7.000){2}{\rule{0.421pt}{0.400pt}}
\multiput(967.00,406.93)(0.950,-0.485){11}{\rule{0.843pt}{0.117pt}}
\multiput(967.00,407.17)(11.251,-7.000){2}{\rule{0.421pt}{0.400pt}}
\multiput(980.00,399.93)(0.950,-0.485){11}{\rule{0.843pt}{0.117pt}}
\multiput(980.00,400.17)(11.251,-7.000){2}{\rule{0.421pt}{0.400pt}}
\multiput(993.00,392.93)(0.950,-0.485){11}{\rule{0.843pt}{0.117pt}}
\multiput(993.00,393.17)(11.251,-7.000){2}{\rule{0.421pt}{0.400pt}}
\multiput(1006.00,385.93)(0.950,-0.485){11}{\rule{0.843pt}{0.117pt}}
\multiput(1006.00,386.17)(11.251,-7.000){2}{\rule{0.421pt}{0.400pt}}
\multiput(1019.00,378.93)(1.026,-0.485){11}{\rule{0.900pt}{0.117pt}}
\multiput(1019.00,379.17)(12.132,-7.000){2}{\rule{0.450pt}{0.400pt}}
\multiput(1033.00,371.93)(1.123,-0.482){9}{\rule{0.967pt}{0.116pt}}
\multiput(1033.00,372.17)(10.994,-6.000){2}{\rule{0.483pt}{0.400pt}}
\multiput(1046.00,365.93)(0.950,-0.485){11}{\rule{0.843pt}{0.117pt}}
\multiput(1046.00,366.17)(11.251,-7.000){2}{\rule{0.421pt}{0.400pt}}
\multiput(1059.00,358.93)(0.950,-0.485){11}{\rule{0.843pt}{0.117pt}}
\multiput(1059.00,359.17)(11.251,-7.000){2}{\rule{0.421pt}{0.400pt}}
\multiput(1072.00,351.93)(0.950,-0.485){11}{\rule{0.843pt}{0.117pt}}
\multiput(1072.00,352.17)(11.251,-7.000){2}{\rule{0.421pt}{0.400pt}}
\multiput(1085.00,344.93)(0.950,-0.485){11}{\rule{0.843pt}{0.117pt}}
\multiput(1085.00,345.17)(11.251,-7.000){2}{\rule{0.421pt}{0.400pt}}
\multiput(1098.00,337.93)(0.950,-0.485){11}{\rule{0.843pt}{0.117pt}}
\multiput(1098.00,338.17)(11.251,-7.000){2}{\rule{0.421pt}{0.400pt}}
\multiput(1111.00,330.93)(0.950,-0.485){11}{\rule{0.843pt}{0.117pt}}
\multiput(1111.00,331.17)(11.251,-7.000){2}{\rule{0.421pt}{0.400pt}}
\multiput(1124.00,323.93)(0.950,-0.485){11}{\rule{0.843pt}{0.117pt}}
\multiput(1124.00,324.17)(11.251,-7.000){2}{\rule{0.421pt}{0.400pt}}
\multiput(1137.00,316.93)(1.026,-0.485){11}{\rule{0.900pt}{0.117pt}}
\multiput(1137.00,317.17)(12.132,-7.000){2}{\rule{0.450pt}{0.400pt}}
\multiput(1151.00,309.93)(0.950,-0.485){11}{\rule{0.843pt}{0.117pt}}
\multiput(1151.00,310.17)(11.251,-7.000){2}{\rule{0.421pt}{0.400pt}}
\multiput(1164.00,302.93)(1.123,-0.482){9}{\rule{0.967pt}{0.116pt}}
\multiput(1164.00,303.17)(10.994,-6.000){2}{\rule{0.483pt}{0.400pt}}
\multiput(1177.00,296.93)(0.950,-0.485){11}{\rule{0.843pt}{0.117pt}}
\multiput(1177.00,297.17)(11.251,-7.000){2}{\rule{0.421pt}{0.400pt}}
\multiput(1190.00,289.93)(0.950,-0.485){11}{\rule{0.843pt}{0.117pt}}
\multiput(1190.00,290.17)(11.251,-7.000){2}{\rule{0.421pt}{0.400pt}}
\multiput(1203.00,282.93)(0.950,-0.485){11}{\rule{0.843pt}{0.117pt}}
\multiput(1203.00,283.17)(11.251,-7.000){2}{\rule{0.421pt}{0.400pt}}
\multiput(1216.00,275.93)(0.950,-0.485){11}{\rule{0.843pt}{0.117pt}}
\multiput(1216.00,276.17)(11.251,-7.000){2}{\rule{0.421pt}{0.400pt}}
\multiput(1229.00,268.93)(0.950,-0.485){11}{\rule{0.843pt}{0.117pt}}
\multiput(1229.00,269.17)(11.251,-7.000){2}{\rule{0.421pt}{0.400pt}}
\multiput(1242.00,261.93)(0.950,-0.485){11}{\rule{0.843pt}{0.117pt}}
\multiput(1242.00,262.17)(11.251,-7.000){2}{\rule{0.421pt}{0.400pt}}
\multiput(1255.00,254.93)(1.026,-0.485){11}{\rule{0.900pt}{0.117pt}}
\multiput(1255.00,255.17)(12.132,-7.000){2}{\rule{0.450pt}{0.400pt}}
\multiput(1269.00,247.93)(0.950,-0.485){11}{\rule{0.843pt}{0.117pt}}
\multiput(1269.00,248.17)(11.251,-7.000){2}{\rule{0.421pt}{0.400pt}}
\multiput(1282.00,240.93)(0.950,-0.485){11}{\rule{0.843pt}{0.117pt}}
\multiput(1282.00,241.17)(11.251,-7.000){2}{\rule{0.421pt}{0.400pt}}
\multiput(1295.00,233.93)(1.123,-0.482){9}{\rule{0.967pt}{0.116pt}}
\multiput(1295.00,234.17)(10.994,-6.000){2}{\rule{0.483pt}{0.400pt}}
\multiput(1308.00,227.93)(0.950,-0.485){11}{\rule{0.843pt}{0.117pt}}
\multiput(1308.00,228.17)(11.251,-7.000){2}{\rule{0.421pt}{0.400pt}}
\multiput(1321.00,220.93)(0.950,-0.485){11}{\rule{0.843pt}{0.117pt}}
\multiput(1321.00,221.17)(11.251,-7.000){2}{\rule{0.421pt}{0.400pt}}
\multiput(1334.00,213.93)(0.950,-0.485){11}{\rule{0.843pt}{0.117pt}}
\multiput(1334.00,214.17)(11.251,-7.000){2}{\rule{0.421pt}{0.400pt}}
\multiput(1347.00,206.93)(0.950,-0.485){11}{\rule{0.843pt}{0.117pt}}
\multiput(1347.00,207.17)(11.251,-7.000){2}{\rule{0.421pt}{0.400pt}}
\multiput(1360.00,199.93)(0.950,-0.485){11}{\rule{0.843pt}{0.117pt}}
\multiput(1360.00,200.17)(11.251,-7.000){2}{\rule{0.421pt}{0.400pt}}
\multiput(1373.00,192.93)(1.026,-0.485){11}{\rule{0.900pt}{0.117pt}}
\multiput(1373.00,193.17)(12.132,-7.000){2}{\rule{0.450pt}{0.400pt}}
\multiput(1387.00,185.93)(0.950,-0.485){11}{\rule{0.843pt}{0.117pt}}
\multiput(1387.00,186.17)(11.251,-7.000){2}{\rule{0.421pt}{0.400pt}}
\multiput(1400.00,178.93)(0.950,-0.485){11}{\rule{0.843pt}{0.117pt}}
\multiput(1400.00,179.17)(11.251,-7.000){2}{\rule{0.421pt}{0.400pt}}
\multiput(1413.00,171.93)(1.123,-0.482){9}{\rule{0.967pt}{0.116pt}}
\multiput(1413.00,172.17)(10.994,-6.000){2}{\rule{0.483pt}{0.400pt}}
\multiput(1426.00,165.93)(0.950,-0.485){11}{\rule{0.843pt}{0.117pt}}
\multiput(1426.00,166.17)(11.251,-7.000){2}{\rule{0.421pt}{0.400pt}}
\sbox{\plotpoint}{\rule[-0.400pt]{0.800pt}{0.800pt}}\put(1415,181){\makebox(0,0){$+$}}
\put(1360,207){\makebox(0,0){$+$}}
\put(1305,235){\makebox(0,0){$+$}}
\put(1249,263){\makebox(0,0){$+$}}
\put(1193,292){\makebox(0,0){$+$}}
\put(1137,320){\makebox(0,0){$+$}}
\put(1081,349){\makebox(0,0){$+$}}
\put(1026,378){\makebox(0,0){$+$}}
\put(970,407){\makebox(0,0){$+$}}
\put(914,437){\makebox(0,0){$+$}}
\put(858,466){\makebox(0,0){$+$}}
\put(802,495){\makebox(0,0){$+$}}
\put(746,524){\makebox(0,0){$+$}}
\put(691,553){\makebox(0,0){$+$}}
\put(635,583){\makebox(0,0){$+$}}
\put(579,612){\makebox(0,0){$+$}}
\put(523,641){\makebox(0,0){$+$}}
\put(467,670){\makebox(0,0){$+$}}
\put(412,700){\makebox(0,0){$+$}}
\put(356,729){\makebox(0,0){$+$}}
\put(300,758){\makebox(0,0){$+$}}
\put(244,788){\makebox(0,0){$+$}}
\end{picture}}
\vspace{3pt}
\caption{Density dependence of the largest Lyapunov exponent
$\lambda_{max}$ for $N=64$, $d=2$. The plusses are DSMC results.
The linear curves is a function of the form~(\ref{lambda}), where the
first two terms are included. For $w_0$, the clock speed from the
simulation is  taken, $w_1$ is a fit parameter.} \label{fig:lyap}
\end{figure}

In this case, simulations are done with $N=64$ particles, again in $d=2$
dimensions. The Lyapunov exponent divided by the collision frequency is plotted
as a function of the density  $\tilden_2$  in \Fig~\ref{fig:lyap}.  To justify
the form of the density expansion given in (\ref{lambda}), we also plotted  the
function  $-w_0\ln\tilden_2 +w_1$,  where $w_0$ is the clock speed found from
the  parallel simulation of the clock values, $w_0=3.4479\pm 0.0016$ and $w_1$
is obtained from fitting (using only points from densities  $\tilde n_2$
smaller or equal to $10^{-5}$). This function describes the simulation points
very well, i.e., the slope is indeed given by the clock speed. The deviations
at higher density may be accounted for by higher powers of $1/\ln\tilde n_2$.

Surprisingly, the value of $w_0$ is far from the predicted value in Eq.
(\ref{approx}). This is due to finite size  effects, which have been well
studied by now\cite{leid,Brunet,Lemarchand}. Corrections due to the finiteness
of $N$ are, for large $N$, of the order of $\ln^{-2}N$, but for $N=64$ we are
not yet in the asymptotic regime for which this scaling holds. The
investigation for larger  numbers of particles and the calculation of the
finite size corrections will be done in future work, but we remark that for a
simplified clock model, in which the collision frequency is taken to be
velocity independent, this was already carried out in \cite{leid}.

Thirdly, we will check the velocity distribution $P_{\mbox{\scriptsize
head}}(\vv)$ in the head of the clock distribution.  This distribution is
determined as follows. After the simulation has run for some time, we determine
the average of the clock values and shift all the clock values by that amount,
so we can access the stationary $F(k,\vv)$ rather than the shifting
$C(k,\vv,t)$. We then pick some lowest clock value $k_0$; all particles with a
lower clock value are discarded. Of the remaining particles, which are those in
the head of the distribution, the velocity distribution is constructed.  We
adjust $k_0$ such that $k>k_0$ can be identified with the tail of the
distribution. To get good statistics (which is difficult because there are not
many particles in the head), we measure this distribution at several times, and
average the result.

As the velocity distribution $P_{\mbox{\scriptsize head}}(\vv)$ only  involves
clock values, there is no density dependence and a simulation  at one density
is enough. We ran a simulation with  $N=128$ particles, and $d=2$. The
distribution of the clock values was obtained, and a value of $k_0=7$ seemed a
reasonable start of the tail  of the distribution. We have checked that the
results do not change much  when instead we take $k_0=6$ or $k_0=8$.

\begin{figure}
\centerline{\setlength{\unitlength}{0.240900pt}
\ifx\plotpoint\undefined\newsavebox{\plotpoint}\fi
\sbox{\plotpoint}{\rule[-0.200pt]{0.400pt}{0.400pt}}\begin{picture}(1500,900)(0,0)
\font\gnuplot=cmr10 at 10pt
\gnuplot
\sbox{\plotpoint}{\rule[-0.200pt]{0.400pt}{0.400pt}}\put(161.0,123.0){\rule[-0.200pt]{4.818pt}{0.400pt}}
\put(141,123){\makebox(0,0)[r]{0}}
\put(1419.0,123.0){\rule[-0.200pt]{4.818pt}{0.400pt}}
\put(161.0,228.0){\rule[-0.200pt]{4.818pt}{0.400pt}}
\put(141,228){\makebox(0,0)[r]{0.1}}
\put(1419.0,228.0){\rule[-0.200pt]{4.818pt}{0.400pt}}
\put(161.0,334.0){\rule[-0.200pt]{4.818pt}{0.400pt}}
\put(141,334){\makebox(0,0)[r]{0.2}}
\put(1419.0,334.0){\rule[-0.200pt]{4.818pt}{0.400pt}}
\put(161.0,439.0){\rule[-0.200pt]{4.818pt}{0.400pt}}
\put(141,439){\makebox(0,0)[r]{0.3}}
\put(1419.0,439.0){\rule[-0.200pt]{4.818pt}{0.400pt}}
\put(161.0,544.0){\rule[-0.200pt]{4.818pt}{0.400pt}}
\put(141,544){\makebox(0,0)[r]{0.4}}
\put(1419.0,544.0){\rule[-0.200pt]{4.818pt}{0.400pt}}
\put(161.0,649.0){\rule[-0.200pt]{4.818pt}{0.400pt}}
\put(141,649){\makebox(0,0)[r]{0.5}}
\put(1419.0,649.0){\rule[-0.200pt]{4.818pt}{0.400pt}}
\put(161.0,755.0){\rule[-0.200pt]{4.818pt}{0.400pt}}
\put(141,755){\makebox(0,0)[r]{0.6}}
\put(1419.0,755.0){\rule[-0.200pt]{4.818pt}{0.400pt}}
\put(161.0,860.0){\rule[-0.200pt]{4.818pt}{0.400pt}}
\put(141,860){\makebox(0,0)[r]{0.7}}
\put(1419.0,860.0){\rule[-0.200pt]{4.818pt}{0.400pt}}
\put(161.0,123.0){\rule[-0.200pt]{0.400pt}{4.818pt}}
\put(161,82){\makebox(0,0){0}}
\put(161.0,840.0){\rule[-0.200pt]{0.400pt}{4.818pt}}
\put(321.0,123.0){\rule[-0.200pt]{0.400pt}{4.818pt}}
\put(321,82){\makebox(0,0){0.5}}
\put(321.0,840.0){\rule[-0.200pt]{0.400pt}{4.818pt}}
\put(481.0,123.0){\rule[-0.200pt]{0.400pt}{4.818pt}}
\put(481,82){\makebox(0,0){1}}
\put(481.0,840.0){\rule[-0.200pt]{0.400pt}{4.818pt}}
\put(640.0,123.0){\rule[-0.200pt]{0.400pt}{4.818pt}}
\put(640,82){\makebox(0,0){1.5}}
\put(640.0,840.0){\rule[-0.200pt]{0.400pt}{4.818pt}}
\put(800.0,123.0){\rule[-0.200pt]{0.400pt}{4.818pt}}
\put(800,82){\makebox(0,0){2}}
\put(800.0,840.0){\rule[-0.200pt]{0.400pt}{4.818pt}}
\put(960.0,123.0){\rule[-0.200pt]{0.400pt}{4.818pt}}
\put(960,82){\makebox(0,0){2.5}}
\put(960.0,840.0){\rule[-0.200pt]{0.400pt}{4.818pt}}
\put(1120.0,123.0){\rule[-0.200pt]{0.400pt}{4.818pt}}
\put(1120,82){\makebox(0,0){3}}
\put(1120.0,840.0){\rule[-0.200pt]{0.400pt}{4.818pt}}
\put(1279.0,123.0){\rule[-0.200pt]{0.400pt}{4.818pt}}
\put(1279,82){\makebox(0,0){3.5}}
\put(1279.0,840.0){\rule[-0.200pt]{0.400pt}{4.818pt}}
\put(1439.0,123.0){\rule[-0.200pt]{0.400pt}{4.818pt}}
\put(1439,82){\makebox(0,0){4}}
\put(1439.0,840.0){\rule[-0.200pt]{0.400pt}{4.818pt}}
\put(161.0,123.0){\rule[-0.200pt]{307.870pt}{0.400pt}}
\put(1439.0,123.0){\rule[-0.200pt]{0.400pt}{177.543pt}}
\put(161.0,860.0){\rule[-0.200pt]{307.870pt}{0.400pt}}
\put(40,491){\makebox(0,0){\rotatebox{90}{$P_{head}$}}}
\put(800,21){\makebox(0,0){$v$}}
\put(161.0,123.0){\rule[-0.200pt]{0.400pt}{177.543pt}}
\put(161.0,123.0){\rule[-0.200pt]{0.400pt}{17.345pt}}
\put(161.0,195.0){\rule[-0.200pt]{15.418pt}{0.400pt}}
\put(225.0,123.0){\rule[-0.200pt]{0.400pt}{17.345pt}}
\put(161.0,123.0){\rule[-0.200pt]{15.418pt}{0.400pt}}
\put(225.0,123.0){\rule[-0.200pt]{0.400pt}{39.748pt}}
\put(225.0,288.0){\rule[-0.200pt]{15.418pt}{0.400pt}}
\put(289.0,123.0){\rule[-0.200pt]{0.400pt}{39.748pt}}
\put(225.0,123.0){\rule[-0.200pt]{15.418pt}{0.400pt}}
\put(289.0,123.0){\rule[-0.200pt]{0.400pt}{89.615pt}}
\put(289.0,495.0){\rule[-0.200pt]{15.418pt}{0.400pt}}
\put(353.0,123.0){\rule[-0.200pt]{0.400pt}{89.615pt}}
\put(289.0,123.0){\rule[-0.200pt]{15.418pt}{0.400pt}}
\put(353.0,123.0){\rule[-0.200pt]{0.400pt}{106.960pt}}
\put(353.0,567.0){\rule[-0.200pt]{15.418pt}{0.400pt}}
\put(417.0,123.0){\rule[-0.200pt]{0.400pt}{106.960pt}}
\put(353.0,123.0){\rule[-0.200pt]{15.418pt}{0.400pt}}
\put(417.0,123.0){\rule[-0.200pt]{0.400pt}{129.363pt}}
\put(417.0,660.0){\rule[-0.200pt]{15.418pt}{0.400pt}}
\put(481.0,123.0){\rule[-0.200pt]{0.400pt}{129.363pt}}
\put(417.0,123.0){\rule[-0.200pt]{15.418pt}{0.400pt}}
\put(481.0,123.0){\rule[-0.200pt]{0.400pt}{126.713pt}}
\put(481.0,649.0){\rule[-0.200pt]{15.177pt}{0.400pt}}
\put(544.0,123.0){\rule[-0.200pt]{0.400pt}{126.713pt}}
\put(481.0,123.0){\rule[-0.200pt]{15.177pt}{0.400pt}}
\put(544.0,123.0){\rule[-0.200pt]{0.400pt}{144.299pt}}
\put(544.0,722.0){\rule[-0.200pt]{15.418pt}{0.400pt}}
\put(608.0,123.0){\rule[-0.200pt]{0.400pt}{144.299pt}}
\put(544.0,123.0){\rule[-0.200pt]{15.418pt}{0.400pt}}
\put(608.0,123.0){\rule[-0.200pt]{0.400pt}{101.901pt}}
\put(608.0,546.0){\rule[-0.200pt]{15.418pt}{0.400pt}}
\put(672.0,123.0){\rule[-0.200pt]{0.400pt}{101.901pt}}
\put(608.0,123.0){\rule[-0.200pt]{15.418pt}{0.400pt}}
\put(672.0,123.0){\rule[-0.200pt]{0.400pt}{111.778pt}}
\put(672.0,587.0){\rule[-0.200pt]{15.418pt}{0.400pt}}
\put(736.0,123.0){\rule[-0.200pt]{0.400pt}{111.778pt}}
\put(672.0,123.0){\rule[-0.200pt]{15.418pt}{0.400pt}}
\put(736.0,123.0){\rule[-0.200pt]{0.400pt}{99.492pt}}
\put(736.0,536.0){\rule[-0.200pt]{15.418pt}{0.400pt}}
\put(800.0,123.0){\rule[-0.200pt]{0.400pt}{99.492pt}}
\put(736.0,123.0){\rule[-0.200pt]{15.418pt}{0.400pt}}
\put(800.0,123.0){\rule[-0.200pt]{0.400pt}{69.620pt}}
\put(800.0,412.0){\rule[-0.200pt]{15.418pt}{0.400pt}}
\put(864.0,123.0){\rule[-0.200pt]{0.400pt}{69.620pt}}
\put(800.0,123.0){\rule[-0.200pt]{15.418pt}{0.400pt}}
\put(864.0,123.0){\rule[-0.200pt]{0.400pt}{67.211pt}}
\put(864.0,402.0){\rule[-0.200pt]{15.418pt}{0.400pt}}
\put(928.0,123.0){\rule[-0.200pt]{0.400pt}{67.211pt}}
\put(864.0,123.0){\rule[-0.200pt]{15.418pt}{0.400pt}}
\put(928.0,123.0){\rule[-0.200pt]{0.400pt}{39.748pt}}
\put(928.0,288.0){\rule[-0.200pt]{15.418pt}{0.400pt}}
\put(992.0,123.0){\rule[-0.200pt]{0.400pt}{39.748pt}}
\put(928.0,123.0){\rule[-0.200pt]{15.418pt}{0.400pt}}
\put(992.0,123.0){\rule[-0.200pt]{0.400pt}{52.275pt}}
\put(992.0,340.0){\rule[-0.200pt]{15.418pt}{0.400pt}}
\put(1056.0,123.0){\rule[-0.200pt]{0.400pt}{52.275pt}}
\put(992.0,123.0){\rule[-0.200pt]{15.418pt}{0.400pt}}
\put(1056.0,123.0){\rule[-0.200pt]{0.400pt}{27.463pt}}
\put(1056.0,237.0){\rule[-0.200pt]{15.418pt}{0.400pt}}
\put(1120.0,123.0){\rule[-0.200pt]{0.400pt}{27.463pt}}
\put(1056.0,123.0){\rule[-0.200pt]{15.418pt}{0.400pt}}
\put(1120.0,123.0){\rule[-0.200pt]{0.400pt}{17.345pt}}
\put(1120.0,195.0){\rule[-0.200pt]{15.177pt}{0.400pt}}
\put(1183.0,123.0){\rule[-0.200pt]{0.400pt}{17.345pt}}
\put(1120.0,123.0){\rule[-0.200pt]{15.177pt}{0.400pt}}
\put(1183.0,123.0){\rule[-0.200pt]{0.400pt}{9.877pt}}
\put(1183.0,164.0){\rule[-0.200pt]{15.418pt}{0.400pt}}
\put(1247.0,123.0){\rule[-0.200pt]{0.400pt}{9.877pt}}
\put(1183.0,123.0){\rule[-0.200pt]{15.418pt}{0.400pt}}
\put(1247.0,123.0){\rule[-0.200pt]{0.400pt}{14.936pt}}
\put(1247.0,185.0){\rule[-0.200pt]{15.418pt}{0.400pt}}
\put(1311.0,123.0){\rule[-0.200pt]{0.400pt}{14.936pt}}
\put(1247.0,123.0){\rule[-0.200pt]{15.418pt}{0.400pt}}
\put(1311.0,123.0){\rule[-0.200pt]{0.400pt}{2.409pt}}
\put(1311.0,133.0){\rule[-0.200pt]{15.418pt}{0.400pt}}
\put(1375.0,123.0){\rule[-0.200pt]{0.400pt}{2.409pt}}
\put(1311.0,123.0){\rule[-0.200pt]{15.418pt}{0.400pt}}
\put(1375,123){\usebox{\plotpoint}}
\put(1375.0,123.0){\rule[-0.200pt]{15.418pt}{0.400pt}}
\put(1375.0,123.0){\rule[-0.200pt]{15.418pt}{0.400pt}}
\put(161,123){\usebox{\plotpoint}}
\multiput(161,123)(6.006,19.867){3}{\usebox{\plotpoint}}
\multiput(174,166)(6.137,19.827){2}{\usebox{\plotpoint}}
\multiput(187,208)(6.137,19.827){2}{\usebox{\plotpoint}}
\multiput(200,250)(6.273,19.785){2}{\usebox{\plotpoint}}
\multiput(213,291)(6.415,19.739){2}{\usebox{\plotpoint}}
\multiput(226,331)(5.964,19.880){2}{\usebox{\plotpoint}}
\multiput(238,371)(6.718,19.638){2}{\usebox{\plotpoint}}
\multiput(251,409)(6.880,19.582){2}{\usebox{\plotpoint}}
\multiput(264,446)(7.227,19.457){2}{\usebox{\plotpoint}}
\put(283.41,497.77){\usebox{\plotpoint}}
\multiput(290,515)(7.812,19.229){2}{\usebox{\plotpoint}}
\multiput(303,547)(8.253,19.044){2}{\usebox{\plotpoint}}
\put(323.62,593.41){\usebox{\plotpoint}}
\multiput(329,605)(9.282,18.564){2}{\usebox{\plotpoint}}
\put(352.05,648.78){\usebox{\plotpoint}}
\put(362.77,666.55){\usebox{\plotpoint}}
\put(373.77,684.14){\usebox{\plotpoint}}
\put(385.53,701.23){\usebox{\plotpoint}}
\put(398.75,717.19){\usebox{\plotpoint}}
\put(413.42,731.85){\usebox{\plotpoint}}
\put(429.83,744.49){\usebox{\plotpoint}}
\put(448.12,754.20){\usebox{\plotpoint}}
\put(467.91,760.29){\usebox{\plotpoint}}
\put(488.50,761.31){\usebox{\plotpoint}}
\put(508.61,756.43){\usebox{\plotpoint}}
\put(527.55,748.01){\usebox{\plotpoint}}
\put(545.11,737.00){\usebox{\plotpoint}}
\put(561.62,724.43){\usebox{\plotpoint}}
\put(576.76,710.24){\usebox{\plotpoint}}
\put(591.27,695.40){\usebox{\plotpoint}}
\put(605.19,680.01){\usebox{\plotpoint}}
\put(618.57,664.15){\usebox{\plotpoint}}
\put(631.45,647.88){\usebox{\plotpoint}}
\put(643.87,631.25){\usebox{\plotpoint}}
\put(655.81,614.28){\usebox{\plotpoint}}
\put(667.51,597.14){\usebox{\plotpoint}}
\put(679.57,580.24){\usebox{\plotpoint}}
\put(691.29,563.12){\usebox{\plotpoint}}
\multiput(703,546)(11.720,-17.130){2}{\usebox{\plotpoint}}
\put(726.45,511.73){\usebox{\plotpoint}}
\put(738.51,494.83){\usebox{\plotpoint}}
\put(750.35,477.79){\usebox{\plotpoint}}
\put(762.07,460.66){\usebox{\plotpoint}}
\put(774.01,443.68){\usebox{\plotpoint}}
\put(786.35,427.00){\usebox{\plotpoint}}
\put(798.53,410.21){\usebox{\plotpoint}}
\put(810.60,393.34){\usebox{\plotpoint}}
\put(823.51,377.10){\usebox{\plotpoint}}
\put(836.44,360.87){\usebox{\plotpoint}}
\put(850.04,345.19){\usebox{\plotpoint}}
\put(863.63,329.50){\usebox{\plotpoint}}
\put(877.47,314.04){\usebox{\plotpoint}}
\put(891.89,299.11){\usebox{\plotpoint}}
\put(906.56,284.44){\usebox{\plotpoint}}
\put(921.68,270.22){\usebox{\plotpoint}}
\put(937.42,256.70){\usebox{\plotpoint}}
\put(953.08,243.10){\usebox{\plotpoint}}
\put(969.85,230.88){\usebox{\plotpoint}}
\put(986.91,219.06){\usebox{\plotpoint}}
\put(1004.58,208.18){\usebox{\plotpoint}}
\put(1022.57,197.84){\usebox{\plotpoint}}
\put(1040.91,188.12){\usebox{\plotpoint}}
\put(1059.97,179.94){\usebox{\plotpoint}}
\put(1078.99,171.67){\usebox{\plotpoint}}
\put(1098.54,164.71){\usebox{\plotpoint}}
\put(1118.27,158.30){\usebox{\plotpoint}}
\put(1138.29,152.86){\usebox{\plotpoint}}
\put(1158.51,148.19){\usebox{\plotpoint}}
\put(1178.89,144.32){\usebox{\plotpoint}}
\put(1199.22,140.20){\usebox{\plotpoint}}
\put(1219.73,137.04){\usebox{\plotpoint}}
\put(1240.34,134.72){\usebox{\plotpoint}}
\put(1261.00,132.77){\usebox{\plotpoint}}
\put(1281.69,131.18){\usebox{\plotpoint}}
\put(1302.39,129.59){\usebox{\plotpoint}}
\put(1323.08,127.99){\usebox{\plotpoint}}
\put(1343.80,127.00){\usebox{\plotpoint}}
\put(1364.52,126.00){\usebox{\plotpoint}}
\put(1385.27,126.00){\usebox{\plotpoint}}
\put(1405.99,125.00){\usebox{\plotpoint}}
\put(1426.74,124.94){\usebox{\plotpoint}}
\put(1439,124){\usebox{\plotpoint}}
\sbox{\plotpoint}{\rule[-0.400pt]{0.800pt}{0.800pt}}\put(161,123){\usebox{\plotpoint}}
\multiput(162.41,123.00)(0.509,1.071){19}{\rule{0.123pt}{1.862pt}}
\multiput(159.34,123.00)(13.000,23.136){2}{\rule{0.800pt}{0.931pt}}
\multiput(175.41,150.00)(0.509,1.071){19}{\rule{0.123pt}{1.862pt}}
\multiput(172.34,150.00)(13.000,23.136){2}{\rule{0.800pt}{0.931pt}}
\multiput(188.41,177.00)(0.509,1.071){19}{\rule{0.123pt}{1.862pt}}
\multiput(185.34,177.00)(13.000,23.136){2}{\rule{0.800pt}{0.931pt}}
\multiput(201.41,204.00)(0.509,1.071){19}{\rule{0.123pt}{1.862pt}}
\multiput(198.34,204.00)(13.000,23.136){2}{\rule{0.800pt}{0.931pt}}
\multiput(214.41,231.00)(0.509,1.071){19}{\rule{0.123pt}{1.862pt}}
\multiput(211.34,231.00)(13.000,23.136){2}{\rule{0.800pt}{0.931pt}}
\multiput(227.41,258.00)(0.511,1.123){17}{\rule{0.123pt}{1.933pt}}
\multiput(224.34,258.00)(12.000,21.987){2}{\rule{0.800pt}{0.967pt}}
\multiput(239.41,284.00)(0.509,1.029){19}{\rule{0.123pt}{1.800pt}}
\multiput(236.34,284.00)(13.000,22.264){2}{\rule{0.800pt}{0.900pt}}
\multiput(252.41,310.00)(0.509,0.988){19}{\rule{0.123pt}{1.738pt}}
\multiput(249.34,310.00)(13.000,21.392){2}{\rule{0.800pt}{0.869pt}}
\multiput(265.41,335.00)(0.509,0.988){19}{\rule{0.123pt}{1.738pt}}
\multiput(262.34,335.00)(13.000,21.392){2}{\rule{0.800pt}{0.869pt}}
\multiput(278.41,360.00)(0.509,0.947){19}{\rule{0.123pt}{1.677pt}}
\multiput(275.34,360.00)(13.000,20.519){2}{\rule{0.800pt}{0.838pt}}
\multiput(291.41,384.00)(0.509,0.947){19}{\rule{0.123pt}{1.677pt}}
\multiput(288.34,384.00)(13.000,20.519){2}{\rule{0.800pt}{0.838pt}}
\multiput(304.41,408.00)(0.509,0.905){19}{\rule{0.123pt}{1.615pt}}
\multiput(301.34,408.00)(13.000,19.647){2}{\rule{0.800pt}{0.808pt}}
\multiput(317.41,431.00)(0.509,0.864){19}{\rule{0.123pt}{1.554pt}}
\multiput(314.34,431.00)(13.000,18.775){2}{\rule{0.800pt}{0.777pt}}
\multiput(330.41,453.00)(0.509,0.823){19}{\rule{0.123pt}{1.492pt}}
\multiput(327.34,453.00)(13.000,17.903){2}{\rule{0.800pt}{0.746pt}}
\multiput(343.41,474.00)(0.509,0.823){19}{\rule{0.123pt}{1.492pt}}
\multiput(340.34,474.00)(13.000,17.903){2}{\rule{0.800pt}{0.746pt}}
\multiput(356.41,495.00)(0.509,0.740){19}{\rule{0.123pt}{1.369pt}}
\multiput(353.34,495.00)(13.000,16.158){2}{\rule{0.800pt}{0.685pt}}
\multiput(369.41,514.00)(0.511,0.807){17}{\rule{0.123pt}{1.467pt}}
\multiput(366.34,514.00)(12.000,15.956){2}{\rule{0.800pt}{0.733pt}}
\multiput(381.41,533.00)(0.509,0.657){19}{\rule{0.123pt}{1.246pt}}
\multiput(378.34,533.00)(13.000,14.414){2}{\rule{0.800pt}{0.623pt}}
\multiput(394.41,550.00)(0.509,0.657){19}{\rule{0.123pt}{1.246pt}}
\multiput(391.34,550.00)(13.000,14.414){2}{\rule{0.800pt}{0.623pt}}
\multiput(407.41,567.00)(0.509,0.574){19}{\rule{0.123pt}{1.123pt}}
\multiput(404.34,567.00)(13.000,12.669){2}{\rule{0.800pt}{0.562pt}}
\multiput(420.41,582.00)(0.509,0.533){19}{\rule{0.123pt}{1.062pt}}
\multiput(417.34,582.00)(13.000,11.797){2}{\rule{0.800pt}{0.531pt}}
\multiput(432.00,597.41)(0.492,0.509){19}{\rule{1.000pt}{0.123pt}}
\multiput(432.00,594.34)(10.924,13.000){2}{\rule{0.500pt}{0.800pt}}
\multiput(445.00,610.41)(0.536,0.511){17}{\rule{1.067pt}{0.123pt}}
\multiput(445.00,607.34)(10.786,12.000){2}{\rule{0.533pt}{0.800pt}}
\multiput(458.00,622.40)(0.654,0.514){13}{\rule{1.240pt}{0.124pt}}
\multiput(458.00,619.34)(10.426,10.000){2}{\rule{0.620pt}{0.800pt}}
\multiput(471.00,632.40)(0.737,0.516){11}{\rule{1.356pt}{0.124pt}}
\multiput(471.00,629.34)(10.186,9.000){2}{\rule{0.678pt}{0.800pt}}
\multiput(484.00,641.40)(0.847,0.520){9}{\rule{1.500pt}{0.125pt}}
\multiput(484.00,638.34)(9.887,8.000){2}{\rule{0.750pt}{0.800pt}}
\multiput(497.00,649.40)(1.000,0.526){7}{\rule{1.686pt}{0.127pt}}
\multiput(497.00,646.34)(9.501,7.000){2}{\rule{0.843pt}{0.800pt}}
\multiput(510.00,656.38)(1.600,0.560){3}{\rule{2.120pt}{0.135pt}}
\multiput(510.00,653.34)(7.600,5.000){2}{\rule{1.060pt}{0.800pt}}
\put(522,660.34){\rule{2.800pt}{0.800pt}}
\multiput(522.00,658.34)(7.188,4.000){2}{\rule{1.400pt}{0.800pt}}
\put(535,663.84){\rule{3.132pt}{0.800pt}}
\multiput(535.00,662.34)(6.500,3.000){2}{\rule{1.566pt}{0.800pt}}
\put(548,665.84){\rule{3.132pt}{0.800pt}}
\multiput(548.00,665.34)(6.500,1.000){2}{\rule{1.566pt}{0.800pt}}
\put(574,665.84){\rule{3.132pt}{0.800pt}}
\multiput(574.00,666.34)(6.500,-1.000){2}{\rule{1.566pt}{0.800pt}}
\put(587,664.34){\rule{3.132pt}{0.800pt}}
\multiput(587.00,665.34)(6.500,-2.000){2}{\rule{1.566pt}{0.800pt}}
\put(600,661.34){\rule{2.800pt}{0.800pt}}
\multiput(600.00,663.34)(7.188,-4.000){2}{\rule{1.400pt}{0.800pt}}
\put(613,657.34){\rule{2.800pt}{0.800pt}}
\multiput(613.00,659.34)(7.188,-4.000){2}{\rule{1.400pt}{0.800pt}}
\multiput(626.00,655.07)(1.244,-0.536){5}{\rule{1.933pt}{0.129pt}}
\multiput(626.00,655.34)(8.987,-6.000){2}{\rule{0.967pt}{0.800pt}}
\multiput(639.00,649.08)(1.000,-0.526){7}{\rule{1.686pt}{0.127pt}}
\multiput(639.00,649.34)(9.501,-7.000){2}{\rule{0.843pt}{0.800pt}}
\multiput(652.00,642.08)(0.774,-0.520){9}{\rule{1.400pt}{0.125pt}}
\multiput(652.00,642.34)(9.094,-8.000){2}{\rule{0.700pt}{0.800pt}}
\multiput(664.00,634.08)(0.847,-0.520){9}{\rule{1.500pt}{0.125pt}}
\multiput(664.00,634.34)(9.887,-8.000){2}{\rule{0.750pt}{0.800pt}}
\multiput(677.00,626.08)(0.654,-0.514){13}{\rule{1.240pt}{0.124pt}}
\multiput(677.00,626.34)(10.426,-10.000){2}{\rule{0.620pt}{0.800pt}}
\multiput(690.00,616.08)(0.654,-0.514){13}{\rule{1.240pt}{0.124pt}}
\multiput(690.00,616.34)(10.426,-10.000){2}{\rule{0.620pt}{0.800pt}}
\multiput(703.00,606.08)(0.536,-0.511){17}{\rule{1.067pt}{0.123pt}}
\multiput(703.00,606.34)(10.786,-12.000){2}{\rule{0.533pt}{0.800pt}}
\multiput(716.00,594.08)(0.589,-0.512){15}{\rule{1.145pt}{0.123pt}}
\multiput(716.00,594.34)(10.623,-11.000){2}{\rule{0.573pt}{0.800pt}}
\multiput(729.00,583.08)(0.492,-0.509){19}{\rule{1.000pt}{0.123pt}}
\multiput(729.00,583.34)(10.924,-13.000){2}{\rule{0.500pt}{0.800pt}}
\multiput(742.00,570.08)(0.492,-0.509){19}{\rule{1.000pt}{0.123pt}}
\multiput(742.00,570.34)(10.924,-13.000){2}{\rule{0.500pt}{0.800pt}}
\multiput(755.00,557.08)(0.492,-0.509){19}{\rule{1.000pt}{0.123pt}}
\multiput(755.00,557.34)(10.924,-13.000){2}{\rule{0.500pt}{0.800pt}}
\multiput(769.41,541.59)(0.509,-0.533){19}{\rule{0.123pt}{1.062pt}}
\multiput(766.34,543.80)(13.000,-11.797){2}{\rule{0.800pt}{0.531pt}}
\multiput(782.41,527.59)(0.509,-0.533){19}{\rule{0.123pt}{1.062pt}}
\multiput(779.34,529.80)(13.000,-11.797){2}{\rule{0.800pt}{0.531pt}}
\multiput(795.41,513.30)(0.511,-0.581){17}{\rule{0.123pt}{1.133pt}}
\multiput(792.34,515.65)(12.000,-11.648){2}{\rule{0.800pt}{0.567pt}}
\multiput(807.41,499.34)(0.509,-0.574){19}{\rule{0.123pt}{1.123pt}}
\multiput(804.34,501.67)(13.000,-12.669){2}{\rule{0.800pt}{0.562pt}}
\multiput(820.41,484.34)(0.509,-0.574){19}{\rule{0.123pt}{1.123pt}}
\multiput(817.34,486.67)(13.000,-12.669){2}{\rule{0.800pt}{0.562pt}}
\multiput(833.41,469.34)(0.509,-0.574){19}{\rule{0.123pt}{1.123pt}}
\multiput(830.34,471.67)(13.000,-12.669){2}{\rule{0.800pt}{0.562pt}}
\multiput(846.41,454.59)(0.509,-0.533){19}{\rule{0.123pt}{1.062pt}}
\multiput(843.34,456.80)(13.000,-11.797){2}{\rule{0.800pt}{0.531pt}}
\multiput(859.41,440.34)(0.509,-0.574){19}{\rule{0.123pt}{1.123pt}}
\multiput(856.34,442.67)(13.000,-12.669){2}{\rule{0.800pt}{0.562pt}}
\multiput(872.41,425.34)(0.509,-0.574){19}{\rule{0.123pt}{1.123pt}}
\multiput(869.34,427.67)(13.000,-12.669){2}{\rule{0.800pt}{0.562pt}}
\multiput(885.41,410.59)(0.509,-0.533){19}{\rule{0.123pt}{1.062pt}}
\multiput(882.34,412.80)(13.000,-11.797){2}{\rule{0.800pt}{0.531pt}}
\multiput(898.41,396.59)(0.509,-0.533){19}{\rule{0.123pt}{1.062pt}}
\multiput(895.34,398.80)(13.000,-11.797){2}{\rule{0.800pt}{0.531pt}}
\multiput(911.41,382.59)(0.509,-0.533){19}{\rule{0.123pt}{1.062pt}}
\multiput(908.34,384.80)(13.000,-11.797){2}{\rule{0.800pt}{0.531pt}}
\multiput(924.41,368.59)(0.509,-0.533){19}{\rule{0.123pt}{1.062pt}}
\multiput(921.34,370.80)(13.000,-11.797){2}{\rule{0.800pt}{0.531pt}}
\multiput(937.41,354.57)(0.511,-0.536){17}{\rule{0.123pt}{1.067pt}}
\multiput(934.34,356.79)(12.000,-10.786){2}{\rule{0.800pt}{0.533pt}}
\multiput(948.00,344.08)(0.492,-0.509){19}{\rule{1.000pt}{0.123pt}}
\multiput(948.00,344.34)(10.924,-13.000){2}{\rule{0.500pt}{0.800pt}}
\multiput(961.00,331.08)(0.492,-0.509){19}{\rule{1.000pt}{0.123pt}}
\multiput(961.00,331.34)(10.924,-13.000){2}{\rule{0.500pt}{0.800pt}}
\multiput(974.00,318.08)(0.536,-0.511){17}{\rule{1.067pt}{0.123pt}}
\multiput(974.00,318.34)(10.786,-12.000){2}{\rule{0.533pt}{0.800pt}}
\multiput(987.00,306.08)(0.536,-0.511){17}{\rule{1.067pt}{0.123pt}}
\multiput(987.00,306.34)(10.786,-12.000){2}{\rule{0.533pt}{0.800pt}}
\multiput(1000.00,294.08)(0.589,-0.512){15}{\rule{1.145pt}{0.123pt}}
\multiput(1000.00,294.34)(10.623,-11.000){2}{\rule{0.573pt}{0.800pt}}
\multiput(1013.00,283.08)(0.589,-0.512){15}{\rule{1.145pt}{0.123pt}}
\multiput(1013.00,283.34)(10.623,-11.000){2}{\rule{0.573pt}{0.800pt}}
\multiput(1026.00,272.08)(0.654,-0.514){13}{\rule{1.240pt}{0.124pt}}
\multiput(1026.00,272.34)(10.426,-10.000){2}{\rule{0.620pt}{0.800pt}}
\multiput(1039.00,262.08)(0.654,-0.514){13}{\rule{1.240pt}{0.124pt}}
\multiput(1039.00,262.34)(10.426,-10.000){2}{\rule{0.620pt}{0.800pt}}
\multiput(1052.00,252.08)(0.654,-0.514){13}{\rule{1.240pt}{0.124pt}}
\multiput(1052.00,252.34)(10.426,-10.000){2}{\rule{0.620pt}{0.800pt}}
\multiput(1065.00,242.08)(0.737,-0.516){11}{\rule{1.356pt}{0.124pt}}
\multiput(1065.00,242.34)(10.186,-9.000){2}{\rule{0.678pt}{0.800pt}}
\multiput(1078.00,233.08)(0.774,-0.520){9}{\rule{1.400pt}{0.125pt}}
\multiput(1078.00,233.34)(9.094,-8.000){2}{\rule{0.700pt}{0.800pt}}
\multiput(1090.00,225.08)(0.847,-0.520){9}{\rule{1.500pt}{0.125pt}}
\multiput(1090.00,225.34)(9.887,-8.000){2}{\rule{0.750pt}{0.800pt}}
\multiput(1103.00,217.08)(0.847,-0.520){9}{\rule{1.500pt}{0.125pt}}
\multiput(1103.00,217.34)(9.887,-8.000){2}{\rule{0.750pt}{0.800pt}}
\multiput(1116.00,209.08)(1.000,-0.526){7}{\rule{1.686pt}{0.127pt}}
\multiput(1116.00,209.34)(9.501,-7.000){2}{\rule{0.843pt}{0.800pt}}
\multiput(1129.00,202.08)(1.000,-0.526){7}{\rule{1.686pt}{0.127pt}}
\multiput(1129.00,202.34)(9.501,-7.000){2}{\rule{0.843pt}{0.800pt}}
\multiput(1142.00,195.07)(1.244,-0.536){5}{\rule{1.933pt}{0.129pt}}
\multiput(1142.00,195.34)(8.987,-6.000){2}{\rule{0.967pt}{0.800pt}}
\multiput(1155.00,189.07)(1.244,-0.536){5}{\rule{1.933pt}{0.129pt}}
\multiput(1155.00,189.34)(8.987,-6.000){2}{\rule{0.967pt}{0.800pt}}
\multiput(1168.00,183.06)(1.768,-0.560){3}{\rule{2.280pt}{0.135pt}}
\multiput(1168.00,183.34)(8.268,-5.000){2}{\rule{1.140pt}{0.800pt}}
\multiput(1181.00,178.06)(1.768,-0.560){3}{\rule{2.280pt}{0.135pt}}
\multiput(1181.00,178.34)(8.268,-5.000){2}{\rule{1.140pt}{0.800pt}}
\multiput(1194.00,173.06)(1.768,-0.560){3}{\rule{2.280pt}{0.135pt}}
\multiput(1194.00,173.34)(8.268,-5.000){2}{\rule{1.140pt}{0.800pt}}
\put(1207,166.34){\rule{2.800pt}{0.800pt}}
\multiput(1207.00,168.34)(7.188,-4.000){2}{\rule{1.400pt}{0.800pt}}
\put(1220,162.34){\rule{2.600pt}{0.800pt}}
\multiput(1220.00,164.34)(6.604,-4.000){2}{\rule{1.300pt}{0.800pt}}
\put(1232,158.34){\rule{2.800pt}{0.800pt}}
\multiput(1232.00,160.34)(7.188,-4.000){2}{\rule{1.400pt}{0.800pt}}
\put(1245,154.34){\rule{2.800pt}{0.800pt}}
\multiput(1245.00,156.34)(7.188,-4.000){2}{\rule{1.400pt}{0.800pt}}
\put(1258,150.84){\rule{3.132pt}{0.800pt}}
\multiput(1258.00,152.34)(6.500,-3.000){2}{\rule{1.566pt}{0.800pt}}
\put(1271,147.84){\rule{3.132pt}{0.800pt}}
\multiput(1271.00,149.34)(6.500,-3.000){2}{\rule{1.566pt}{0.800pt}}
\put(1284,145.34){\rule{3.132pt}{0.800pt}}
\multiput(1284.00,146.34)(6.500,-2.000){2}{\rule{1.566pt}{0.800pt}}
\put(1297,142.84){\rule{3.132pt}{0.800pt}}
\multiput(1297.00,144.34)(6.500,-3.000){2}{\rule{1.566pt}{0.800pt}}
\put(1310,140.34){\rule{3.132pt}{0.800pt}}
\multiput(1310.00,141.34)(6.500,-2.000){2}{\rule{1.566pt}{0.800pt}}
\put(1323,138.34){\rule{3.132pt}{0.800pt}}
\multiput(1323.00,139.34)(6.500,-2.000){2}{\rule{1.566pt}{0.800pt}}
\put(1336,136.84){\rule{3.132pt}{0.800pt}}
\multiput(1336.00,137.34)(6.500,-1.000){2}{\rule{1.566pt}{0.800pt}}
\put(1349,135.34){\rule{3.132pt}{0.800pt}}
\multiput(1349.00,136.34)(6.500,-2.000){2}{\rule{1.566pt}{0.800pt}}
\put(1362,133.34){\rule{2.891pt}{0.800pt}}
\multiput(1362.00,134.34)(6.000,-2.000){2}{\rule{1.445pt}{0.800pt}}
\put(1374,131.84){\rule{3.132pt}{0.800pt}}
\multiput(1374.00,132.34)(6.500,-1.000){2}{\rule{1.566pt}{0.800pt}}
\put(1387,130.84){\rule{3.132pt}{0.800pt}}
\multiput(1387.00,131.34)(6.500,-1.000){2}{\rule{1.566pt}{0.800pt}}
\put(1400,129.84){\rule{3.132pt}{0.800pt}}
\multiput(1400.00,130.34)(6.500,-1.000){2}{\rule{1.566pt}{0.800pt}}
\put(1413,128.84){\rule{3.132pt}{0.800pt}}
\multiput(1413.00,129.34)(6.500,-1.000){2}{\rule{1.566pt}{0.800pt}}
\put(1426,127.84){\rule{3.132pt}{0.800pt}}
\multiput(1426.00,128.34)(6.500,-1.000){2}{\rule{1.566pt}{0.800pt}}
\put(561.0,668.0){\rule[-0.400pt]{3.132pt}{0.800pt}}
\end{picture}}
\vspace{3pt}
\caption{Comparison of the velocity distribution in the leading edge. The
histogram results from the simulations for $N=128$, in $d=2$ dimensions.
The solid curve is a plot of the prediction~(\ref{predA}). The dotted curve
is the
two-dimensional Maxwell
distribution, $v\exp(-v^2/2)$.
}
\label{fig:predA}
\end{figure}
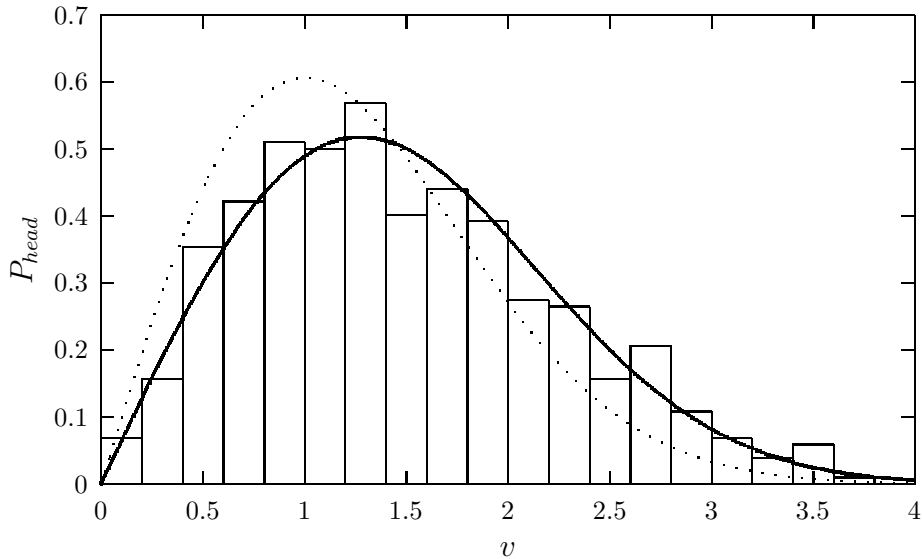

Combining \Eqs~(\ref{A}) and (\ref{headdistr}) gives the explicit prediction
for the velocity distribution in the head:
\begin{equation}
        P_{\mbox{\scriptsize head}}(v) = \left(
        0.63943 + 0.16289 v^2 + 0.004351 v^4\right)v\,e^{-v^2/2}.
\eql{predA}
\end{equation}
This prediction has been plotted in \Fig~\ref{fig:predA} together with the
distribution found from the simulations, and the velocity distribution in the
whole gas. Even though the statistics isn't perfect, there is a very good
agreement between the two. It should be noted that leaving out the $v^4$ term
in \eq{predA} gives only a small difference.  It is also evident that the
velocity distribution in the head of the clock distribution  has higher mean
velocity than the velocity distribution characterizing  the full gas, which is
also plotted in figure \eq{fig:predA}.

\section{The Dilute, Random Hard-Ball Lorentz Gas}

A very useful and simple example of a dynamical system which is a dispersing
billiard, rather than a semi-dispersing billiard is provided by the Lorentz gas
model. In this model one places fixed, hard  balls of radius $a$ in a
$d$-dimensional space, and then considers the  motion of a point particle of
mass $m$ and with initial velocity $\vv$, in the free space between the
scatterers~\cite{chapcow}. The particle moves  freely between collisions and
makes specular, energy conserving collisions with the scatterers . We consider
here only the case where the scatterers are not allowed to overlap each other.
One can imagine that the scatterers are placed with their centers on the sites
of a regular lattice, or that the scatterers are placed  randomly in space.
Here we consider the case where $N$ scatterers are placed randomly in space in
a volume $V$ in such a way that the average distance between scatterers is much
greater than their radii, $a$. That is, the scatterers form a quenched, dilute
gas with  $na^d \ll 1$.

In this section we show that the simplifications in the dynamics of this model
produce simplifications in the calculations of the Lyapunov exponents and KS
entropies of the moving particle, due primarily to the dispersing nature of the
collisions of the moving particle with the scatterers. A Lorentz gas in $d$
dimensions may have at most $d-1$ positive Lyapunov exponents. That is, the
phase space dimension for the moving particle is $2d$, the requirement of
constant energy removes one degree of freedom, and the direction in phase space
has a zero Lyapunov exponent associated with it, since two points on the same
trajectory will not separate or approach each other in the course of time. Thus
there can be at most $2d-2$ nonzero Lyapunov exponents, of which only $d-1$ can
be positive, since the exponents come in plus-minus pairs for this Hamiltonian
system.

To analyze the Lyapunov spectrum we consider two infinitesimally close
trajectories in phase space and follow the spatial and velocity deviation
vectors $\vdr,\vdv$ separating the two trajectories, in time. By arguments
almost  identical to (but simpler than) those presented in Section  II for the
case where all particles move, we have the following equations of motion for
the position, $\vec{r}$, velocity, $\vec{v}$ spatial deviation vector, $\vdr$,
and velocity deviation vector, $\vdv$, of the moving particle, between
collisions
\begin{eqnarray}
\dot{\vec{r}} & = & \vv, \nonumber \\
\dot{\vv} & = &  0,  \nonumber \\
\delta\dot{\vec r} & = & \vdv, \nonumber \\
\delta\dot{\vec v} & = & 0.
\label{lg1}
\end{eqnarray}
At a collision with a scatterer, these dynamical quantities change according to
\begin{eqnarray}
\vvr' & = & \vvr, \nonumber \\
\vv' & = & \vv-2(\vv\cdot\hat{\sigma})\hat{\sigma}, \nonumber \\
\vdr' & = & {\bf{M}}_{\hat{\sigma}}\cdot\vdr , \nonumber \\
\vdv' & = &  {\bf{M}}_{\hat{\sigma}}\cdot\vdv -2
{\bf{Q}}_{\hat\sigma}
\cdot\vdr .
\label{lg2}
\end{eqnarray}
Again, the primed variables denote values immediately after a collision. while
the unprimed  ones denote values immediately before the collision. The distance
from the center of the scatterer to the moving particle, at collision,  is
$a\hat{\sigma}$, and the tensor  ${\bf{Q}}_{\hat\sigma}$  is given by
\begin{equation}
{\bf{Q}}_{\hat\sigma}
= \frac{[(\hat\sigma\cdot\vec v)\identity
			+\hat\sigma\vec v]\cdot
			 [(\hat\sigma\cdot\vec v)\identity
			-\vec v\hat\sigma]
	}{a(\hat\sigma\cdot\vec v)}.
\label{lg3}
\end{equation}
It should be noted that if the velocity deviation vector  and the velocity
vector  are orthogonal before collision, then  they will be orthogonal after
collision as well. That is, if $\vv\cdot\vdv =0$, then $\vv'\cdot\vdv'=0$,
also. Thus we can take $\vdv$ to be perpendicular to $\vv$ for all time, and
without loss of generality, we can take $\vdr$ to be perpendicular to $\vv$ as
well. Of course, the condition that $\vv\cdot\vdv=0$ is simply the statement
that the two trajectories are on the same constant energy surface. We will
denote the spatial and velocity deviation vectors for the Lorentz gas with a
subscript $\perp$ to indicate that they are defined in a plane
perpendicular\footnote{Of  course,  the components  of $\vdv,\vdr$ in the
direction of $\vv$  are not related to the  non-zero Lyapunov exponents or the
KS entropy, since these components do not grow exponentially.} to $\vv$. It
follows that we may replace the $d\times d$ ROC matrix defined by
\begin{equation}
\vdr =\ro\cdot
\vdv
\end{equation}
by a $(d-1) \times (d-1)$ matrix $\ro_{\perp}$ defined by
\begin{equation}
\vdr_{\perp} = \ro_{\perp}\cdot\vdv_{\perp}.
\label{lg4}
\end{equation}
For the two dimensional case $\ro_{\perp}$ is a simple scalar which we denote
by $\rho$. One easily finds that between collisions $\rho$ grows with time as
\begin{equation}
\rho(t) = \rho(0) + t,
\label{lg5}
\end{equation}
The change in $\rho$ at collision satisfies the ``mirror" equation~\cite{sinai}
\begin{equation}
\frac{1}{\rho'}=\frac{1}{\rho} + \frac{2v}{a\cos\phi}.
\label{lg6}
\end{equation}
Here the primes denote values immediately after a collision, and $v$ is the
magnitude of the velocity of the particle. If the radius of curvature $\rho$ is
positive, initially, it will always remain positive, and  it also follows from
Eq. (\ref{lg6}) that the value of $v\rho$ after collision is less than half of
the radius of the scatterers. Consequently the radius of curvature typically
grows to be of the order of a mean free time, and it becomes much smaller
immediately after a collision with a scatterer. For three dimensional systems a
similar situation results. Now $\ro_{\perp}$ is a $2 \times 2$ matrix which
satisfies the free motion equation
\begin{equation}
\ro_{\perp}(t) = \ro_{\perp}(0) +
t\identity,
\label{lg7}
\end{equation}
and changes at collision according to
\begin{equation}
\left[ \ro'_{\perp}\right]^{-1} ={\bf{M}}_{\hat{\sigma}}\cdot\left\{
\left[\ro_{\perp}\right]^{-1} +\frac{2}{a}\left[\vv\hat{\sigma}+\hat{\sigma}\vv
-\frac{v^2\hat{\sigma}\hat{\sigma}}{(\vv\cdot\hat{\sigma})}-(\vv\cdot\hat{\sigma
})\identity\right]\right\}\cdot{\bf{M}}_{\hat{\sigma}}.
\label{lg8}
\end{equation}
Here the inverse radius of curvature matrices $[\ro'_{\perp}]^{-1}$ and
$[\ro_{\perp}]^{-1}$ are defined in planes perpendicular to $\vv'$ and to
$\vv$, respectively.  In the hybrid notation of Eq. (\ref{lg8}), in which both
$d\times d$  matrices and $d-1\times d-1$ matrices figure, the inverse matrices
in  the directions along $\vv'$ and $\vv$, respectively, may be defined by
$[\ro'_{\perp}]^{-1}\vv'=0$ and $[\ro_{\perp}]^{-1}\vv=0$. The final  matrix in
the right-hand side of Eq. (\ref{lg8}) can then be restriced to  the plane
perpendicular to $\vv'$ straightforwardly.

It is worth pointing out some important differences between the ROC matrices
defined here for the Lorentz gas and those defined earlier for the regular gas
of moving particles. Here the ROC matrices are defined in a subspace orthogonal
 to the velocity of the moving particle. Further the change in the matrix
elements at collision is from a typically large value on the order of a mean
free time to an always small value, on the order of the time it takes to move a
distance equal to half the radius of a scatterer. This latter property is a
property of dispersing billiards. For the regular gas case,  only a few of the
elements of the ROC matrices become small after a collision, which means that
one cannot find an accurate approximation to the ROC  matrices by considering
only one collision. This latter property is associated with  semi-dispersing
billiards where a reflection from  a scatterer does not change the  diagonal
components of the ROC matrix that correspond to the flat directions of the
scatterer, at all.

\subsection{Informal Calculation of the KS Entropy and Lyapunov Exponents for
the Dilute, Random Lorentz Gas}

Here we show that simple kinetic theory methods allow us to compute the
Lyapunov exponents and KS entropies of Lorentz gases in two and three
dimensions~\cite{vbld}. To do that we use methods similar to those in Section
III. That is, we consider the equations for the deviation vectors, Eqs.
(\ref{lg1} - \ref{lg3}) above.  The velocity deviation vector changes only upon
collision with a scatterer. We will base our calculation on the exponential
growth rate of the magnitude of the velocity deviation vector, and for three
dimensional systems, on the exponential growth rate of the volume element in
velocity space.

We begin by writing the spatial deviation vector just before collision as
\begin{equation}
\vdr = t\vdv + \vdr(0),
\label{lg9}
\end{equation}
where $\vdr(0)$ is the spatial deviation vector just after the previous
collision with a scatterer. This equation is essentially the same as Eq.
(\ref{ks9}), but now it is a good approximation to neglect the spatial
deviation vector vector $\vdr(0)$ since  $\vdr(0)$ is of relative order $a/vt$
compared to the term $t\vdv$, in all directions of $\vdr$, perpendicular to
$\vv$. Thus we neglect this term and insert Eq. (\ref{lg9}) into the last
equality of Eq. (\ref{lg2}) to obtain\footnote{In principle, the term
${\bf{M}}_{\hat{\sigma}}\cdot\vdv$ in  \Eq~(\ref{lg10}) can be neglected also,
but only for directions  perpendicular to the velocity. If one is careful to
consider only deviations $\vdv$ in the subspace perpendicular to the  velocity
of the particle, it is possible to carry out the calculation with  this term
neglected.}
\begin{equation}
\vdv' = {\bf{M}}_{\hat{\sigma}}\cdot\vdv -2t
{\bf{Q}}_{\hat\sigma}
\cdot\vdv \equiv
{\bf{a}}\cdot\vdv,
\label{lg10}
\end{equation}
where we have defined a  matrix ${\bf{a}}$ that gives the change in the
velocity deviation vector at collision. Then we can express the velocity
deviation vector at some time $t$ in terms of its initial value as
\begin{equation}
\vdv(t) = \ma_{\cal{N}}\cdot\ma_{{\cal{N}}-1}\cdots\ma_{1}\cdot\vdv(0),
\label{lg11}
\end{equation}
where we have labeled the successive collisions by the subscripts $1,2,...,
{\cal{N}}$.  We can determine the largest Lyapunov exponent by examining the
growth of the magnitude of the velocity deviation vector with time, and the KS
entropy as the growth of the volume element with time. Therefore, with the
approximations mentioned above,
\begin{eqnarray}
\lambda_{max}
& = & \lim_{t\rightarrow\infty}\frac{1}{t}\ln\frac{|\vdv(t)|}{|\vdv(0)|}
\nonumber \\
& = &
\lim_{t\rightarrow\infty}\frac{{\cal{N}}}{t}\frac{1}{{\cal{N}}}\sum_{1}^{{\cal{N
}}}\ln\frac{|\vdv_{i}^{+}|}
   {|\vdv_{i-1}^{+}|},
\label{lg12}
\end{eqnarray}
where $\vdv_{i}^{+}$ is the velocity deviation vector immediately after the
collision labeled by the subscript $i$. Similarly, the sum of the positive
Lyapunov exponents is given by
\begin{equation}
\sum_{\lambda_{i}>0}\lambda_{i} = h_{KS} =
\lim_{t\rightarrow\infty}\frac{{\cal{N}}}{t}\frac{1}{{\cal{N}}}\sum_{1}^{\cal{N}
}\ln|\det\ma_{i}|.
\label{lg13}
\end{equation}

To evaluate the sums appearing in Eqs. (\ref{lg12}, \ref{lg13}), we note that
to leading order in the density none of the  collisions are correlated with any
previous collision, that is, the leading contribution to the Lyapunov exponents
 comes from collision sequences where the moving particle does not encounter
the same scatterer more than once in the sequence \footnote{In two dimensions
the particle will hit the same scatterer an infinite number of times. However
the effects of such processes are of higher density, and can be neglected here
since the  number of collisions between successive collisions with the same
scatterer become typically very large as the density of scatterers approaches
zero.}.  Therefore we can treat each term in the sums in Eqs. (\ref{lg12},
\ref{lg13}) as being independent of the other terms  in the sum. We have
expressed $\lambda_{max}$ and the sum of the positive Lyapunov exponents as
arithmetic averages, but for long times and with independently distributed
terms in the average, we can replace the arithmetic averages by ensemble
averages over a suitable equilibrium ensemble. That is
\begin{equation}
\lambda_{max}=\nu \left<
\ln \left[\frac{|\delta\vec{v}^{+}|}{|\delta\vec{v}^{-}|}\right]\right>
\label{lg14}
\end{equation}
and
\begin{equation}
\sum_{\lambda_i >0}\lambda_i = \nu \left< \ln |\det{\bf a}|\right>
\label{lg15}
\end{equation}
where $\vdv^{-}$ and $\vdv^{+}$ are the velocity deviation vectors before and
after collision, respectively, $\nu$ is the (low  density)value of the
collision frequency, ${\cal N}/t$ as $t$ becomes large, and the angular
brackets denote an equilibrium average.

We now consider a typical collision of the moving particle with one of the
scatterers. The free time between one collision and the next is sampled from
the normalized equilibrium distribution of free  times\cite{chapcow}, $P(\tau)$
given at low densities by
\begin{equation}
P(\tau)=\nu e^{-\nu\tau}.
\label{lg16}
\end{equation}
The construction of the matrix ${\bf a}$ requires some geometry and depends on
the number of dimensions of the system. In any case we take the velocity vector
before collision, $\vec{v}$ to be directed along the $z$-axis, and take
$\hat{\sigma}\cdot\vec{v}=-v\cos\phi$, where $-\pi/2 \leq \phi \leq \pi/2$. The
velocity deviation before collision $\delta\vec{v}^-$  is perpendicular to the
$z$-axis. Then it is a simple matter to compute
$|\delta\vec{v}^{+}|/|\delta\vec{v}^-|$ and $|\det{\bf a}|$. For
two-dimensional systems $\delta \vec{v}$ and the matrix ${\bf a}$  are given in
this representation by\footnote{In contrast to the ROC matrices $\ro$, $\bf a$
is a $d\times d$ matrix.  If one chooses one of the basis vectors of $\bf a$
perpendicular to $\vv$, the remaining ones are the basis of the corresponding
$d-1\times d-1$  matrix, from which one can also obtain \Eq\ (\ref{lg18}), and,
in  the three dimensional case, \Eqs\ (\ref{lg19}) and (\ref{lg20}).}
\begin{equation}
\delta {\vec{v}}^-=\left(\begin{array}{c}1\\0  \end{array}\right)
|{\delta \vec{v}}^-| \, ; \ \ \ \
{\bf a}=\left( \begin{array}{cc}
(1+ \Lambda) \cos 2\phi
 & \sin2\phi
\\
(1+\Lambda) \sin2\phi &
-\cos2\phi
\end{array}\right),
\label{lg17}
\end{equation}
where we introduced $\Lambda = (2 v \tau)/(a \cos \phi)$. To leading order in
$v\tau/a$ we find that
\begin{equation}
\frac{|\delta\vec{v}^+|}{|\delta\vec{v}^-|}=
\Lambda; \ \ \ \ \ \ \ \ \ \ \ |\det{\bf a}|
=
\Lambda.
\label{lg18}
\end{equation}
For three dimensional systems the unit vector $\hat{\sigma}$ can be represented
as $\hat{\sigma}=-\cos\phi \; \hat{z}+\sin\phi \cos\alpha \; \hat{x} + \sin\phi
\sin\alpha \; \hat{y}$.  Now the  ranges of the angles $\phi$ and $\alpha$ are
$0 \leq \phi \leq \pi/2$ and  $0 \leq \alpha \leq 2\pi$. There is an additional
angle  $\psi$ in the $ x,y$ plane such that the  velocity deviation before
collision $\delta\vec{v}^-=|\delta\vec{v}^-|[\hat{x}\cos\psi +
\hat{y}\sin\psi]$. It is somewhat more convenient to use a symmetric matrix,
$\tilde{\bf{a}} = (\bf{1}-2\hat{\sigma}\hat{\sigma})\cdot\bf{a}$, given by
\begin{equation}
\tilde{\bf{a}}=
\left( \begin{array}{ccc}
 1+\Lambda(\cos^{2} \phi + \sin^{2} \phi \cos^{2} \alpha ) &
\Lambda \sin^{2} \phi \cos \alpha \sin \alpha & 0 \\
\Lambda \sin^{2} \phi \cos \alpha \sin \alpha & 1 + \Lambda (\cos^{2}
\phi + \sin^{2}\phi \sin^{2} \alpha) & 0 \\
0 & 0 & 1
\end{array} \right),
\label{lg18a}
\end{equation}
One easily finds
\begin{equation}
\frac{|\delta\vec{v}^+|}{|\delta\vec{v}^-|}=\frac{2\tau
v}{a}\left[\frac{\cos^2(\alpha-\psi)}{\cos^2\phi}+\sin^2(\alpha-\psi)
\cos^2\phi\right]^{1/2},
\label{lg19}
\end{equation}
and
\begin{equation}
|\det{\tilde{\bf a}}|=|\det{\bf a}|=\left(\frac{2v\tau}{a}\right)^2.
\label{lg20}
\end{equation}
to leading order in $v\tau/a$.

\begin{figure}[htb]
\centerline{\psfig{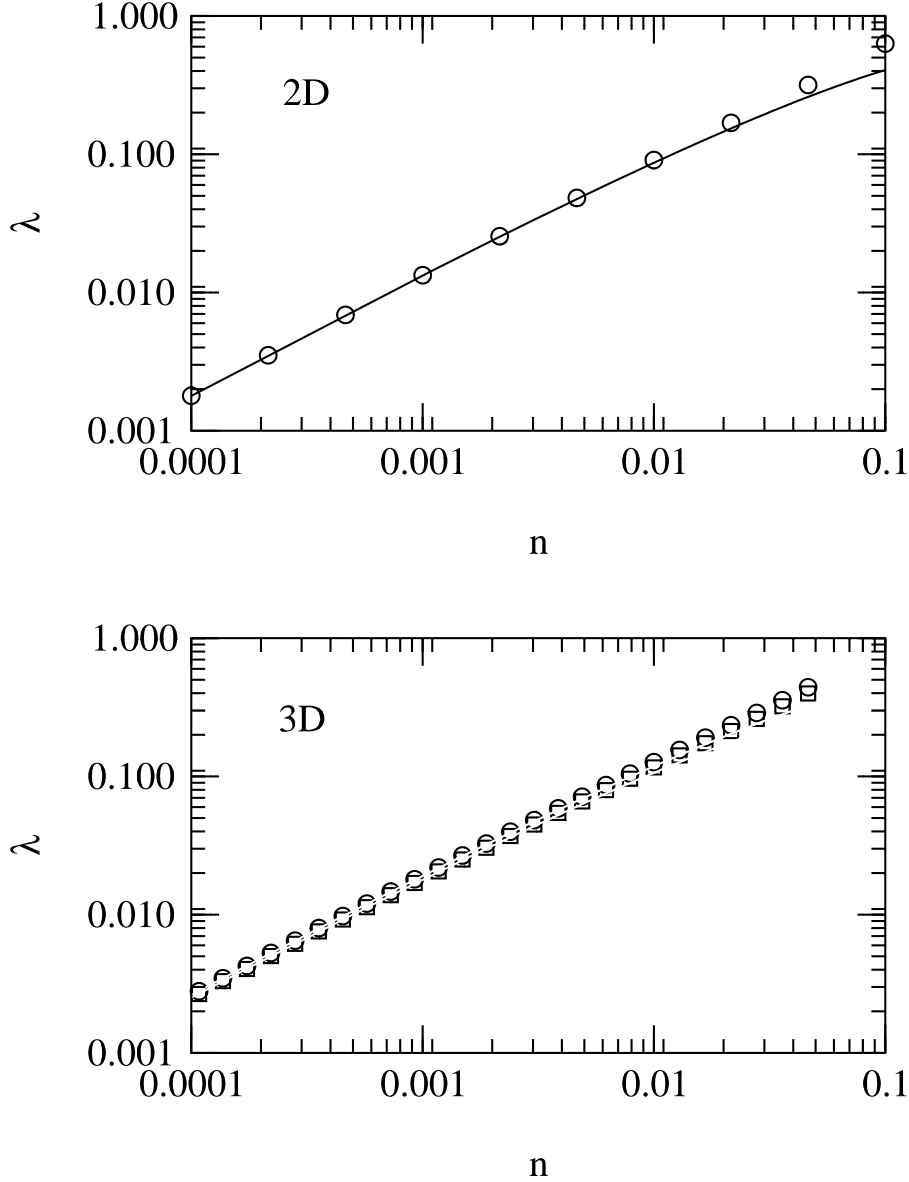}}
\caption{A plot of the Lyapunov exponents, in units of $v/a$, for the
moving particle in a random, dilute Lorentz gas in two dimensions (top)
and three dimensions (bottom), as functions of the density $n$, in units
of $a^{-d}$. The solid lines are the results given by kinetic theory,
\Eq\ (\ref{lg22}), respectively, \Eqs\ (\ref{lammax}) and (\ref{lg24}),
and the data points are the numerical results of Dellago and Posch.}
 \label{lya}
\end{figure}

To complete the calculation we must evaluate the averages appearing in Eqs.
(\ref{lg12}, \ref{lg13}). That is we average over the distribution of  free
times and over the rate at which scattering events are taking place with the
various scattering angles. Additionally in 3 dimensions an average over a
stationary distribution of angles $\psi$ has to be performed in general. Due to
the isotropy of the scattering geometry $\psi$ can here be  absorbed in a
redefinition $\alpha' =\alpha -\psi$ of the azimuthal angle  $\alpha$. This
will not be true any more if the  isotropy of velocity  space is broken (e.g.
by an external field).  The appropriate average of a quantity $F$ takes the
simple form
\begin{equation}
\left< F \right> = \frac{
1}{J}\int_0^\infty d\,\tau\int\,
d\hat{\sigma}
\cos\phi P(\tau)F,
\label{lg21}
\end{equation}
where $P(\tau)$ is the free time distribution given by Eq. (\ref{lg16}) and $J$
is a normalization factor obtained by setting $F=1$ in the numerator. The
integration over the unit vector $\hat{\sigma}$, i.e., over the appropriate
solid angle, ranges over $-\pi/2 \leq \phi \leq \pi/2$ in two dimensions and
over  $0 \leq \phi \leq \pi/2 $ and $ 0 \leq \alpha \leq 2\pi$ in three
dimensions. After carrying out the required integrations we find that
\begin{equation}
\lambda^+ = \lambda_{max} = 2nav[-\ln(2na^2) +1 -{\cal C}]+\cdots
,
\label{lg22}
\end{equation}
for two dimensions. Here ${\cal C}$ is Euler's constant, and the terms not
given explicitly in Eq. (\ref{lg22}) are higher order in the density.
Similarly, for the three dimensional Lorentz gas we obtain
\begin{equation} \label{lammax}
\lambda^{+}_{max} = na^{2}v\pi[-\ln(\tilde{n}/2) +\ln 2 -\frac{1}{2}-{\cal
C}]+\cdots,
\end{equation}
\begin{equation}
\lambda^{+}_{max}+\lambda^{+}_{min} = 2na^{2}v\pi[-\ln(\tilde{n}/2) -{\cal
C}]+\cdots
,
\label{lg23}
\end{equation}
from which it follows that
\begin{equation}
\lambda^{+}_{min}= na^{2}v\pi[-\ln(\tilde{n}/2) -\ln 2
+\frac{1}{2}-{\cal C}]+\cdots,
\label{lg24}
\end{equation}
where $\tilde{n}=na^3\pi$.  We have therefore determined the Lyapunov spectrum
for the equilibrium Lorentz gas at low densities in both two and three
dimensions  \cite{vbd,vbld}.  There is good agreement with simulations, as
shown in figure \ref{lya}. We note that the two positive Lyapunov exponents for
three  dimensions differ slightly, and that we were able to get individual
values because we could calculate the largest exponent and the sum of the two
exponents. We could not determine all of the Lyapunov exponents for a $d > 3$
dimensional Lorentz gas this way. Moreover, for a spatially inhomogeneous
system, such as those considered in the application of escape-rate methods, the
simple kinetic arguments used here are not sufficient and Boltzmann-type
methods are essential for the determination of the Lyapunov exponents and KS
entropies.

In Fig. (\ref{lya}) we illustrate the results obtained above for the Lyapunov
exponents of the dilute Lorentz gas in both two and three dimensions, as
functions of the reduced density of the scatterers, and  compare them with the
numerical simulations of Dellago and Posch~\cite{lorpos,dpef}. As one can see
the agreement is excellent.

\subsection{Formal Kinetic Theory for the Low Density Lorentz Gas}

The formal theory for the KS entropy of the regular gas  is easily applied to
the Lorentz gas, which is, of course, considerably simpler. Thus by following
the arguments leading to Eq. (\ref{ks6}) for the sum of the positive Lyapunov
exponents for the regular gas, we find that the KS entropy for the equilibrium
Lorentz gas is given by
\begin{eqnarray}
\sum_{\lambda_{i}>0}\lambda_{i} =  a^{d-1}\int dx\,d\ro \,d\vec{R}
d\hat{\sigma}\Theta(-\vv\cdot\hat{\sigma})|\vv\cdot\hat{\sigma}|
\delta(\vvr - \vec{R} -a\hat{\sigma}) \times \nonumber \\
\ln|\det\left[{\bf{M}}_{\hat{\sigma}}+2
{\bf{Q}}_{\hat\sigma}
\cdot\ro
\right]|{\cal{F}}_{2}(x,\vec{R},\ro).
\label{lg25}
\end{eqnarray}
Here $\vec{R}$ denotes the location of the scatterer with  which the moving
particle is colliding, and $\F_2$ is the pair distribution function for the
moving particle to have coordinate, $\vvr$, velocity, $\vv$, ROC matrix, $\ro$,
while the center of the scatterer is located  at $\vec{R}$. At low densities we
may assume that the moving particle and the scatterer are uncorrelated, so that
the density expansion for $\F_2$, immediately before collision  would have the
form
\begin{equation}
{\cal{F}}_{2}(x,\vec{R},\ro) = n\F_{1}(x,\ro) + \ldots,
\label{lg26}
\end{equation}
where $n$ is the number density of the scatterers and $\F_{1}(x,\ro)$ is the
equilibrium single particle distribution function for the moving particle.

We may easily construct an extended Lorentz-Boltzmann equation (ELBE) for
$\F_1$ along the lines used previously for the extended Boltzmann equation in
Section III. The ELBE is given by~\cite{vbd,vbld}
\begin{eqnarray}
\left[\frac{\partial}{\partial t} +{\cal{L}}_0\right]\F_1(x,\ro) =
a^{d-1}\int
d{\vec{R}}d\hat{\sigma}\Theta(-\vv\cdot{\hat{\sigma}})|(\vv\cdot\hat{\sigma})\times
\nonumber \\
\left[\delta(\vvr -\vec{R}-a\hat{\sigma})\int d\ro'\delta(\ro
-\ro(\ro'))
{\cal{P}}'_{\hat\sigma}
-\delta(\vvr
-\vec{R}+a\hat{\sigma})\right]\F_1(x,\ro).
\label{lg27}
\end{eqnarray}
Here the operator ${\cal{P}'_{\hat\sigma}}$  is a substitution operator that
replaces velocities, $\vv$ and ROC matrices, $\ro$, by their restitution
values, i.e., the values needed before collision to produce the values $\vv$,
and $\ro$ after collision with a scatterer with collision vector
$\hat{\sigma}$. Also, the free particle streaming operator on the left-hand
side of Eq. (\ref{lg27}) is given by
\begin{equation}
{\cal{L}}_0 = \vv\cdot\frac{\partial}{\partial \vvr} +\sum_{\alpha
=1}^{d}\frac{\partial}{\partial \ro_{\alpha\alpha}},
\label{lg28}
\end{equation}
since between  collisions, $\vvr$ varies as $\vvr(0)+t\vv$ and $\ro$ varies as
$\ro(0)+t\identity$.

Returning for the moment to Eq. (\ref{lg25}) for the KS entropy, we see by
evaluating the determinant in the  integrand on the right side, that not all
components of $\ro$ contribute. In fact, only the components of $\ro$ in the
plane perpendicular to $\vv$ contribute to $h_{KS}$. Here we will work out the
details of the calculation of $h_{KS}$ for the two dimensional case, leaving
the details of the three dimensional case to the literature~\cite{vbld}. For
$d=2$ we can easily evaluate the determinant in Eq. (\ref{lg25}), and find that
it is
\begin{equation}
|\det\left[{\bf{M}}_{\hat{\sigma}}+
2{\bf{Q}}_{\hat\sigma}
\cdot\ro \right]|
=1 +\left(\frac{2v\rho}{a\cos\phi}\right).
\label{lg29}
\end{equation}
Here the scalar $\rho$ is the component of the $\ro$ matrix given by
$\rho=\hat{v}_{\perp}\cdot\ro\cdot\hat{v}_{\perp}$,  where $\hat{v}_{\perp}$ is
a unit vector orthogonal to $\vv$. Moreover, for low densities we can use the
approximation $n\F_{1}(x,\ro)$ for $\F_2$. Then the expression for $h_{KS}$
becomes, at low densities
\begin{equation}
h_{KS}=an\int dx\,d\rho
d\hat{\sigma}\Theta(-\vv\cdot\hat{\sigma})|\vv\cdot\hat{\sigma}|
\ln\left[1 +\left(\frac{2v\rho}{a\cos\phi}\right)
\right]{\cal{F}}_{1}
(x,\rho),
\label{lg30}
\end{equation}
where $x=\vvr,\vv$, and we have to determine $F_{1}(x,\rho)$ as the solution of
the ELBE where we integrate over all components of $\ro$, except the one
diagonal component $\rho$. The ELBE then becomes, in the spatially homogeneous,
equilibrium case
\begin{eqnarray}
\frac{\partial}{\partial \rho}\F_{1}(\vvr,\vv,\rho)
=nav\int_{-\pi/2}^{\pi/2}d\phi \,\cos\phi \left[
\int_0^{\infty}d\rho'\delta
\Bigg
(\rho-\frac{a\cos\phi}{2v
+\frac{a\cos\phi}{v\rho'}}
\Bigg
)\F_{1}(\vvr,\vv',\rho')-\F_{1}(\vvr,\vv,\rho)
\right].
\label{lg31}
\end{eqnarray}
The argument of the $\delta$ function is simply obtained by using the mirror
formula given by Eq. (\ref{lg6}), but now the unprimed variable is the value
{\em after} collision, and the primed variable is the value of $\rho$ before
collision. A further, and useful simplification results from the observation
that $\rho'$ is typically of the order of the mean free time between
collisions, which, for low density, is much larger than $a/v$. Therefore the
delta function can be replaced by
\begin{equation}
\delta
\Bigg
(\rho-\frac{a\cos\phi}{2v+\frac{a\cos\phi}{v\rho'}}
\Bigg
)
\approx \delta
\left
(\rho-\frac{a\cos\phi}{2v}
\right
)
{}.
\label{lg32}
\end{equation}
In a spatially homogeneous and isotropic equilibrium state, $\F_1$ has to be a
function of the norm of the velocity vector $|\vv|$ and the radius of
curvature $\rho$ only. Using that the magnitude of the velocity, $|\vv|$,
always stays the same, we know that there is a solution of the form
\begin{equation}
\F_{1}(\vvr,\vv,\rho) =
\varphi(\vv)\psi(\rho),
\label{lg33}
\end{equation}
with $\varphi(\vv)=(2\pi V)^{-1}\delta(|\vv|-v_0)$  is the normalized
equilibrium spatial and velocity distribution function for the moving particle,
$v_0$ is its constant speed, and $V$ is the volume of the system. Now all we
have to do is to determine $\psi(\rho)$. An inspection of Eq. (\ref{lg31}),
with the approximation, Eq. (\ref{lg32}) shows that $\psi(\rho)$ satisfies the
equation
\begin{equation}
\frac{\partial \psi(\rho)}{\partial \rho} +2nav\psi(\rho) =0,
\label{lg34}
\end{equation}
for $\rho \geq a/(2v)$, with solution
\begin{equation}
\psi(\rho)= (1/t_0)e^{-t/t_0}\,\,\,\,{\rm for}\,\,\rho \geq a/(2v).
\label{lg35}
\end{equation}
Here $t_0$ is the mean free time given by $t_0=(2nav)^{-1}$. In the case that
$\rho < a/(2v)$, one can easily solve the full equation with the delta function
to find
\begin{equation}
\psi(\rho) =
(1/t_0)\left\{1-\left[1-\left(\frac{2v\rho}{a}\right)\right]^{1/2}\right\}\,\,\,
{\rm
for}\,\,\rho < a/(2v).
\label{lg36}
\end{equation}
Combining these results with Eq. (\ref{lg33}) and inserting them in
Eq.~(\ref{lg30}) we recover the result, Eq. (\ref{lg22}) for the KS entropy,
equivalently the positive Lyapunov exponent, for the low density, equilibrium,
random Lorentz gas. A similar, but somewhat more elaborate calculation can be
carried out, for the three dimensional case, to obtain exactly the same result
as obtained in the informal theory for  the KS entropy. To obtain the largest
Lyapunov exponent by more formal methods, one has to resort to methods for
determining the largest eigenvalue for products of random matrices. This is
well described in the literature~\cite{vbld} and we will not pursue this issue
further here.

\section{Conclusions and Open Problems}

In this article we have given a survey of the applications of the kinetic
theory of dilute, hard-ball gases to the calculation of quantities that
characterize the chaotic behavior of such systems. Results have been obtained
for the KS entropy per particle and for the largest Lyapunov exponent of dilute
hard-ball systems, as well as for the Lyapunov spectrum for the moving particle
in the dilute, random Lorentz gas, with nonoverlapping, fixed,  hard-ball
scatterers. All of these results are in good to excellent agreement with the
results of computer simulations. In the study of the largest Lyapunov exponent,
we have developed a very interesting clock model which seems to explain many
features of the behavior of this exponent. Moreover, the method for treating
the clock model reveals a deep and, perhaps unexpected, connection between the
theory of Lyapunov exponents and the theory of hydrodynamic fronts.

We emphasize that the results given here apply to a dilute gas in equilibrium,
but nonequilibrium situations have been treated by these methods as well. For
example, it is possible to calculate the Lyapunov spectrum for a dilute, random
Lorentz gas in a nonequilibrium steady state produced by a thermostatted
electric field, at least for small fields, and to obtain results that are in
excellent agreement with computer simulations~\cite{vbdcpd,vbldef,dpef}.
Calculations are currently underway for the largest Lyapunov exponent for a
hard-ball gas, when the gas is subjected to a thermostatted, external force
that maintains a steady shear flow in the gas. Furthermore, one can use kinetic
theory to calculate the Lyapunov spectrum for trajectories on the fractal
repeller for a Lorentz gas with open, absorbing boundaries~\cite{vbd,vbld2}.
Such results are useful for understanding escape-rate methods for relating
chaotic quantities for trajectories on a fractal repeller of an open system to
the transport properties of the system, as described by Gaspard and
Nicolis~\cite{gaspbook,jrdbook,gaspnic}. These results will eventually be
extended to hard-ball gases as well.

Of course, many problems remain to be solved. Here we mention some of the most
immediate problems: \begin{enumerate}

\item All of the results described here apply to hard-core systems. That is the
particles interact with a potential energy that is either zero, beyond a given
separation, or infinite, below that separation. It is worth studying the
properties of dilute systems with smoother potential energies for a number of
reasons: (a) The results of Rom-Kedar and Turaev~\cite{RomKedar} suggest that
the dynamics of particles with short ranged but smooth potentials may exhibit
regions of nonhyperbolic behavior. It is important to know more about these
regions and to assess their effect on the overall chaotic behavior of gases
interacting with such potentials. (b) We know very little about the chaotic
properties of dilute gases interacting with long range forces, such as Maxwell
molecules and Coulomb gases.

\item An open problem, even for dilute gases, is to obtain the complete
spectrum of Lyapunov exponents for a hard-ball gas. This is a very challenging
problem in mathematical physics, and no easy approach is in sight. There is a
very tantalizing set of numerical results by Posch and
coworkers~\cite{poschetal} showing that the smallest nonzero Lyapunov exponents
have a hydrodynamic structure, in that the  exponents themselves seem to scale
as the inverse of the linear size of the system, and that the spatial deviation
vectors seem to form collective modes of both transverse and longitudinal
types. Eckmann and Gat~\cite{gateck} have proposed an explanation of this
hydrodynamic-like behavior of the lowest Lyapunov modes using techniques from
the theory of random matrices. It would be useful to understand and to extend
their results using kinetic theory methods.

\item The extension of these results to gases at higher densities remains an
open, challenging problem. The kinetic theory of dense gases has exposed a
number of effects caused by long range dynamical correlations between the
particles. Such effects include nonanalytic terms in the density expansion of
transport coefficients and long time tail phenomena in time correlation
functions, among others~\cite{jrdhvb}. It is of some interest to see the effect
of these dynamical correlations on the chaotic properties of the gas, as well.
Furthermore, a high density hard-ball system  may form a glass, and there would
be useful information obtained about the glassy state if one could study the
chaotic behavior of the gas through the glass transition.

\item The extension of the clock model to higher density systems can also be
expected to reveal new and interesting phenomena connected to the effects of
density and other fluctuations in the gas upon the clock speed and related
quantities. At present there are some weak numerical indications~\cite{evsea}
that the largest Lyapunov exponent might diverge in the limit of large numbers
of particles as $\ln N$, where $N$ is the number of particles in the system. It
might be possible to confirm or to rule out this possibility by extending the
clock model so as to include the effects of fluctuations in the fluid.

\item Lyapunov exponents and KS entropies are not the only properties that
characterize chaotic systems. There are many more quantities, such as
topological pressures, fractal dimensions, etc., that remain to be explored by
the methods outlined here.

\item A subject of considerable interest and activity is the physics of gases
that make inelastic collisions, {\it i.e.} the physics of granular
materials~\cite{ernst}. One would like to know what the chaotic properties of
such systems might be, or, more generally, how to define such properties in
nonstationary systems.

\end{enumerate}

In conclusion, we have described in this article only the first steps, taken
over the last few years, to develop useful methods for the calculations of
Lyapunov exponents and KS entropies for the systems of particles that can be
treated by kinetic theory. We are particularly delighted that kinetic theory
has something to contribute to the field of the chaotic behavior of large
systems of particles, and that the ideas of Maxwell and Boltzmann are still
having new and fruitful applications.

Acknowledgements: The authors would like to thank  Professor Harald A.\ Posch,
Professor Christoph Dellago, and Dr.\ Arnulf Latz for many helpful
conversations,  and for collaborations on much of the research described here.
here;  We are also grateful to Professor Christoph Dellago and Professor Harald
A.\  Posch  for supplying Figs. \ref{hks2}, \ref{hks3} and \ref{lya}. H.\ v.\
B.\ and R.\ v.\ Z. are supported by FOM, SMC and by the NWO Priority Program
Non-Linear Systems, which are financially supported by the ``Nederlandse
Organisatie  voor Wetenschappelijk Onderzoek (NWO)". J.\ R.\ D.\ would like to
thank the National Science Foundation for support under grant PHY-9600428.

\end{document}